\documentclass[fleqn,10pt]{wlscirep}
\usepackage[utf8]{inputenc}
\usepackage[T1]{fontenc}
\usepackage[english]{babel}
\usepackage{subfig}
\usepackage{newlfont}
\usepackage{bibentry}
\usepackage{hyperref}
\usepackage{graphicx}
\usepackage{url}
\usepackage{amsmath}
\usepackage{array}
\usepackage{enumitem}
\usepackage{subfig}
\usepackage{mwe}
\usepackage{makecell}
\usepackage{listings}
\usepackage[justification=centering]{caption}
\usepackage{geometry}
\usepackage{tikz}
\usepackage{lipsum}
\usepackage{environ}

\title{A Multi-Modal Machine Learning Approach to Detect Extreme Rainfall Events in Sicily}

\author[1]{Eleonora Vitanza}
\author[1]{Giovanna Maria Dimitri}
\author[1,*]{Chiara Mocenni}
\affil[1]{Department of Information Engineering and Mathematics, University of Siena, Via Roma, 56, 53100, Siena (Italy)}
\affil[*]{To whom correspondence should be addressed; E-mail: chiara.mocenni@unisi.it.}


\begin{abstract}
In 2021 300 \textit{mm} of rain, nearly half the average annual rainfall, fell near Catania (Sicily island, Italy). Such events took place in just a few hours, with dramatic consequences on the environmental, social, economic, and health systems of the region. This is the reason why, detecting extreme rainfall events is a crucial prerequisite for planning actions able to reverse possibly intensified dramatic future scenarios. In this paper, the Affinity Propagation algorithm, a clustering algorithm grounded on machine learning, was applied, to the best of our knowledge, for the first time, to identify excess rain events in Sicily. This was possible by using a high-frequency, large dataset we collected, ranging from 2009 to 2021 which we named RSE (the Rainfall Sicily Extreme dataset). Weather indicators were then been employed to validate the results, thus confirming the presence of recent anomalous rainfall events in eastern Sicily. We believe that easy-to-use and multi-modal data science techniques, such as the one proposed in this study, could give rise to significant improvements in policy-making for successfully contrasting climate changes.
\end{abstract}
\begin{document}

\flushbottom
\maketitle
\NewEnviron{myequation}{%
    \begin{equation}
    \scalebox{0.86}{$\BODY$}
    \end{equation}
    }
\thispagestyle{empty}
\section*{Introduction}
Is it possible to detect extreme rainfall events areas by clustering spatio-temporal data?
The intensification of weather extremes, which is dramatically changing the climate scenario worldwide, is currently thought to be as one of the most important factors related to green-house effect and climate change \cite{trenberth2,lavell,karl,mitchell,bolin,gordon,moss,baker}. The increase in the frequency and intensity of daily temperatures has contributed to a widespread escalation of daily precipitation \cite{stott,cavicchia}. Moreover, severe weather and climate events, interacting with exposed and vulnerable human and natural systems, can lead to disasters which require an extraordinary adaptation ability \cite{lavell}. It is therefore mainly for this reason that, nowadays, the study of climate change is not only about temperature increase, but it also focuses on catastrophic rainfall extreme events and drought \cite{moss,jentsch}.
The concept of \textit{extreme} precipitation and its changes in response to warming are well described in \cite{pendergrass}. For this reason, the scientific community faces an increasing demand for regularly updated estimations of evolving climate conditions and extreme weather events \cite{trenberth2,jentsch,knapp}. Moreover, a correlation between changes in heavy precipitation and landslides in several regions has been found in \cite{lavell}. More specifically, it is possible to identify 3 examples of extreme weather events, that have raised the question of a potential link to climate change: more intense precipitation events, increased summer drying over most mid-latitude areas and increase in tropical cyclone peak winds intensities \cite{van}. These results show that rainfall extreme events are related to climate change and represent the triggers of a chain of reactions involving several human activities. The change in temperatures will, in fact, have serious long-term effects \cite{trenberth,lenton, sherwood}, although extreme rainfall events will also cause a short-term danger to the environment and the population. 
In more recent years several extreme events all over the world caused large losses of lives, as well as a tremendous increase in economic losses from weather hazards \cite{jongman}. Such disasters have forced public opinion to consider climate change as the main cause of these events \cite{easterling} and to deeply analyse the economic consequences of climate change in terms of investments and productivity \cite{stern, batten}. A relevant example of this regards wine industry. For instance, in the past two decades, Sicilian winemakers have enhanced the biological production of wine all around the island, especially on the slopes of Mount Etna. Although wine is not essential to human survival, it is an important product of human ingenuity and its economy is rapidly growing \cite{viana}. Agricultural activities depend on climate and are interconnected to weather changes. Any shift in climate and weather patterns may potentially affect the entire local wine industry \cite{mozell} and the stability of many crops, thus undermining the related economies \cite{batten}. Any shift in climate and weather patterns may potentially affect the entire local wine industry \cite{mozell}. Abnormal climate changes might also undermine the stability of crops and might be critical for the related economy \cite{batten}.  Considering all of these aspects, in Mediterranean areas, rainfall is probably the most important climatic variable due to its manifestation as a deficient resource (dryness) or a catastrophic agent, such as water bombs \cite{cannarozzo}. Therefore, many challenges arise during the measurement of the precipitation. For instance, \textit{in situ} measurements are especially affected by wind effects on the gauge catch, particularly for snow but also for light rain \cite{trenberth}. Moreover, to reduce this uncertainty, it is crucial to analyze spatio-temporal data in the most efficient way \cite{mocenni,atluri}.

In this regard, over the last decades scientists conducted several studies on rainfall time series. These studies investigated potential trends in different rainfall indicators, such as total and maximum annual precipitation and mean daily intensity \cite{arnone}. 
A tendency toward higher frequencies of heavy and extreme rainfalls emerged for some areas \cite{bonaccorso2016}. In most of these areas, an increase in total precipitation has also been observed, for instance in \cite{cannarozzo}, thanks to the analysis of 247 stations over the 1921-2000 period.
However, the correlation between the increase of total precipitation and extreme events is not always clear, as in other areas (i.e. Italy) several authors have observed an increase in heavy precipitation, together with a tendency towards a decrease in the total amount of precipitations \cite{brunetti}.  Among the studies mentioned, a few of them were specifically focused on the Mediterranean areas, given their peculiar climate, which is affected by interactions between mid-latitude and tropical processes, lying between the arid climate of North Africa and the temperate and rainy climate of central Europe. For these reasons, even relatively minor modifications of the general circulation can lead to substantial changes in the Mediterranean climate \cite{arnone, diodato}, including rainfall frequency \cite{gabriele}, thus making these areas vulnerable to climatic changes and in particular to  catastrophic precipitations.

In this setting, scientists analysed the region of Sicily to identify climate change signals, as for instance in \cite{maugeri}. In most of those studies, the authors analysed annual, seasonal and monthly rainfall data in the entire Sicilian region, showing a global reduction of total amount of annual rainfall \cite{maugeri}.  For example, in \cite{arnone} the annual maximum rainfall for fixed time duration  of 1, 3, 6, 12 and 24 h, and the daily rainfall series recorded from 1956 to 2005 in approximately 60 stations were analyzed using the non-parametric Mann–Kendall test \cite{lanzante,mcleod}.

Results of this study, confirmed an increasing trend for rainfall of short duration, in particular for the 1 hour rainfall length. On the other hand, time-persistent rainfalls exhibited a decreasing course \cite{lanzante,mcleod}. 
In particular, heavy–torrential precipitation have been reported to be more frequent at a regional scale, while light rainfall have shown negative trends at some sites.
In \cite{bonaccorso} the presence of linear and non-linear trends in 16 series from rain gauge stations, mostly placed in the eastern Sicily, was studied. The results indicated a different behaviour according to the time scale: for short duration, historical series generally presented increasing trends, that switched to decreasing for longer time courses.

A total of 67 sites of daily precipitation records over the 1951–1996 period in Italy were also analyzed in \cite{brunetti} considering seasonal and yearly total precipitation, number of wet days and precipitation intensity with the aim of evaluating the trends both from the single-station records, and for larger areas by using averaged series. Results showed that the trend for the number of wet days in the year was significantly negative throughout Italy, particularly stronger in the north than in the south, especially in  winter. A tendency towards an increase in precipitation intensity, which was globally less strong and significant than the decrease in the number of wet days was also found. 

In \cite{forestieri2015} the authors identified the presence of homogeneous areas over Sicily using the Regional Frequency Analysis (RFA), which is a procedure estimating the frequency of rare events at one site by using data from several sites  \cite{hosking}, used frequently in the analysis of environmental data \cite{sahu}. They also developed Principal Component Analysis (PCA) followed by a clustering analysis, performed by applying the K-Means method, to identify regional groups, starting from annual maximum series for rainfall duration of 1, 3, 6, 12 and 24 h over about 130 rain gauges.

One of the most interesting papers studying different rainfall time series in Sicily is \cite{bonaccorso2016}, where the authors investigated temporal changes in extreme rainfall by performing a regional study. In particular, a regional frequency analysis based
on L-moments approach \cite{noto} was applied to 1, 3, 6, 12 and 24 h annual maxima rainfall (AMR) series grouped per homogeneous regions, identified through a hierarchical cluster analysis \cite{murtagh}. Changes were investigated in a long-term dynamic (from 1928 to 2009) with special reference to the last forty years.
The study \cite{murtagh} detected an increasing trend on rainfall extreme events between 2003 and 2009 with several heavy localized storms all over Sicily and a remarkable tendency towards more intense storm events during the 2000’s affecting mainly the outer western part of the region. On the contrary, the increasing trend in extreme rainfall detected in eastern Sicily, have been considered only apparent, as related to a few severe local storms.\\

In our work we present a spacial and temporal clustering analysis on rainfall data over Sicily, performed by the Affinity Propagation clustering algorithm.
The research interest towards this topic lies in new alarming violent rainfall events occurred in East Sicily at the end of 2021 \cite{wp,nature}.
The main objective of the research is to introduce a machine learning methodology to detect extreme rainfall events. In particular, this work analyzes where extreme events took place in the region of Sicily, highlighting the differences and the similarities both between stations and years, i.e. in a spatio-temporal analysis. 
Differently from \cite{bonaccorso2016}, in our study clustering is not only used for identifying homogeneous sub-regions, but also to detect critical rainfall sites. Moreover, while in \cite{bonaccorso2016} the authors focused on finding long-term trends, we concentrated our attention on short-term changes between 2009 and 2021, analyzing high-frequency data, so as to obtain clusters specifically related to extreme events.

The present work faces several steps:
\begin{itemize}
    \item Consider the RSE (Rainfall Sicily Extreme) dataset, using originally recorded data, which allowed us to highlight different climate behaviours respect to previous studies.
    \item Apply clustering to the Sicilian data in order to cluster regions and detect \textit{extreme} sites according to rainfall data observations. To further validate the clusters obtained we defined rainfall indicators, which allowed us to validate and understand further the meaning of the obtained clusters. 
    \item Use a multi-modal approach to merge both geographical and temporal information gathered from the dataset.
    \item Highlight significant characteristics of rainfall distribution over Sicily, detecting an increasing trend on extreme events in East Sicily, that reinforces the results of the state of the art in \cite{bonaccorso2016}. 
\end{itemize}

According to this scheme, the paper is structured as follows: in Section 1 the regional dataset used in the analysis, including the data pre-processing, is presented. In Section 2 the methods applied in the study, in particular the adopted clustering algorithm, and the statistical validation methods, are introduced. In Section 3, we describe the application of methods to the considered rainfall datasets. In Section 4 we report the discussion of the results, concerning each analyzed variable, and the most relevant conclusions drawn. Furthermore, we report in the supplementary material the analysis concerning the annual histograms of specific rain gauges and local data plots at different levels, as well as the complete annual clustering results. 

\section{RSE: the Rainfall Sicily Extreme dataset}\label{sec.dat}
The dataset used in this analysis consists in geographical rainfall records with a 10 minutes periodicity from 2009 to 2021, provided by SIAS, the Servizio Informativo Agrometeorologico Siciliano \cite{sias}. The dataset together with the code is available at the following \href{https://github.com/elevitanz/Extreme_Events_Sicily.git}{GitHub Repository} \cite{github}.

The most common rainfall measurement gathered from the database is the number of millimeters ($mm$) of rain in a given period. Accordingly, six collections were considered, as described in Table \ref{tab:d}. C.$A$ and C.$B$ contain 13 datasets per station - one per year - with the original data and the weekly mean data, respectively. C.$C$ and C.$D$ include one full dataset per station - involving all the records from 2009 to 2021 - with the original data and the weekly mean data, respectively. C.$A_{s}$ and C.$B_{s}$ are subsets of C.$A$ and C.$B$, respectively, since one station per time is considered, so that each of them includes 13 datasets. 
\begin{center}
\begin{tabular}{|m{1cm}|m{3.6cm}|m{2.15cm}|m{1.85cm}|m{2.25cm}|}
\hline
\textbf{\makecell{Name}} & \textbf{\makecell{Description}} & \textbf{\makecell{\# Datasets\\per station}} & \textbf{\makecell{\# Total\\datasets}} & \textbf{\makecell{\# Records\\per dataset}} \\
\hline

\makecell{C.$A$} & \makecell{Annual\\ Collection} & \makecell{13} & \makecell{442} & \makecell{52560*}\\   
\hline
\makecell{C.$B$} & \makecell{Annual Collection\\Weekly mean} & \makecell{13}& \makecell{442} & \makecell{53}  \\
\hline

\makecell{C.$C$} & \makecell{Full\\ Collection} & \makecell{1}& \makecell{34} & \makecell{683713} \\
\hline
\makecell{C.$D$} & 
\makecell{Full Collection\\Weekly mean} & \makecell{1}& \makecell{34}& \makecell{679} \\
\hline

\makecell{C.$A_{s}$} & \makecell{Single stations\\ Collection} & \makecell{13} & \makecell{13} & \makecell{52560*}\\ 
\hline

\makecell{C.$B_{s}$} & \makecell{Single stations Collection\\Weekly mean} & \makecell{13}& \makecell{13} & \makecell{53}  \\
\hline
\end{tabular}
\captionof{table}[Dataset Collections.]{Dataset Collections. The number of considered stations is 34, except for the Single stations Collections. * 52704 for leap years.}\label{tab:d}
\end{center}

\subsection*{Data preprocessing}\label{sec.prep}
We will now describe the initial data selection process, obtained through the analysis of annual data.
On the basis of an initial graphical analysis reported in the SI document, we decided to select the most extreme stations. A station is considered \textit{extreme} if it is possible to observe a high amount of rain in a relatively short time interval. We implemented this concept of "extremeness" using the following strategy. 

First, we considered the following data for all the 96 available stations in Sicily and for all the years:
\begin{itemize}
    \item The total annual precipitation in $mm$ ($tot$).
    \item The percentage of rainy days over the year ($rd$). This measure indicates the percentage of days with more than 1 $mm$ of rain.
    \item The $mm$ of rain during the rainiest day in the year ($dmax$).
\end{itemize}

Afterwards, a selection strategy has been applied. Extreme rainfall events are generally characterized by the increasing of either drought and/or excessive wetness \cite{cannarozzo}. The logical rule below highlights precisely such characteristics:

\begin{itemize}
    \item[1)] Fix a station.
    \item[2)] Compute $\mu_{1}$: the mean over years of the $rd$ annual indicator.
    \item[3)] Compute $\mu_{2}$: the mean over years of the $dmax$ annual indicator.
    \item[4)] Fix a year $y$.
    \item[5)] If the $rd$ value in the year $y$ is less than $\mu_{1}$ and the $dmax$ value in the year $y$ is grater than $\mu_{2}$, then the year $y$ is considered as \textit{extreme}. Otherwise no.
\end{itemize}

\begin{figure}[h!]
    \centering
    \includegraphics[width=.68\textwidth]{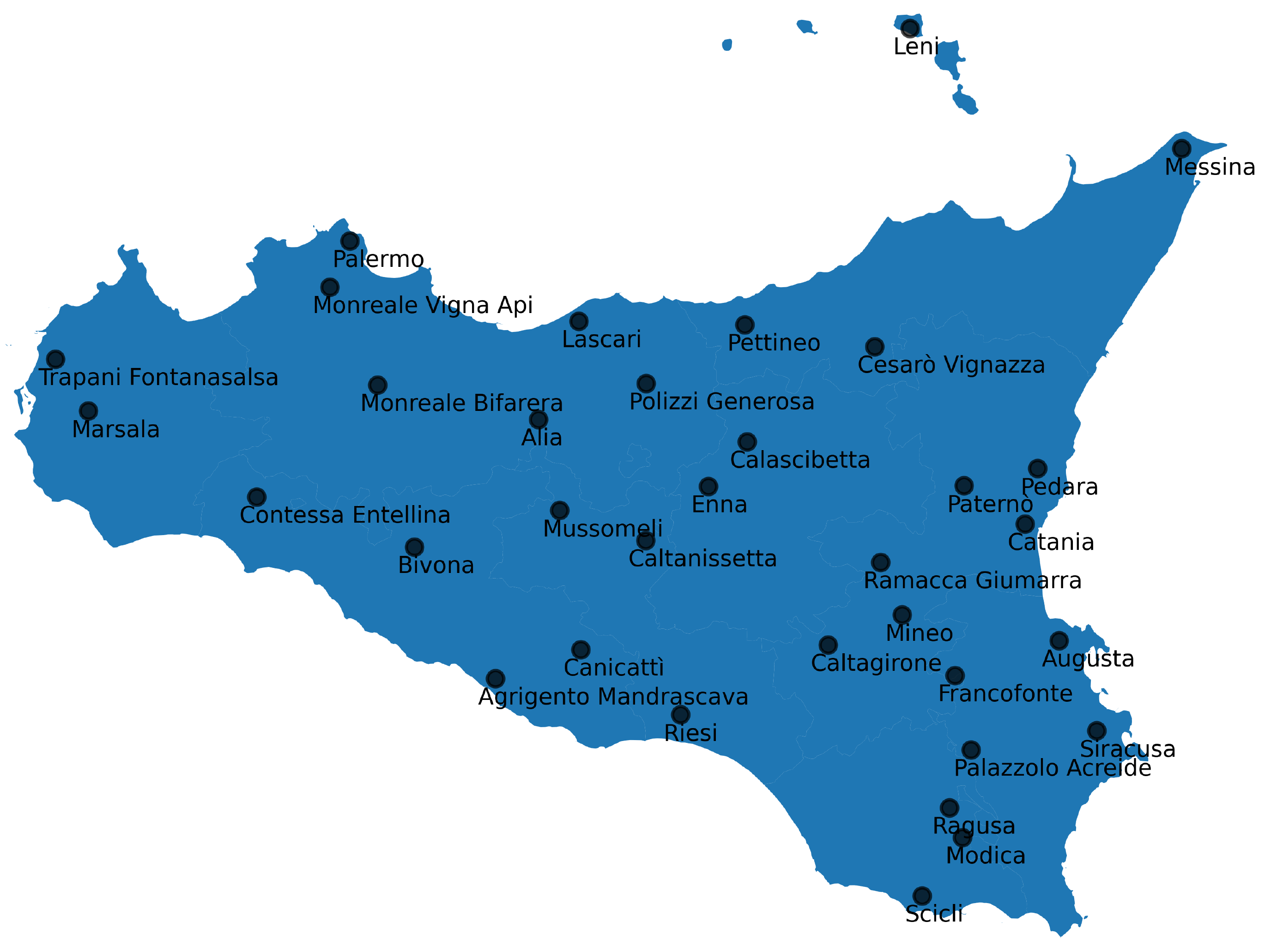}
    \caption{Location of rainfall gauging stations in Sicily.}
    \label{fig:sicily}
\end{figure}

Since the procedure works year by year, we selected the stations satisfying the \textit{extreme events} detection rule for at least 3 years (the stations respecting this condition for at least one year were 85 out of 96, almost all). In this way, we obtained 32 stations out of the 96 rain gauges. Furthermore, we decided to include all of the provincial capitals in the region, thus obtaining the 34 stations shown in Fig. \ref{fig:sicily}.

After the selection, we observed rainfall data time series, by fixing a station and using full, annual, and monthly data plots, as well as mean data graphics (all details regarding these initial observations are reported in the SI document). This preliminary analysis lead to different reasoning. The full plots proved the necessity of quantifying and understanding variation in the stations time series behavior. In contrast, the annual plots showed a typical seasonality pattern. Moreover, the graphics observation led to the idea of comparing annual time series. Finally, a similar reasoning has been done with regard to the monthly view. 

All of the above considerations suggested us to highlight the differences and the similarities both among stations and years, in order to identify multi-modal (geographical and historical) rainfall changes. Instead of performing classical time series analysis, we proceeded by applying the suitable clustering algorithms described in the following sections. 

\section{Methods}
Clustering is an unsupervised machine learning methodology \cite{kaufman, madhulatha}. Its goal is to detect groups of observations sharing similar characteristics. More precisely, it consists in the partitioning of a dataset into subsets, so that the data in each subset are characterized by a higher similarity than elements in different sets, according to some defined distance measure.

Two main types of clustering techniques can be defined: methods in which the number of clusters needs to be established a priori, as, for instance, the K-Means algorithm \cite{likas}, and algorithms in which, instead, the number of clusters is inherently estimated during the optimization phase, such as the Affinity Propagation \cite{frey}. The last one has been used in this work, since no prior knowledge on the number of expected clusters was available.

\subsection*{Affinity Propagation algorithm}\label{sec.ap}
Affinity Propagation (AP), introduced by Frey and Dueck in 2007 \cite{frey}, and its extension to Hierarchical Affinity Propagation \cite{givoni}, are nowadays becoming extremely popular due to their simplicity, general applicability, and performance.

AP takes as input the measures of similarity between pairs of data points, and simultaneously considers all of them as potential exemplars. The number of clusters does not need to be defined in advance, indeed the algorithm is based on the hypothesis that the so called "real-valued messages" are exchanged between data points until a high-quality set of exemplars, together with the corresponding clusters, gradually emerges.
Given that no assumption on the number of clusters was requested in our case, AP has been a natural choice.\\

The algorithm requires two inputs parameters \cite{frey}:
\begin{itemize}
    \item Similarities $s(i,k)$ between data points, representing how similar a point is to be another one’s exemplar. If there is no similarity between two points, as in this case they cannot belong to the same cluster, this similarity can be omitted or set to -$\infty$ depending on the implementation.
    \item Preferences $s(k,k)$, indicating each data point’s suitability to be an exemplar. Since some prior information which points could be favored for being an exemplar can be available, it can be represented through preferences.
\end{itemize}

Similarity is usually defined starting from the negative Euclidean distance or the Pearson correlation coefficient, depending on the considered situation.

If all data points are supposed to be equally suitable as exemplars, the preferences should be set to a common value, such as for example the median of the input similarities, thus resulting in a moderate number of clusters, or their minimum, thus resulting in a small number of clusters \cite{frey}. In this work we initialized the preferences to the median and the availabilities to zero, $a(i,k) = 0$. 
Each iteration step of the optimization performance is composed by 2 main message-passing steps:
\begin{itemize}
    \item[1.] Computing responsibilities: 
    \begin{equation}
    r(i, k) \leftarrow s(i, k)-\max _{k^{\prime} \text { s.t. } k^{\prime} \neq k}\left\{a\left(i, k^{\prime}\right)+s\left(i, k^{\prime}\right)\right\},
    \end{equation}
    where $s(i,k)$ and $s(i,k')$ are similarities, while $a(i,k')$ are availabilities.
    \item[2.] Computing availabilities
    \begin{myequation}
    a(i, k) \leftarrow \min \left\{0, r(k, k)+\sum_{i' \text { s.t. } i' \notin\{i, k\}} \max \{0, r(i', k)\}\right\},
    \end{myequation}
    where $r(k,k)$ are the self-responsibilities, while $r(i',k)$ are general responsibilities. To limit the influence of strong incoming positive responsibilities, the total sum is lower bounded, so that it cannot be negative.
\end{itemize}
 
The “self-availability”, $a(k,k)$ is updated differently, as follows:
\begin{equation}
a(k, k) \leftarrow \sum_{i' \text { s.t } i' \neq k} \max \{0, r(i', k)\}.
\end{equation}

The way for calculating how suitable a point is for being an exemplar is that it is favored more if the initial preference was higher, but the responsibility gets lower when there is a similar point that considers it as a good candidate, so there is a ‘competition’ between the two, until one of the two options is chosen in some iteration. The above procedure may be terminated after a fixed number of iterations, after changes in the messages fall below a threshold, or after the local decisions stay constant for a given number of iterations \cite{frey}.

\subsection*{Statistical validation}
To assess the presence of statistical differences between 2 communities we made use of the well known Kruskal-Wallis test \cite{belouafa,bhatta,miller}.
\subsubsection*{Kruskal-Wallis test}
It is a non parametric statistical test that assesses the differences among three or more independently sampled groups \cite{mckight}. Kruskal-Wallis test is used to determine whether or not there is a statistically significant difference between the medians of three or more independent groups. It does not assume normality in the data and is much less sensitive to outliers than the standard analysis of variance (ANOVA) \cite{hecke}. The test is based on the null hypothesis $H_{0}$ \cite{ostertagova}, which allows one to state whether the considered samples are realizations of identical populations. The application of the test returns a p-value which confirms or rejects the null hypothesis. If $p < 0.05$, then the null hypothesis is rejected, on the contrary, if $p \geq 0.05$, then the null hypothesis is confirmed  \cite{hecke}. The related p-value for the test is computed using the assumption that $H$ has a $\chi^{2}$ distribution.

\subsection*{Experiments}\label{sec.app}
This section explains the experimental procedure followed to design and apply the clustering algorithm. Global and local clustering analysis have been preformed by means of high frequency (measurements collected every 10 minutes) and weekly averaged data. The reason of this choice lied in the need to reduce the dataset dimension for minimizing sensitivity to outliers and oscillations in the original time series of observations.

Two main streams of experiments were performed: 
\begin{enumerate}
    \item \textbf{Geographical (or spacial) clustering}: it consists of grouping similar geographical stations together along different time horizons.
    \item \textbf{Local (or temporal) clustering}: it consists of grouping similar years together on each single location.
\end{enumerate}

For the first category, we ran the algorithm four times, according to the four Collections of datasets C.$A$, C.$B$, C.$C$ and C.$D$ described in Table \ref{tab:d}.
In contrast, the second category involves C.$A_{s}$ and C.$B_{s}$ of Table \ref{tab:d}.
The Affinity Propagation algorithm has been implemented in Python programming language (version 3), making use of with the Scikit-learn library (V. 1.0.2), which is a free software machine learning library for Python \cite{sklearn}, designed to inter-operate with the Python numerical and scientific libraries NumPy (V. 1.21.4) \cite{numpy}, SciPy (V. 1.8.0) \cite{scipy} and Pandas (V. 1.3.5) \cite{pandas}. The algorithm was implemented using default hyper-parameters. For instance, \textit{convergence\_iter} was set to $15$, that is the number of iterations with no change in the number of estimated clusters, that stops the convergence. Moreover, the \textit{preference} value was set to the median of the input similarities. 

\subsubsection*{Metrics for Affinity Propagation}
Two different metrics of similarities were used in the Affinity Propagation algorithm: the Euclidean Metrics and the Correlation Metrics. We present in the following subsections the results obtained for both. 

\paragraph{Euclidean Metrics}\label{eucl}
We first conducted clustering using the Euclidean \textit{affinity} metric, which resulted in a principal large cluster and few smaller communities consisting in one element each. For this reason, we decided to apply an iterated version of the AP algorithm in order to detect new geographical clusters, at first glance hidden by the anomalies. To this aim, the AP algorithm based on a particular multi-step structure was implemented as follows:
\begin{itemize}
    \item[1)] AP is applied to the whole considered collection of datasets.
    \item[2)] The exceptions found at level one from the data are removed, and the AP algorithm reiterated over the remaining datasets.
    \item[3)] The process is repeated from Step 1.
\end{itemize}

\paragraph{Correlation Metrics}
On the basis of the theoretical arguments reported in the \textit{Affinity Propagation algorithm} subsection, the Correlation distance was also chosen as a \textit{affinity} metric for a second exploration analysis. In this case no multi-step procedure was needed.

\subsubsection*{Clusters validation procedure}\label{sec.cvp}
In order to understand the rainfall phenomena that mostly characterize the clusters, several rainfall indicators over the time series were introduced, as reported in Table \ref{tab:inidicators}.

We assembled the original 10 minutes records according to specific needs: naturally an \textit{hour} data includes six consecutive records summed up together, whereas a \textit{day} consists of the sum of 144 consecutive data.
We also computed the total number of rainy hours, where one hour is considered "rainy" if its amount of rain is higher than zero. Hence, some of the introduced indicators are the percentages of light ($l$), moderate ($m$), and heavy ($h$) rainy hours over the total. Moreover, we considered the absolute number of violent rainy hours \textit{v}, which is not expressed in percentage since it represents very rare events.

\begin{center}
\begin{tabular}{ | m{1cm}| m{4.07cm}| m{6.5cm}| }
\hline
\makecell{Variable} & \makecell{Indicator} & \makecell{Description}  \\ 
\hline
$wh$ & Wet hours (\%) & percentage of rainy hours over the total number of hours. \\
\hline
$mh$ & Maximum per hour & maximum amount of rain of the data series grouped by hours. \\ 
\hline
$i$ & Intensity (mm/h) & quotient between the total amount of rain and the number of wet hours. \\ 
\hline
$t$ & Total rain & total amount of rain in the time series. \\
\hline
$mv$ & Maximum daily variation & maximum rainfall variation between two consecutive days over the total time series. \\
\hline
$wd$ & Wet days (\%) & percentage of rainy days over the total number of days. \\ 
\hline
$md$ & Maximum per day & maximum amount of rain of the data series grouped by days. \\
\hline
$l$ & Light rain (\%) & percentage of light (0 - 2.5 mm) rainy hours over the total number of rainy hours. \\
\hline
$m$ & Moderate rain (\%) & percentage of moderate (2.6 - 7.5 mm) rainy hours over the total number of rainy hours. \\ 
\hline
$h$ & Heavy rain (\%) & percentage of heavy (7.6 - 50 mm) rainy hours over the total number of rainy hours. \\ 
\hline
$v$ & Violent rain & number of violent ($>50$ mm) rainy hours in the time series. It was not reported in percentage since it represented very rare events. \\
\hline
\end{tabular}
\captionof{table}{Description of the Indicators.}\label{tab:inidicators}
\end{center}

The Kruskal-Wallis test was then applied to the indicators, in order to understand which of them better characterize clusters. To this aim, the SciPy scientific library has been used. Indeed, it provides algorithms for many classes of problems, extends standard tools of array computing, wraps up highly-optimized implementations, is easy to use, and enlarges NumPy \cite{scipy}. 

The Kruskal Wallis test was applied to each indicator and to all the experiments described above, according to the following logical evaluation steps:

\begin{itemize}
    \item[1)] Fix an indicator $i$.
    \item[2)] Run the clustering algorithm.
    \item[3)] Create an array $k$ with one element for cluster. Every element of $k$ is in turn an array $a_{c}$, containing the indicator values of the stations belonging to that cluster.
    \item[4)] Run the Kruskal-Wallis test on $k$.
    \item[5)] If the p-value is less than $0.05$: $i$ is considered as characterizing for the clusters. Otherwise no.
\end{itemize}

\section{Results and discussion}\label{sec.5}
In this section we report the main results obtained from the study for both geographical and single station investigations. Additional details on the whole dataset results are reported in the supplementary information document.
\subsection*{Geographical investigation}
Results of this set of experiments are visualized in the Sicily map of Fig. \ref{fig:full_1}, where the 34 stations with names or symbols coloured according to their relative clusters are drawn. When the multi-step version of the algorithm is applied, different shapes for the points are used. Specifically, circle, squares and diamond markers represent clusters resulting from the first, second, and third iterations, respectively.
\subsubsection*{Annual clustering}\label{annual}
The results of the geographical clustering year by year for C.$A$ and C.$B$ (Table \ref{tab:d}), both with the Euclidean and Correlation similarities are included in the SI document. Additionally, a video showing the clustering results proceeding in years is available for each collection. We hereby report the main results drawn from the several performed experiments: 
\begin{itemize}
    \item Euclidean metrics - C.$A$ (\href{https://drive.google.com/file/d/1z7sIRxP5v9IvlRcwzQWj00Nf_veFax7P/view?usp=sharing}{Video\_1}): in this case the annual results consist mostly of a principal cluster (at most two) and some anomalous stations. Proceeding in the years, a flow in anomalies that goes from western to eastern Sicily is detectable. We believe that anomalous clusters seem to be more susceptible to extreme events. In fact, since them change drastically in intensity according to the specific location, this legitimizes the algorithm in finding exceptions, namely only one station in one cluster. In order to validate this observation, a case by case analysis has been carried out and reported in the SI document for the year 2021.
    \item Euclidean metrics - C.$B$ (\href{https://drive.google.com/file/d/1zItompeu_qW7SvQ7_y7O__vpRvfRxJX_/view?usp=sharing}{Video\_2}): the reduction in the dataset size led the clusters to be more uniform and referred to geographical divisions. However, there are  some exceptions, mainly in the South-East Sicily, and in the neighbourhood of \textit{Palermo}. As in the previous case, this represents a trend on extreme events, more diffused in the East side of the island.
    \item Correlation metrics - C.$A$ (\href{https://drive.google.com/file/d/1Hqo7mUFFvDggzgazpjsVSddYTwdwNO20/view?usp=sharing}{Video\_3}): in this case a geographical clustering pattern was obtained, identifying eastern and western Sicily. This is coherent with the fact that the Correlation metrics finds shape similarities and it is less sensitive to the micro-climatic differences.
    \item Correlation metrics - C.$B$ (\href{https://drive.google.com/file/d/1Hqo7mUFFvDggzgazpjsVSddYTwdwNO20/view?usp=sharing}{Video\_4}): here the combination between dimensionality reduction and correlation metrics brings to a rough splitting of the island. The number of clusters does not exceed 3 and often very far away stations are grouped together in the same cluster.
\end{itemize}

The results of the 4 settings for the year 2021 are in line with our initial research hypothesis, for which the anomalies correspond to \textit{extreme} stations. This behaviour was further confirmed by the Kruskal-Wallis test, as reported in the SI document.

Moreover, East Sicily emerges as the most \textit{extreme} zone of the island, confirming the real occurred events reported in \cite{wp} and discussed in \cite{nature}.

Finally, \href{https://drive.google.com/file/d/1zItompeu_qW7SvQ7_y7O__vpRvfRxJX_/view?usp=sharing}{Video\_2} shows that the use of weekly averaged data (C.$B$) reduces noise, thus giving rise to balanced presence of both anomalies and territorial clusters in the Euclidean case. On the other hand, \href{https://drive.google.com/file/d/1Hqo7mUFFvDggzgazpjsVSddYTwdwNO20/view?usp=sharing}{Video\_4} shows that C$B$, differently from C.$A$, does not provide clear geographical splitting in the Correlation case.

In conclusion, in the annual case the use of Euclidean metrics led to detect the anomalies, while in contrast, the use of the Correlation metrics as a similarity measure allowed us to identify more uniform clusters.

\subsubsection*{Full clustering}
\begin{figure*}[h!]
\begin{minipage}[c]{.5\linewidth}
\centering
\subfloat[]{\label{fig:c_1}\includegraphics[width = 7.5cm]{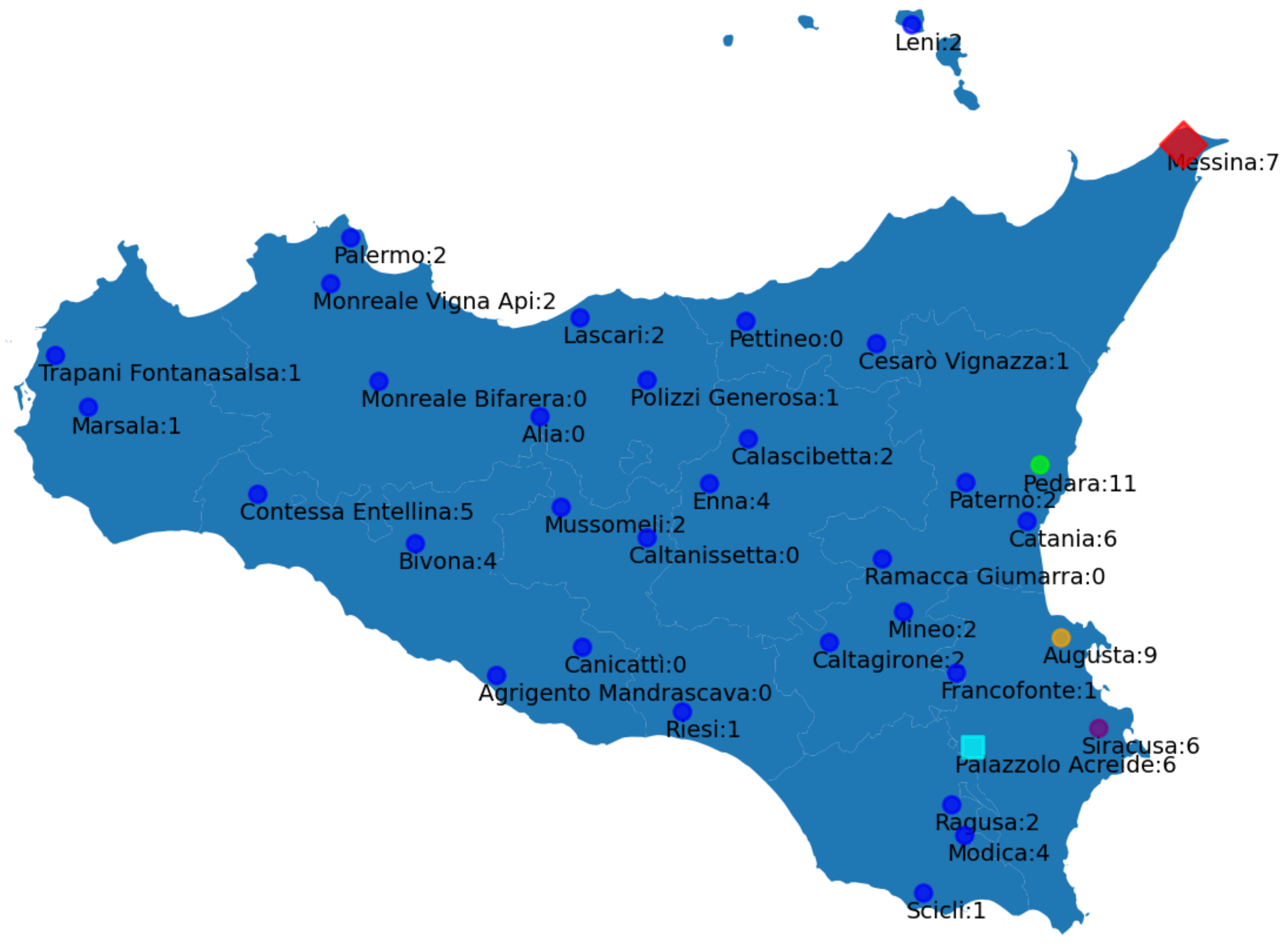}}
\end{minipage}
\begin{minipage}[c]{.5\linewidth}
\centering
\subfloat[]{\label{fig:best}\includegraphics[width = 7.5cm]{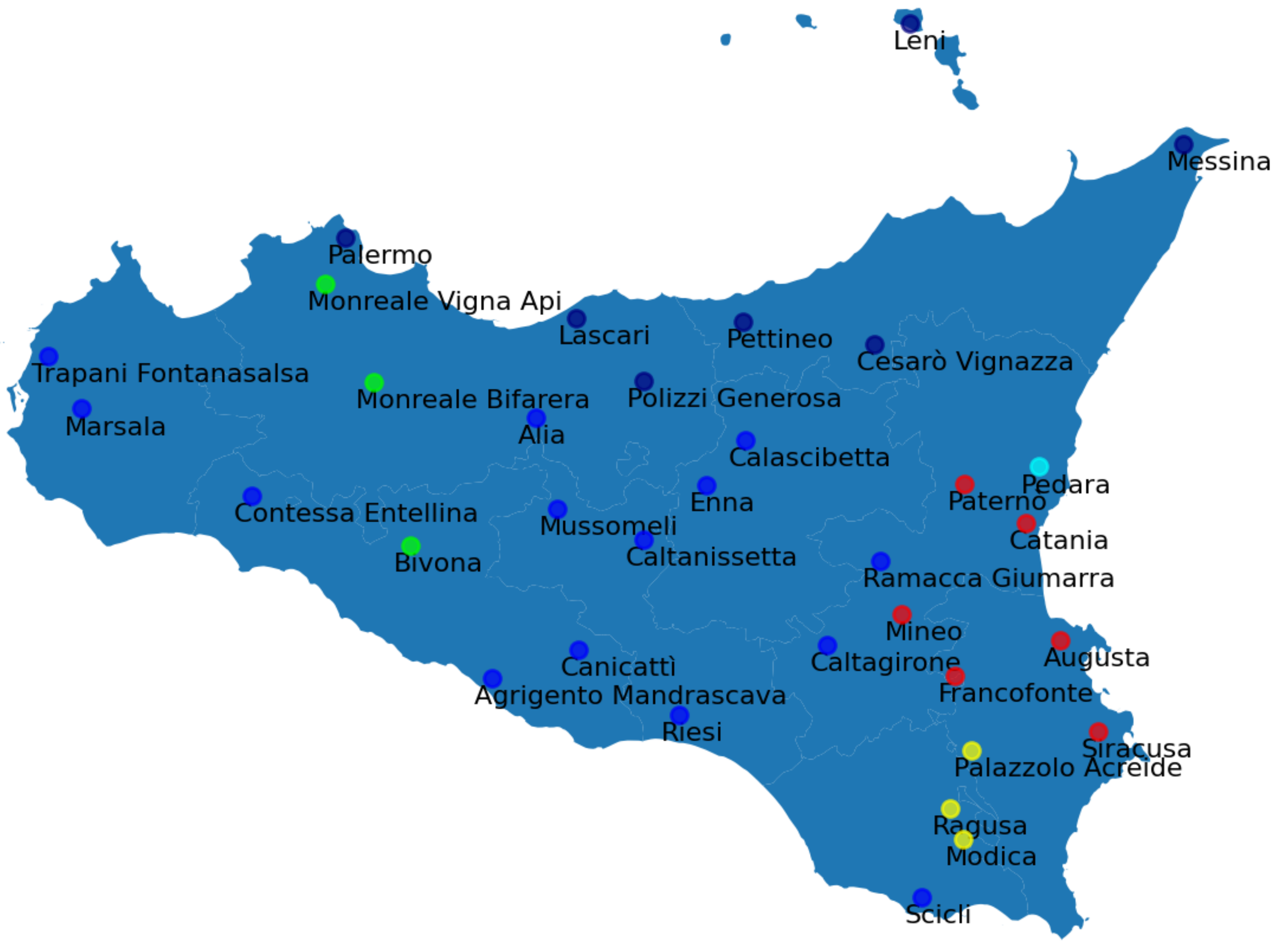}}
\end{minipage}\par
\begin{minipage}[c]{.5\linewidth}
\subfloat[]{\label{main:c}\includegraphics[width = 9.2cm]{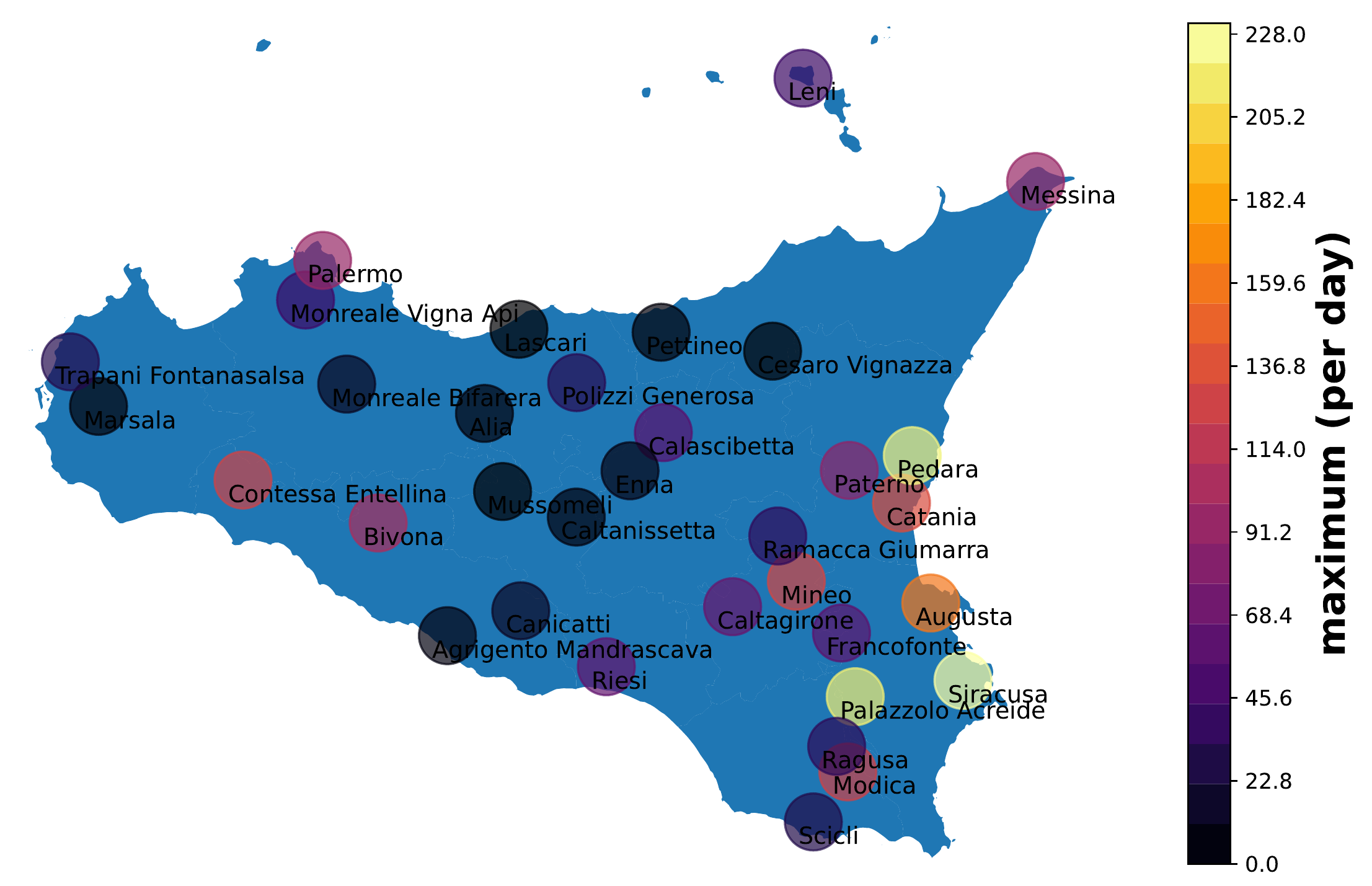}}\\
\end{minipage}
\begin{minipage}[c]{.5\linewidth}
\subfloat[]{\label{main:d}\includegraphics[width = 9.2cm]{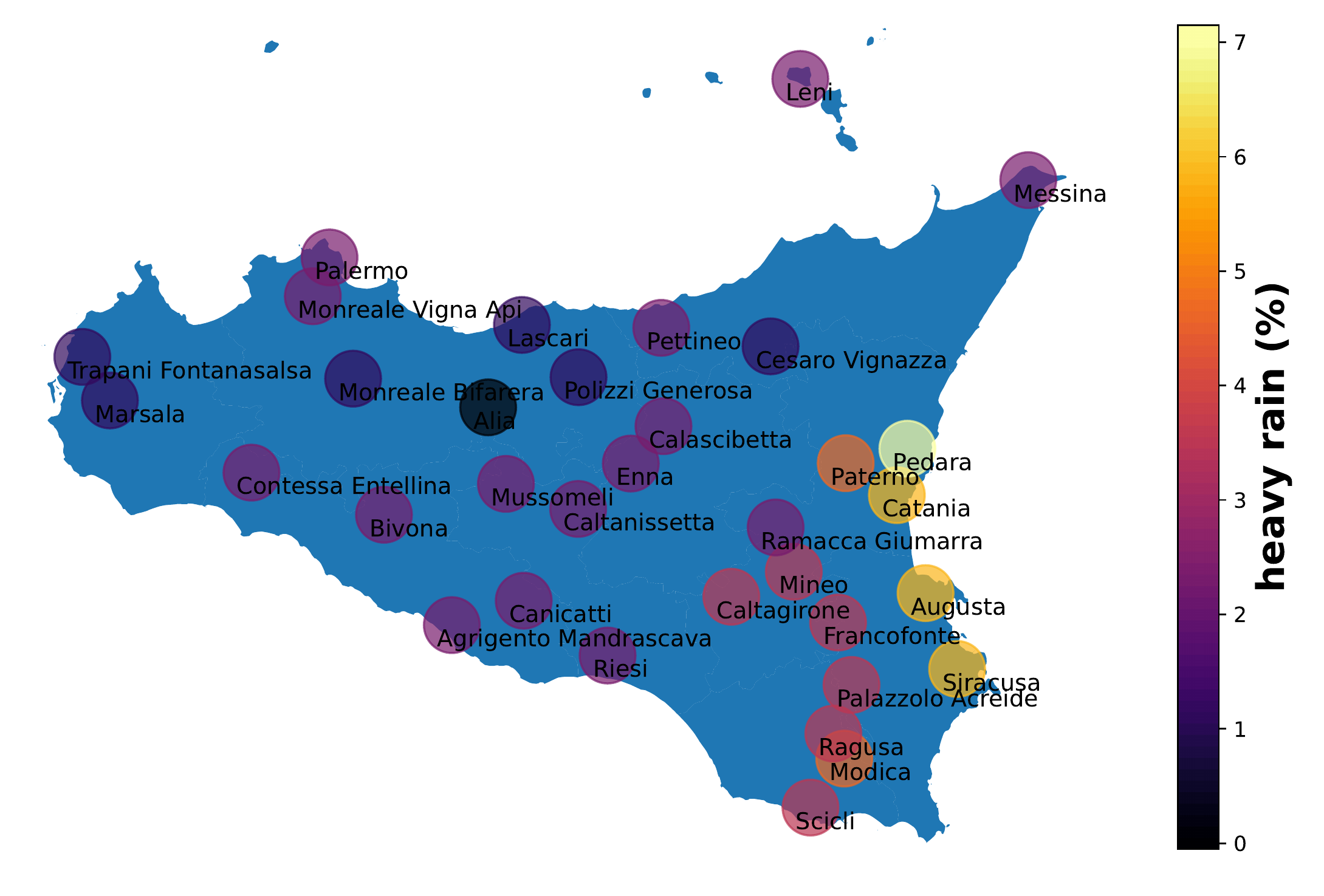}}\\
\end{minipage}
\caption[Full case - Euclidean metrics.]{Full case - Euclidean metrics.
In \textbf{(a)} and \textbf{(b)} different colors represent different clusters, both in the maps and in the histograms. Square and diamond points represent results from, respectively, the second and the third iteration of the algorithm.  \textbf{(a)} C.$C$. The principal cluster is reported in blue. The numbers indicate how many times the stations has been clustered as anomalous in the annual case.
\textbf{(b)} C.$D$. The five main clusters are reported in green, blue, dark blue, red and yellow.
\textbf{(c)} C.$C$. Maximum per day ($md$) heatmap.
\textbf{(d)} C.$D$. Heavy rain (\%) ($h$) heatmap.}\label{fig:full_1}
\end{figure*}

We report here the full clustering results, obtained using C.$C$ and C.$D$ of Table \ref{tab:d}. Similarly to the annual case, the use of Euclidean metrics brings to exceptions detection, by highlighting the presence of clusters composed by a unique site.
We believe that the reason why anomalous clusters are independent lies on the fact that extreme events intensities are very different among sites. Fig. \ref{fig:c_1} shows the presence of one principal cluster and many anomalies, such as \textit{Pedara}, \textit{Augusta} and \textit{Siracusa}. In contrast, Fig. \ref{fig:best} reports different principal clusters - geographically distributed - and only one exception: \textit{Pedara}.\par In order to validate results, we carried out a case by case analysis. First of all, the full and the annual clustering results in the case of Euclidean metrics (C.$A$ and C.$C$, respectively) are compared in Fig. \ref{fig:c_1} by counting how many times the stations has been clustered as anomalous in the annual case. It turns out that the stations with an higher counter are the ones clustered as anomalies in the full case as well, except for \textit{Catania}.

In any case, Fig. \ref{fig:c_1} confirms the results consistency, since East Sicily emerges as the most \textit{extreme} side of the island.

The Kruskal-Wallis test was applied to the full case, providing similar results to the annual case. Among the characterizing indicators (subsection \textit{Clusters validation procedure}), $md$ (Maximum per day) and $h$ (Heavy rain (\%)) are particularly relevant in the full case.

Fig. \ref{main:c} and \ref{main:d} show the $md$ and the $h$ heat-maps, in the full case. In contrast, Fig. \ref{fig:c_1} and \ref{fig:best} show the clustering results.
Several similarities among the maximum values of the indicators and the anomalies can be observed. Therefore, also in the full case, the \textit{extreme} stations coincide with the anomalous clusters. Moreover, the \textit{red} cluster in Fig. \ref{fig:best} represents a cluster of \textit{extreme} stations, confirmed by Fig. \ref{main:d}. In fact, apart for the anomaly of \textit{Pedara}, these stations retain the highest values of the $h$ indicator. \par

\begin{figure*}[h!]
\begin{minipage}{.5\linewidth}
\centering
\subfloat[]{\label{fullcasec1}\includegraphics[width = 7.5cm]{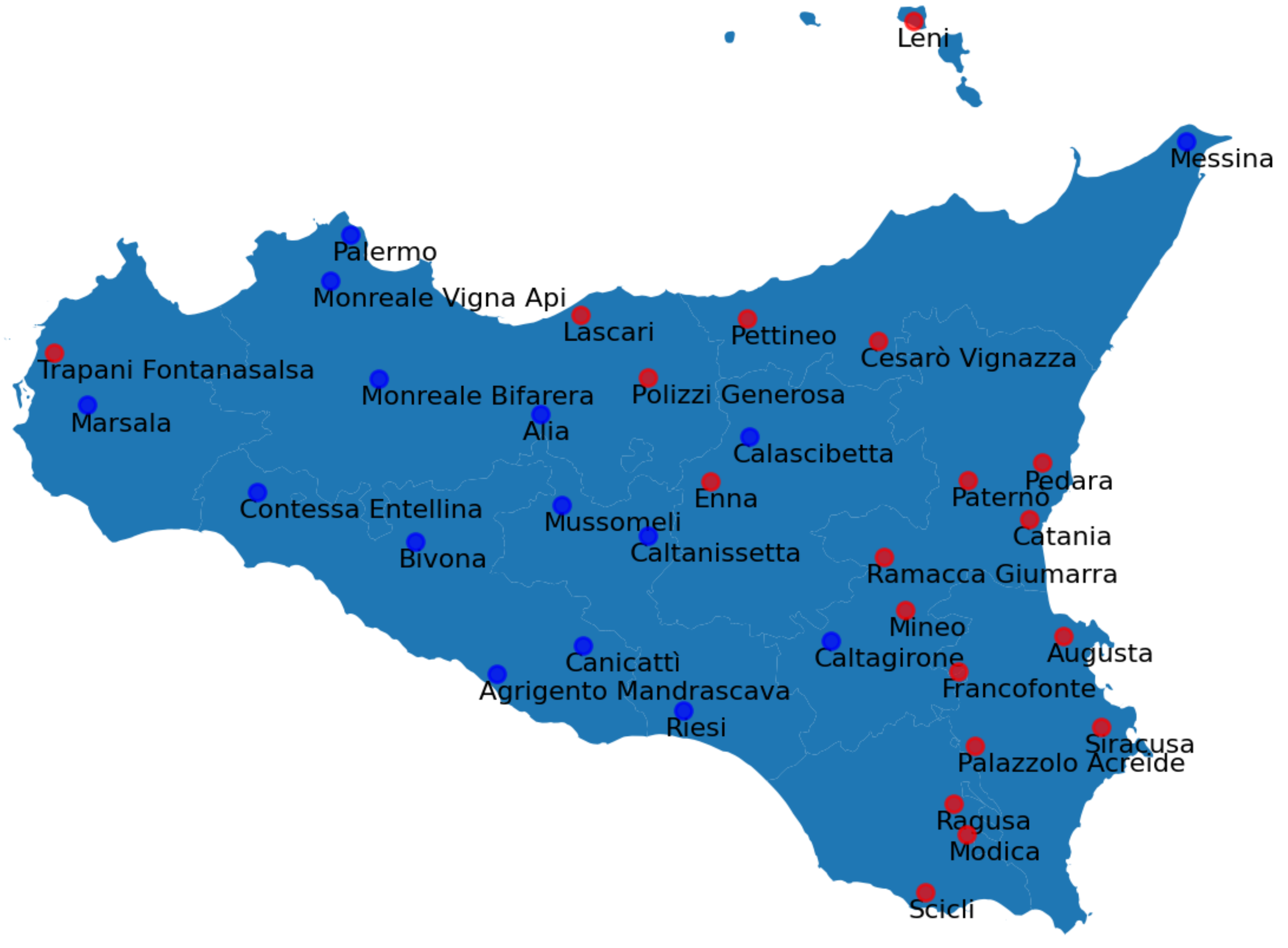}}
\end{minipage}%
\begin{minipage}{.5\linewidth}
\centering
\subfloat[]{\label{fullcasec2}\includegraphics[width = 7.5cm]{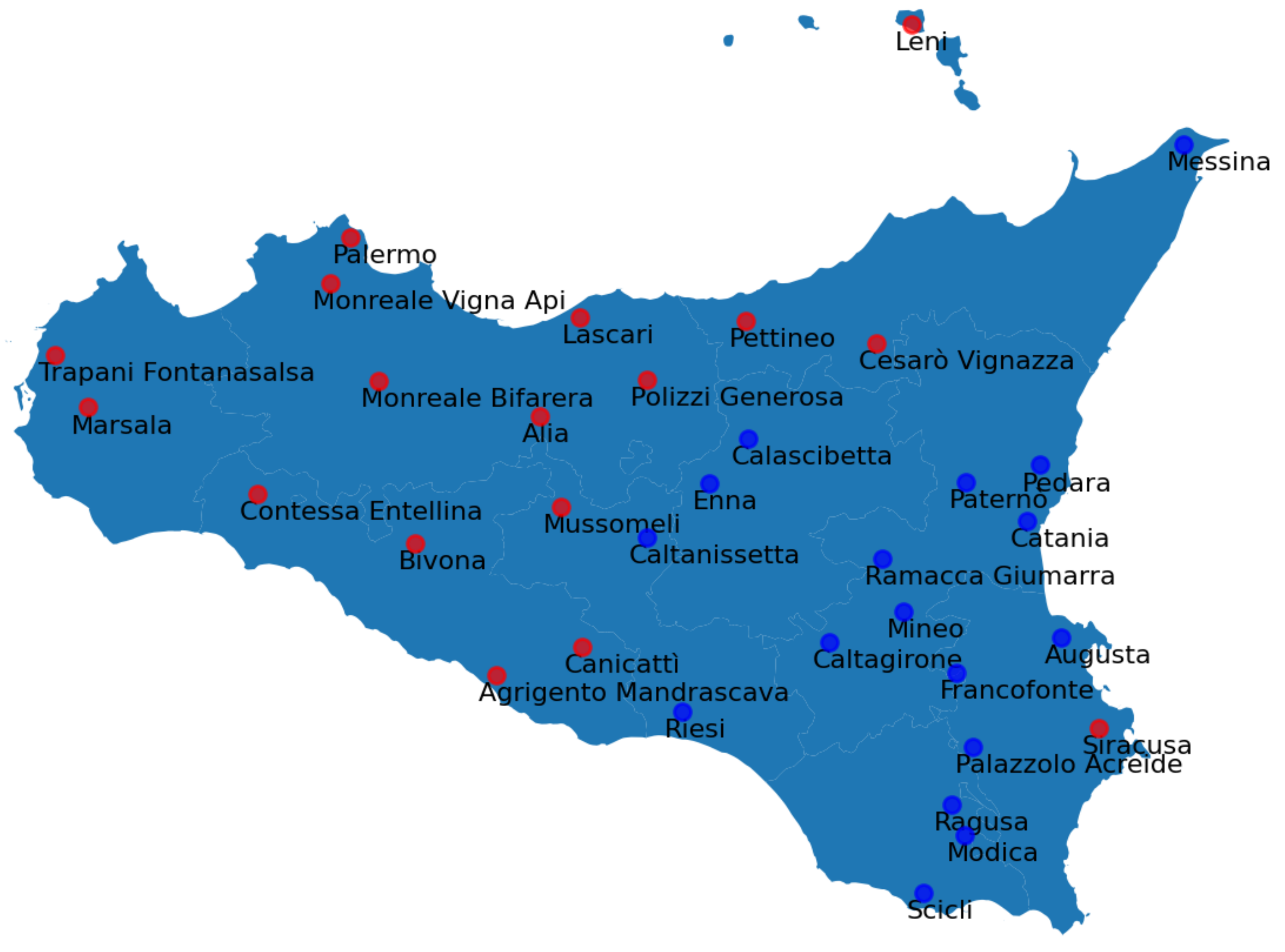}}
\end{minipage}
\caption[Full case - Correlation metrics.]{Full case - Correlation metrics.  \textbf{(a)} C.$C$. \textbf{(b)} C.$D$. The two clusters are reported in red and blue in both the panels.}\label{fig:full_2}
\end{figure*}
\par
In conclusion, the different implemented experimental settings allowed us to discover several different aspects of extreme events. Certainly, the presence of these phenomena in eastern Sicily emerges both from the annual and the full clustering, especially when the Euclidean metrics is used as a similarity measure in the AP algorithm. On the other hand, the use of Correlation metrics brings to consider Sicily composed of two different climatic areas: West side and East side, as shown in Fig. \ref{fig:full_2}, where there are only 2 large clusters. Moreover, in this case no similarities between characterizing indicators and clusters are found (compare Fig. \ref{fullcasec1} and Fig. \ref{main:c}, Fig. \ref{fullcasec2} and Fig. \ref{main:d}).

Eventually, the clustering involving C.$D$ of Table \ref{tab:d} with the use of the Euclidean metrics seems to be the most suitable setting among those tested.
\subsection*{Local investigation}
In the local case we investigated the temporal evolution of rainfall events. In particular, similar years in the entire observed period were detected. To this aim, the AP algorithm was applied only to C.$A_{s}$ and C.$B_{s}$, analysing one station per time. As in the geographical investigation, we chose to use both the Euclidean and the Correlation metrics.

\begin{figure}[h!]
\begin{minipage}{.5\linewidth}
\subfloat[]{\label{ea1}\includegraphics[width = 3.3in]{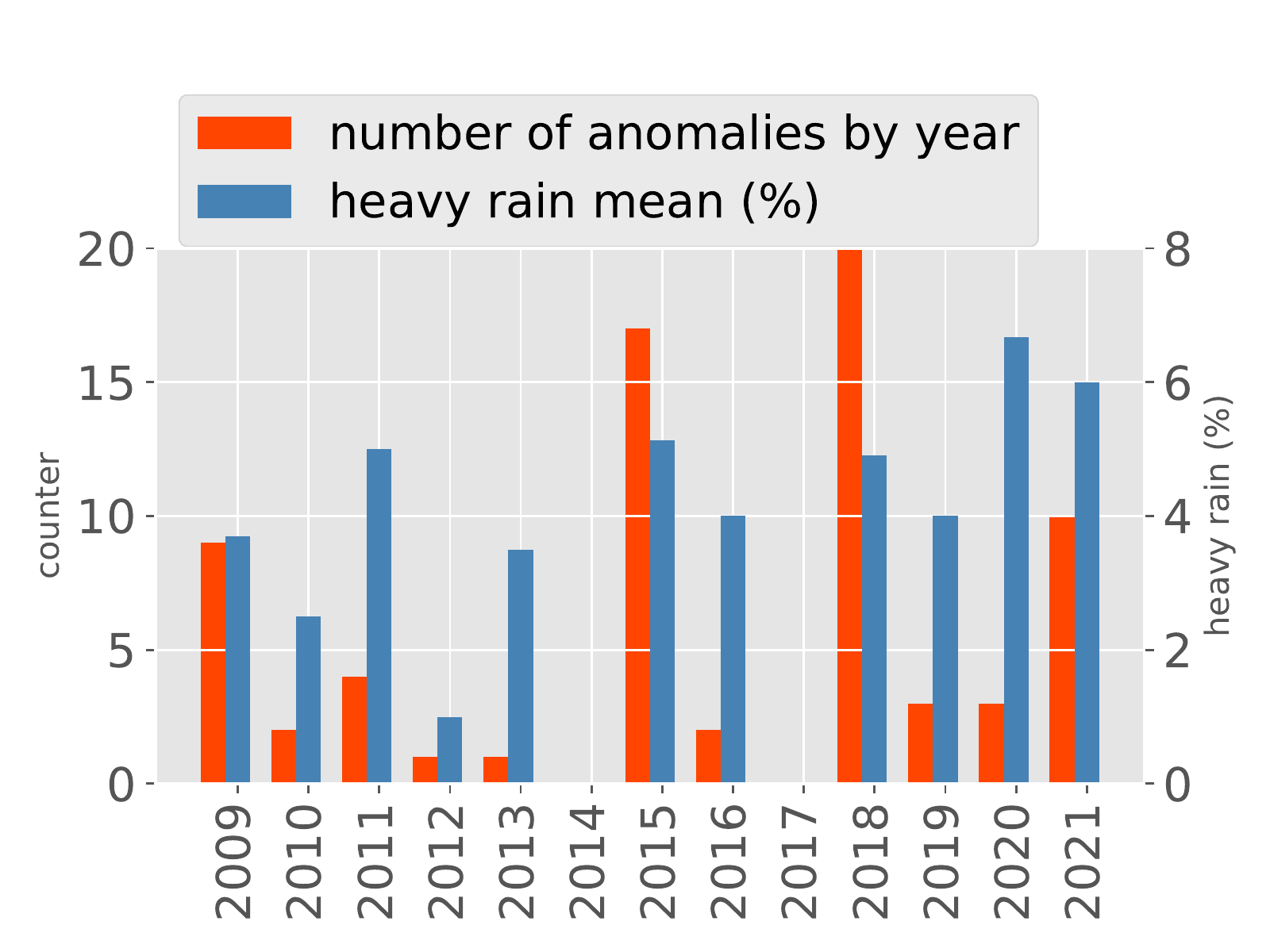}}
\end{minipage}
\begin{minipage}{.5\linewidth}
\subfloat[]{\label{eb1}\includegraphics[width = 3.3in]{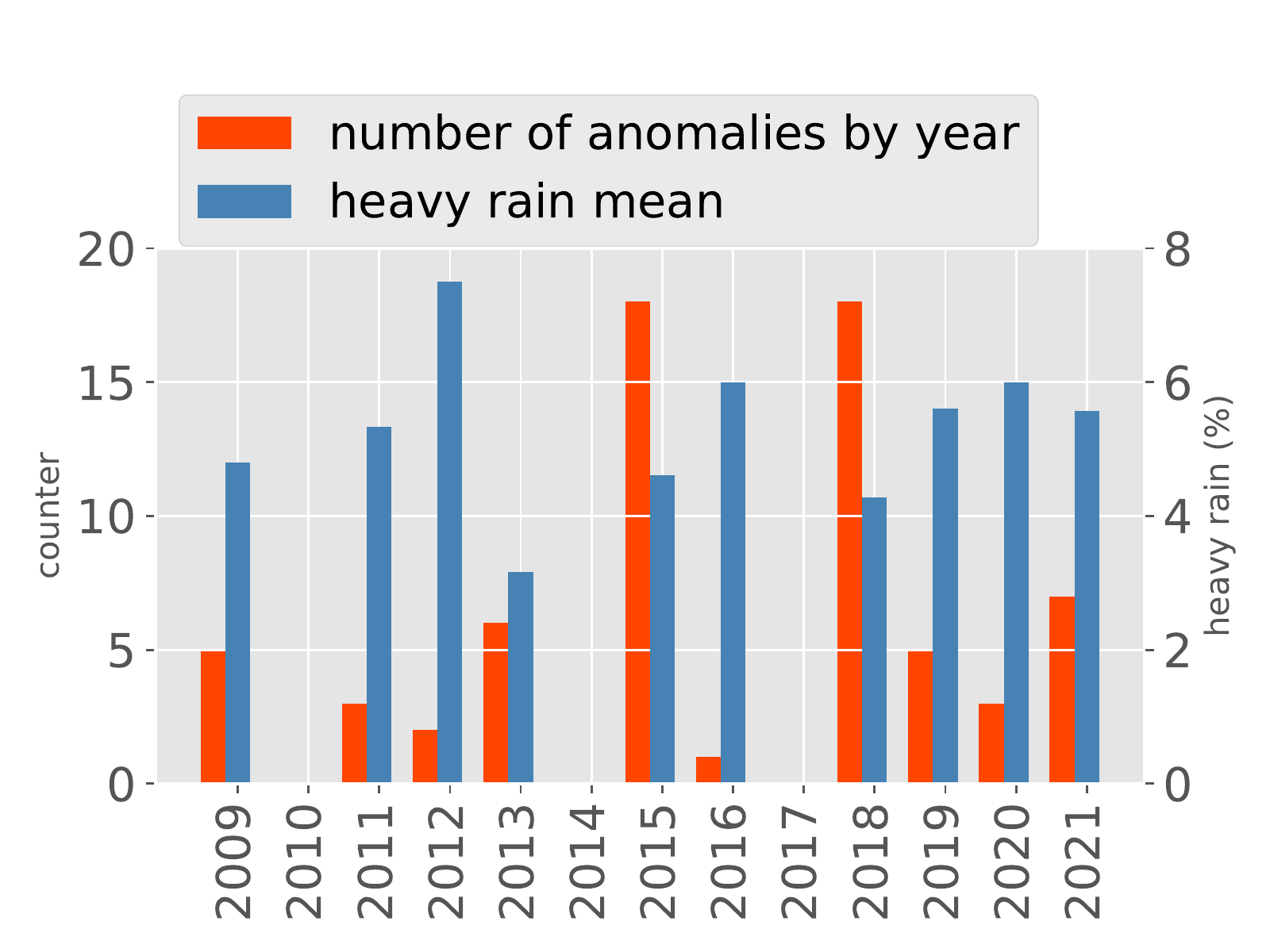}}
\end{minipage}
\caption[Anomalous years - Euclidean metrics.]{Anomalous years - Euclidean metrics. C.$A_{s}$ \textbf{(a)}. C.$B_{s}$ \textbf{(b)}. The heavy rain mean of the year $y$ consists of the mean of the \textit{heavy rain (\%)} values for all the stations that cluster the year $y$ as anomalous.
}\label{fig:local_counters}
\end{figure}

In order to understand the most \textit{anomalous} years, we counted (over stations) how many times one year appears as exception when using the Euclidean distance. Fig. \ref{ea1} and \ref{eb1} report the years counters in respectively C.$A_{s}$ and C.$B_{s}$ (in red).
In both cases the most anomalous years were 2015 and 2018. This means for instance that 2018 is clustered as anomalous in about 20 over 34 stations for Euclidean distance and C.$A_{s}$. We see that also the 2021 counter increases after the years 2019 and 2020. Additionally, Fig. \ref{ea1} and \ref{eb1} also report the \textit{heavy rain (\%)} mean values (in blue). In this case, we fix a year $y$ and we compute the mean of the \textit{heavy rain (\%)} values for all the stations that cluster the year $y$ as anomalous, thus obtaining, for instance, that 2020 and 2021 have the highest mean values.
Summarizing, in the case of C.$A_{s}$ and Euclidean distance, an increasing trend on anomalous years was found concerning the heavy rain mean indicator (see Fig. \ref{ea1} in blue). On the other hand, in C.$B_{s}$ and Euclidean distance, the trend is less detectable and the highest value of heavy rain mean is measured in 2012 (see Fig. \ref{eb1} in blue).

\subsection*{Conclusions}
The main goal of this work was to introduce a clustering approach detecting extreme rainfall events occurred in Sicily, from 2009 to 2021 and to identify communities of sites with similar behaviors. 

To the best of our knowledge, we are presenting for the first time in the literature, the use of multi-modal clustering analysis to detect extreme rainfall events in Sicily. With this approach, we were able to confirm and expand some preliminary observations presented in \cite{bonaccorso2016}, where a statistical approach has been applied to analyse rainfall trends. 
Specifically, in our work a clustering technique, the Affinity Propagation algorithm, was employed to confirm and discover geographical and historical rainfall changes.

In order to understand the rainfall phenomena mostly characterizing the geographical communities identified, several rainfall indicators were introduced and evaluated over the available time series. In addition, the obtained results were validated by means of the Kruskal-Wallis statistical test.

Three types of clustering analysis were conducted - full, annual and local: firstly applied to the entire high frequency time series and then applied to the weekly averaged data. The reason of this choice lied in the need of reducing the dataset size in order to minimize sensitivity to outliers. Eventually, we investigated both the Euclidean and the Correlation metrics as distance measures for the AP algorithm.

The paper presents several significant findings:
\begin{itemize}
    \item East Sicily is increasingly becoming a protagonist of \textit{extreme} events, both in the full period of recordings and in the annual cases. This result is more evident choosing the Euclidean metrics in the implementation of the Affinity Propagation algorithm. 
    \item High frequency data (C.$A$) with the Euclidean metrics brings to the detection of an increasing trend over years of  \textit{extreme events}; in contrast, C.$B$ of weakly averaged data does not provide the same evidence.
    \item 2021 emerges as one of the most anomalous years in the local investigation over time. Moreover, we found from the geographical analysis that it is characterized by \textit{extreme} events in the East side of the island, particularly in the cities of \textit{Catania}, \textit{Siracusa} and \textit{Augusta}.
    \item Using a statistical validation approach, we found out that three indicators describe the anomalous clusters finely: the maximum per day ($md$), the maximum daily variation ($mv$) and the heavy rain percentage ($h$).
    This entails that most of the time anomalous clusters are characterized by the presence of \textit{extreme} events.
    \item The Affinity Propagation algorithm allowed to detect anomalies, namely \textit{extreme} stations, considering the full dataset. Specifically, using the Euclidean metrics, the cities of \textit{Augusta}, \textit{Siracusa}, and \textit{Pedara} were identified clearly as anomalous at the first iteration of the algorithm. In contrast, \textit{Palazzolo Acreide} and \textit{Messina} have been detected at the second and third run of the algorithm, respectively. \textit{Catania} does not emerge as an anomalous cluster, however, similarly to the previous sites, it presents 6 out of 13 anomalous years.
    \item The Euclidean metrics is sensitive to micro-climatic changes, i.e. geographically close stations are clustered in different groups. This is consistent with what can be experimentally observed since there are actually rainfall events of different frequency and magnitude a few kilometers apart.
    \item The Correlation metrics allowed to identify uniform clusters. This is particularly evident in the full case, in which the algorithm splits Sicily in East and West parts.
    \item The dataset size reduction by weekly means is successful in the geographical clusterings. In particular, it merges together anomalies and territorial clusters in a balanced way, finding appropriately \textit{extreme} clusters, such as the one in eastern Sicily.
\end{itemize}

We are aware that those obtained results are no more robust than in previous before-mentioned studies, given the short observational period and therefore potentially affected by the large unforced internal/natural variability. It is clear that a wider spatial and temporal range would be needed to fully validate changes in heavy rainfall trends \cite{li,fischer,westra,ribes}.
However, this paper has a methodological focus aimed to introduce an elegant approach to detect extreme events. In fact, the clustering approach is promising to interpret spatial patterns of heavy rainfall.

Further research is necessary to determine which dimensionality reduction procedure is the best for having a more precise local investigation. For instance, the \textit{total per hour} datasets, the \textit{daily averaged} datasets or some features refinement techniques, as for instance the Principal Component Analysis, can be advantageous to reduce the features-set dimensionality. More accurate rainfall indicators could be applied and derived, in order to entirely characterize rainfall extreme events. Additionally, in order to have more robust results, a more general analysis can be conducted, merging stations by provinces, considering a temporal clustering over the entire region, or even increasing the temporal range of investigation. In this way, the methodology introduced may potentially also yield robust findings in terms of a climate change signal. 

This study could contribute significantly to the development of the decision support systems  based on multi-modal and easy to use data science tools for policy makers, stakeholders, and social actors. We are currently working on the possible development of a demo based on the proposed method, to be later distributed to potential investors and researchers interested in the field.

\section*{Acknowledgment}
We thank \href{http://www.sias.regione.sicilia.it/NHXNAJ235.php}{SIAS} \cite{sias} for the kind permission of the rainfall data.
\section*{Data availability}
The dataset together with the code is available at the following \href{https://github.com/elevitanz/Extreme_Events_Sicily.git}{GitHub Repository} \cite{github}.

\section*{Authors Contributions}
The authors have equally contributed to the work.
\section*{Additional information}
\textbf{Competing interests} The authors do not declare any competing interests for this work.
\end{document}


\maketitle

\SItext

\section*{Annual variables analysis}\label{sec.a}
\begin{figure}[h!]
\begin{minipage}{.16\linewidth}
\centering
\subfloat[]{\label{fig:hist_a}\includegraphics[width = 1in]{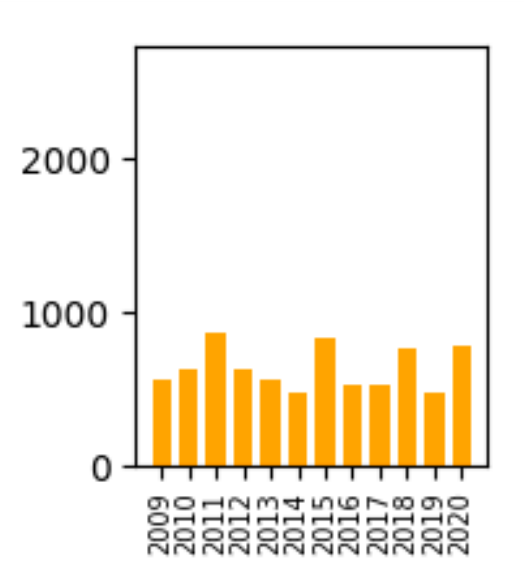}}
\end{minipage}
\begin{minipage}{.16\linewidth}
\centering
\subfloat[]{\label{fig:hist_b}\includegraphics[width = 1in]{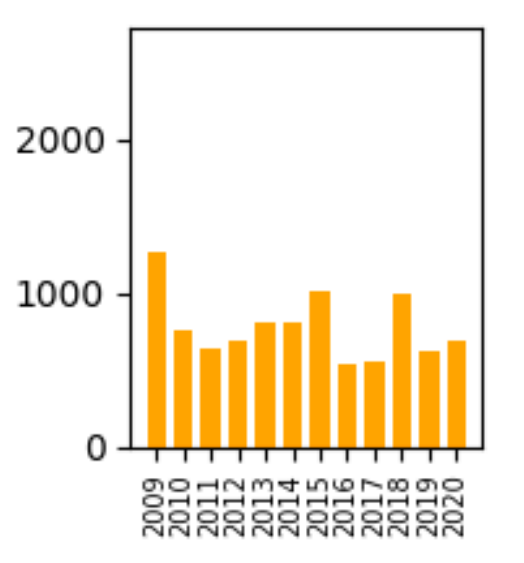}}
\end{minipage}
\begin{minipage}{.16\linewidth}
\centering
\subfloat[]{\label{fig:hist_c}\includegraphics[width = 1in]{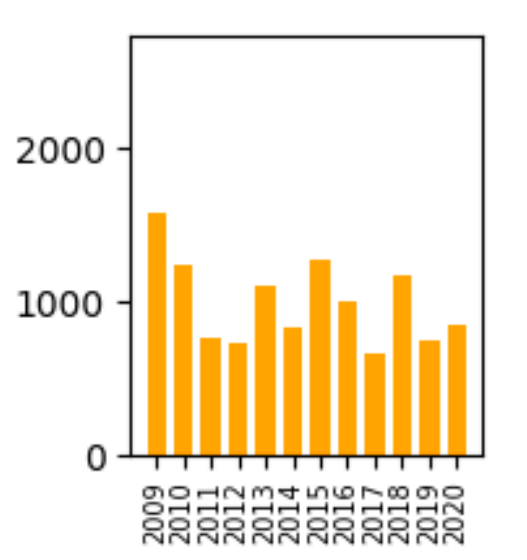}}
\end{minipage}
\begin{minipage}{.16\linewidth}
\centering
\subfloat[]{\label{fig:hist_d}\includegraphics[width = 1in]{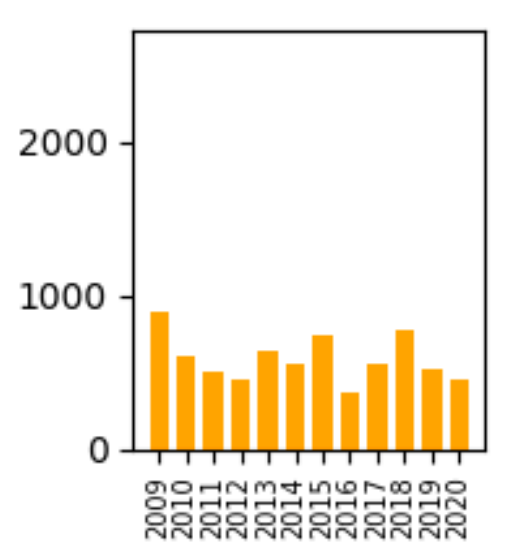}}
\end{minipage}
\begin{minipage}{.16\linewidth}
\centering
\subfloat[]{\label{fig:hist_e}\includegraphics[width = 1in]{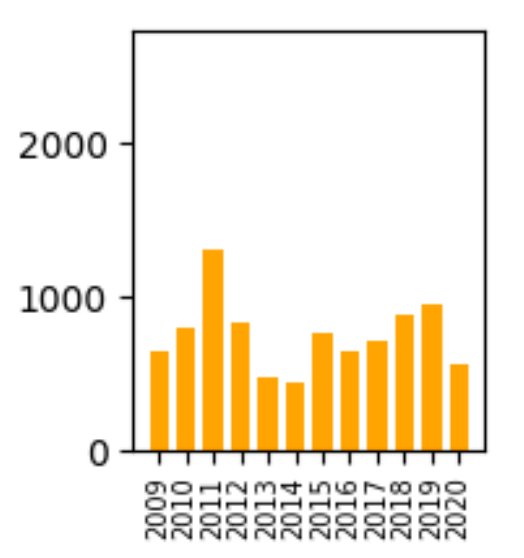}}
\end{minipage}
\begin{minipage}{.16\linewidth}
\centering
\subfloat[]{\label{fig:hist_f}\includegraphics[width = 1in]{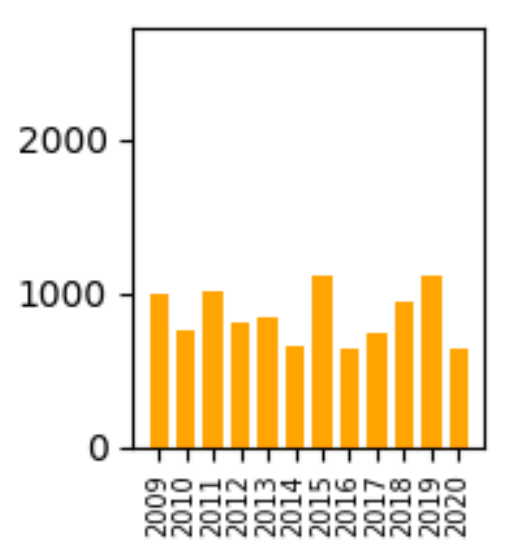}}
\end{minipage}\par\medskip
\begin{minipage}{.16\linewidth}
\centering
\subfloat[]{\label{fig:hist_g}\includegraphics[width = 1in]{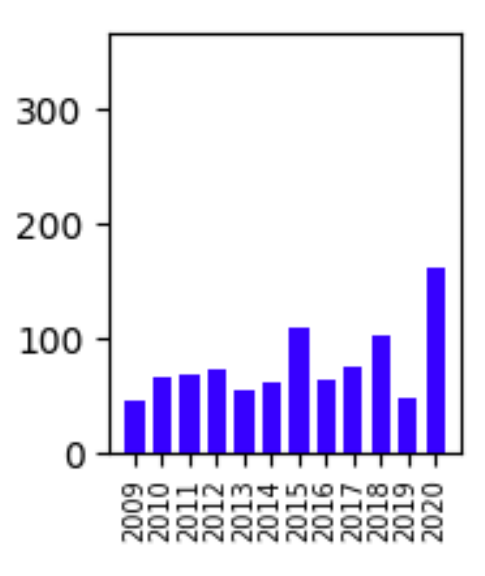}}
\end{minipage}
\begin{minipage}{.16\linewidth}
\centering
\subfloat[]{\label{fig:hist_h}\includegraphics[width = 1in]{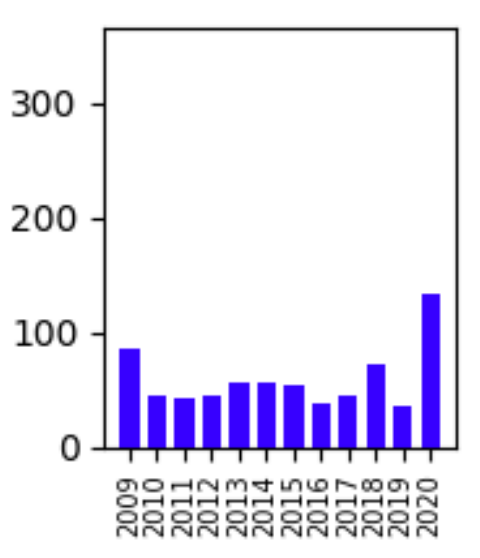}}
\end{minipage}
\begin{minipage}{.16\linewidth}
\centering
\subfloat[]{\label{fig:hist_i}\includegraphics[width = 1in]{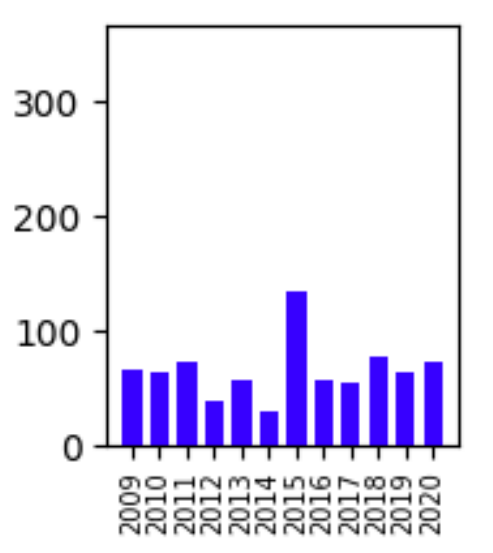}}
\end{minipage}
\begin{minipage}{.16\linewidth}
\centering
\subfloat[]{\label{fig:hist_j}\includegraphics[width = 1in]{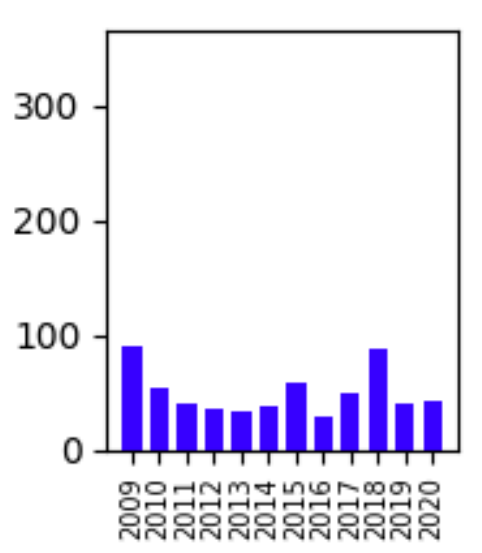}}
\end{minipage}
\begin{minipage}{.16\linewidth}
\centering
\subfloat[]{\label{fig:hist_k}\includegraphics[width = 1in]{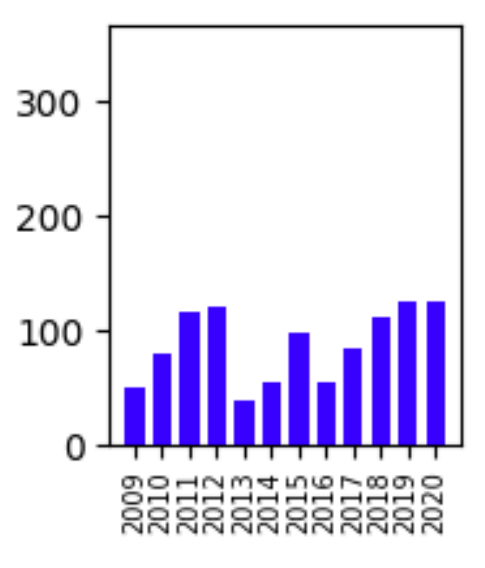}}
\end{minipage}
\begin{minipage}{.16\linewidth}
\centering
\subfloat[]{\label{fig:hist_l}\includegraphics[width = 1in]{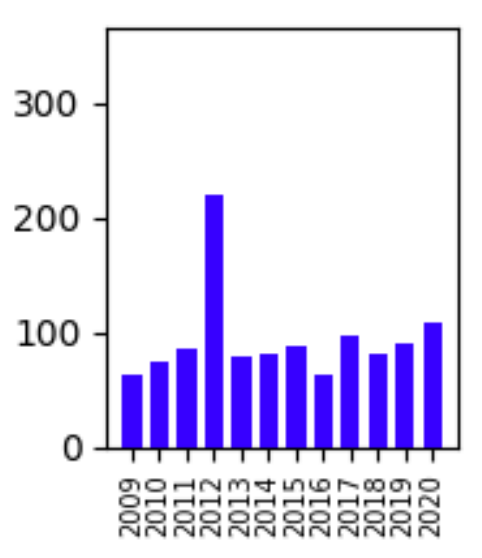}}
\end{minipage}\par\medskip
\begin{minipage}{.16\linewidth}
\centering
\subfloat[]{\label{fig:hist_m}\includegraphics[width = 1in]{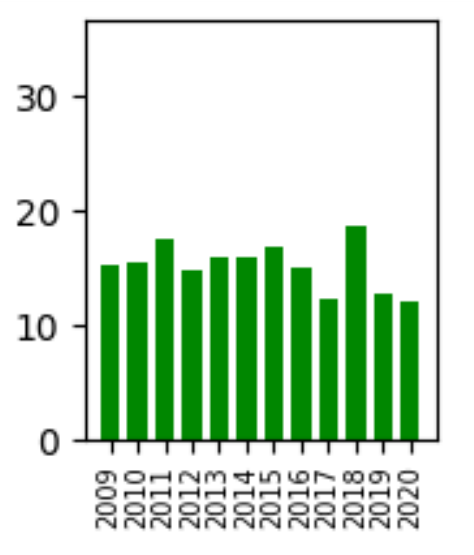}}
\end{minipage}
\begin{minipage}{.16\linewidth}
\centering
\subfloat[]{\label{fig:hist_n}\includegraphics[width = 1in]{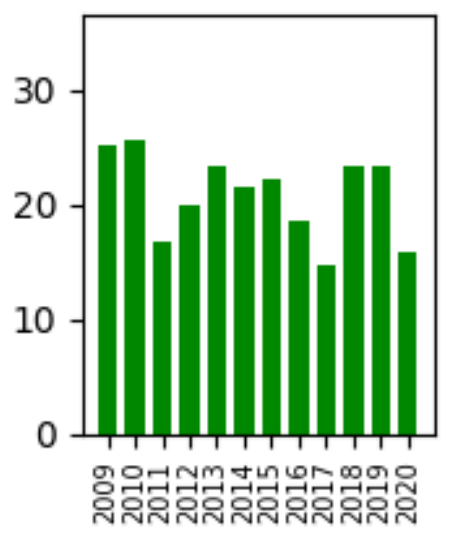}}
\end{minipage}
\begin{minipage}{.16\linewidth}
\centering
\subfloat[]{\label{fig:hist_o}\includegraphics[width = 1in]{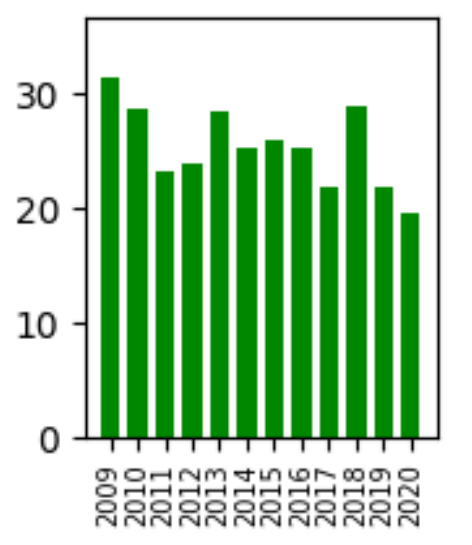}}
\end{minipage}
\begin{minipage}{.16\linewidth}
\centering
\subfloat[]{\label{fig:hist_p}\includegraphics[width = 1in]{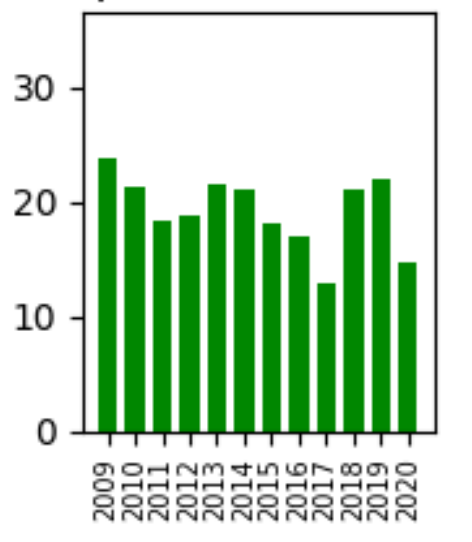}}
\end{minipage}
\begin{minipage}{.16\linewidth}
\centering
\subfloat[]{\label{fig:hist_q}\includegraphics[width = 1in]{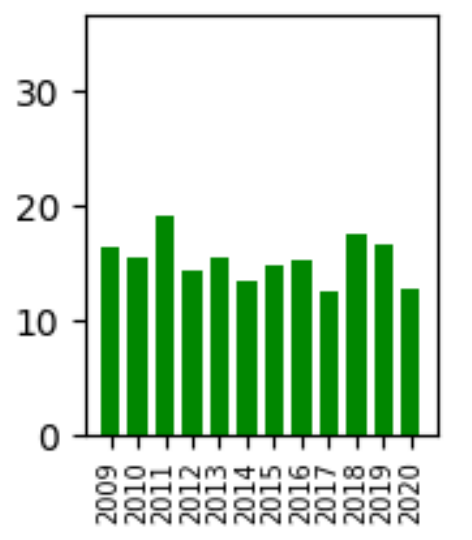}}
\end{minipage}
\begin{minipage}{.16\linewidth}
\centering
\subfloat[]{\label{fig:hist_r}\includegraphics[width = 1in]{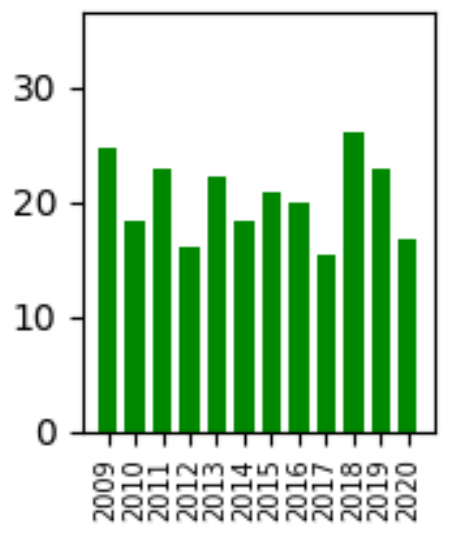}}
\end{minipage}\par\medskip
\caption{Annual variables of \textit{Catania} (first column), \textit{Palermo} (second column), \textit{Messina} (third column), \textit{Trapani Fontanasalsa} (fourth column), \textit{Siracusa} (fifth column) and \textit{Palazzolo Acreide} (sixth column). Orange histograms represent the \textit{tot} annual variable; blue histograms represent the \textit{dmax} annual variable; green histograms represent the \textit{rd} annual variable. The definition of those variables is reported in the main document. In the x axis of every histogram there are the years between 2009 and 2020.}
\label{fig:examples}
\end{figure}

In this section the annual histograms will be displayed, in order to better understand which stations are more affected by extreme events. Figure \ref{fig:examples} shows some of those histograms, in particular the ones relative to \textit{Catania}, \textit{Palermo}, \textit{Messina}, \textit{Trapani Fontanasalsa}, \textit{Siracusa} and \textit{Palazzolo Acreide}.\\
Figure \ref{fig:hist_g} shows an increasing trend in Catania for the $dmax$ variable, with a peak of about 160 $mm$ in 2020. At the same time, we can appreciate in figure \ref{fig:hist_m} a lower $rd$ annual value. Same behaviour can be seen in \textit{Palermo} for the same year (figure \ref{fig:hist_h} and \ref{fig:hist_n}).\par Since this happened in 2020, and not other years, this observation led us to hypothesize the presence of an increasing in extreme rainfall events, in terms of frequency and intensities. This was the rationale we used in our work to select the stations, counting how many years have this property in a station. Moreover, in \textit{Palermo} we could appreciate a decreasing trend of the $tot$ variable (Figure \ref{fig:hist_b}).\par This means that an increasing in the maximum values comes with a decreasing in the total annual rainfall. This in some way seemed to confirm the observations made in the city of \textit{Palermo}.\par Figure \ref{fig:examples} shows different behaviours for the cities of \textit{Messina} and \textit{Trapani Fontanasalsa}. In \textit{Messina} the $dmax$ peak lies in 2015, which is associated with high $rd$  and $tot$ values. Despite such high values, the $dmax$ value of about 135 $mm$ is not negligible (figure \ref{fig:hist_i}). In fact, it represents an extreme year for what concerns the presence of rainfall events.\par In 2020, in contrast, there was the lowest local $rd$ value, associated with a low $tot$ value and a medium/high $dmax$ (figures \ref{fig:hist_c}, \ref{fig:hist_i} and  \ref{fig:hist_o}). This suggests a global reduction of rainfall events in that year but with a considerable maximum per day. Again this case represents the typical extreme rainfall events setting.\par In \textit{Trapani Fontanasalsa} we see maximum peaks in 2009 and 2018 (figure \ref{fig:hist_j}), together with high values on total rainfall and rainy days in those years (figures \ref{fig:hist_d} and \ref{fig:hist_p}). Moreover we can see a notable local trend both in \textit{Messina} and \textit{Trapani Fontanasalsa} for a decreasing on the total annual amount of rain and on the percentage of rainy days. \par Finally, figure \ref{fig:examples} shows also \textit{Siracusa} and \textit{Palazzolo Acreide} annual data. In \textit{Siracusa} we have maximum per day peaks of about 100 $mm$ (figure \ref{fig:hist_k}). For instance, 2019 and 2020 share a similar $dmax$ value, but they represent two different cases. In 2019 we have high $rd$ and $tot$ values, whereas in 2020 we have low values for them, which means again presence of extreme events (figures \ref{fig:hist_e} and \ref{fig:hist_q}). A similar reasoning could be done for 2012 and 2013. In \textit{Palazzolo Acreide} we have a huge $dmax$ peak of more than 200 $mm$ during 2012, associated with low values of the other variables (figures  \ref{fig:hist_f}, \ref{fig:hist_l} and \ref{fig:hist_r}). We observed the same for 2020, even if with a much lower peak.
\section{Data visualization}\label{sec.b}
We report here an increasingly zoom of \textit{Augusta} rainfall time series, with the aim of better understanding rainfall behaviour over time. \textit{Augusta} is located in the province of \textit{Siracusa}, in the South-East of Sicily. We considered it as an example, but we could do the same for any rain gauge.
\subsection{Full view}
\begin{figure}[h!]
    \centering
    \includegraphics[width=.9\textwidth]{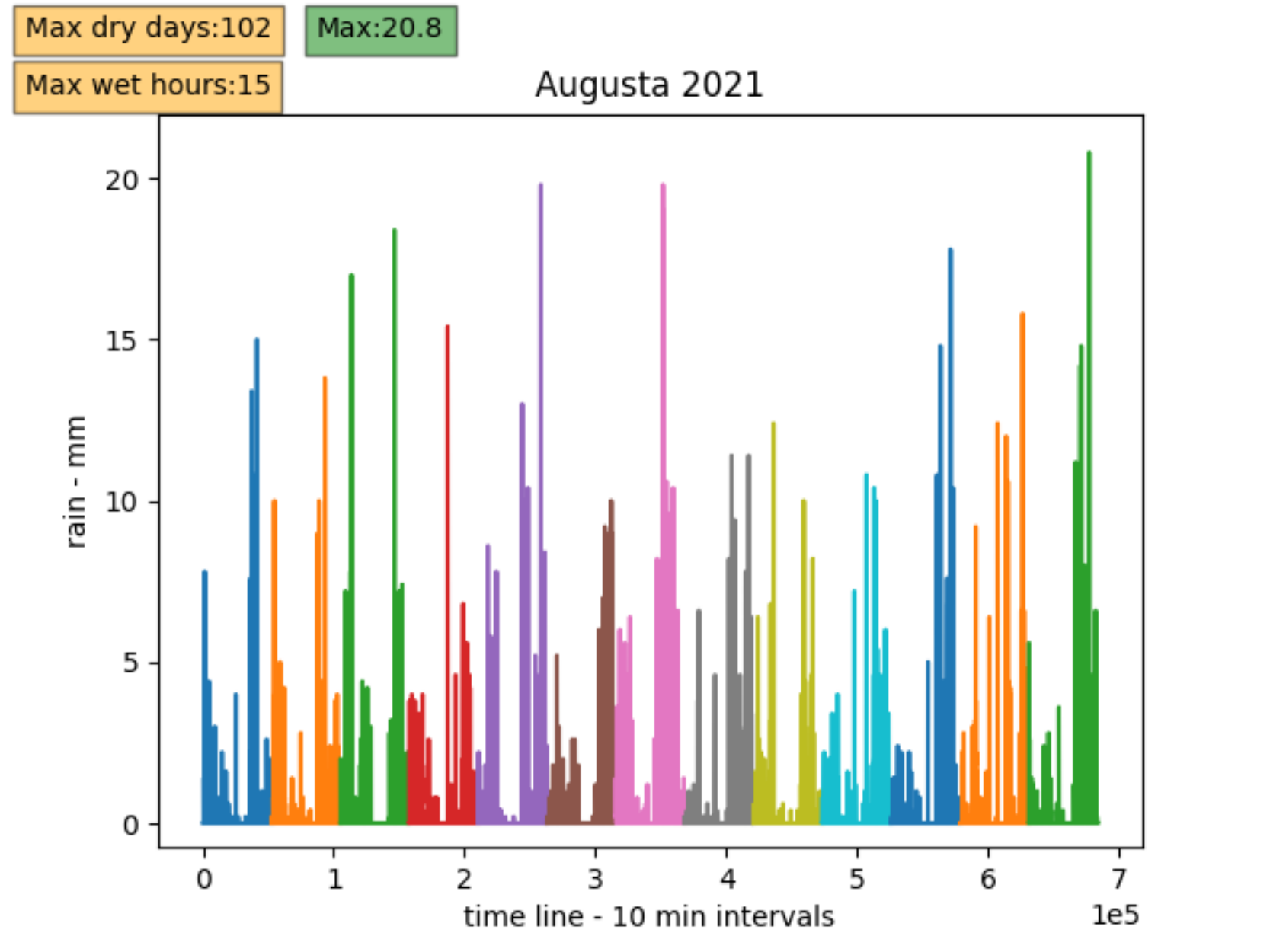}
    \caption{Full view - Augusta}
    \label{fig:augusta}
\end{figure}
\begin{figure}[h!]
    \centering
    \includegraphics[width=.9\textwidth]{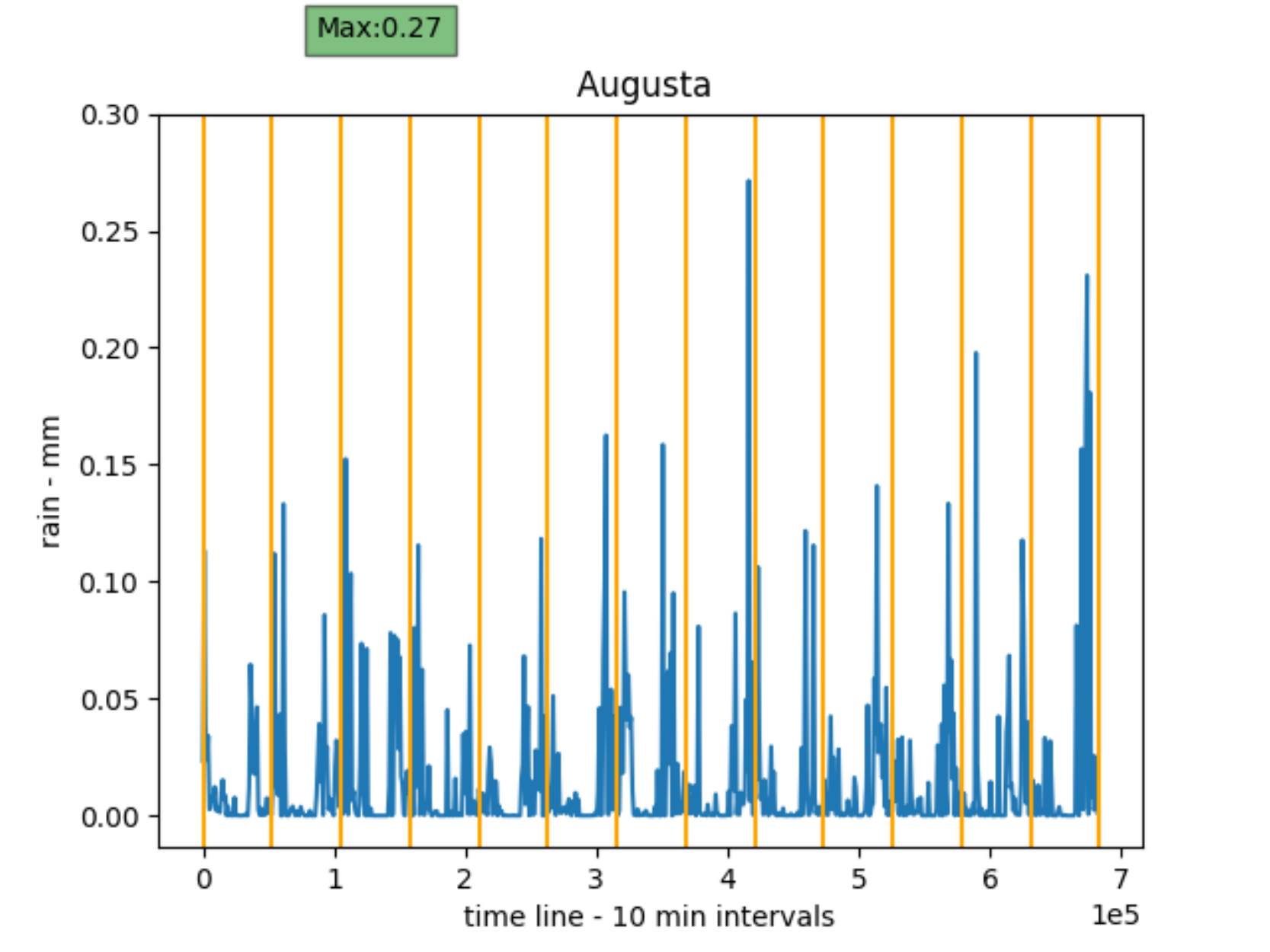}
    \caption{Weekly mean - full view - Augusta}
    \label{fig:mean_augusta}
\end{figure}
\noindent
Figure \ref{fig:augusta} shows all the available data of \textit{Augusta} over time. In the x axis there are all the 10 minutes steps from 01/01/2009 at 00:00 to 01/01/2022 at 00:00, while in the y axis there are the amounts of rain - in $mm$ - for each temporal step. Colors allow to distinguish between different annual records collections.\par. We consider three variables in the top-left of the figure:
\begin{enumerate}
    \item \textit{Max dry days}: it counts how many days of dryness there are. In other words, the longest dry period from 2009 to 2021 in \textit{Augusta}  consists of $102$ consecutive days.
    \item \textit{Max wet hours}: it counts how many hours of wetness there are. In other words, the longest wet period from 2009 to 2021 in \textit{Augusta} lasts $15$ consecutive hours.
    \item \textit{Max}: it counts the maximum amount of rain in 10 minutes over the entire time series.
\end{enumerate}
The \textit{Max} variable indicates 20.8 $mm$ of rain in 10 minutes. The plot shows clearly that this maximum value lies in the end of 2021. This is a confirmation of what we actually observed before choosing this study topic. Nevertheless, we observed a general trend over years to have a peak at about 20 $mm$. Namely, 2011 (the first in green), 2013 (in purple) and 2015 (in pink) have a similar peak to 2021.\par Therefore, globally, we cannot think of a general increasing trend in peaks. In fact, we also notice a reduction in 2016, 2017 and 2018. Another observation is that all of those peaks lies, as expected, in the rainy season, that is in the last part of the years.\par Moreover, the \textit{Max dry days} variable lies in 2011, while the \textit{Max wet hours} one lies in 2012.\par Therefore, at first glance there does not seem to be an increasing dryness associated with a decreasing wetness trend in Augusta. However, considering only the extreme consecutive periods is not sufficient to claim that. In confirmation of this, extreme events main characteristic is represented by water bombs.\par Consequently, we could have an increasing of dryness periods without continuity, observing short-duration rainfall events among a general drought. The idea is that those short-duration rainfall events have actually more and more higher intensity and frequency.\par During all these observations, we also look at \textit{weekly mean data}. We plotted only one value per week, which is the mean of the 10 minutes intervals. Figure \ref{fig:mean_augusta} does not seem to lose relevant information respect to figure \ref{fig:augusta}, except for the maximum mean value founded in 2016. Besides, this case deals with far fewer data and avoids noise. This suggests the idea developed in the main document to use mean data instead of the original ones, in order to reduce computation complexity during the run of the algorithms and to reduce outliers. We computed for every week the mean
over the $1008$ weekly data.
The resulting dataset composed by $679$ values is then reported in the figure \ref{fig:mean_augusta}.\par
It is particularly evident the difference between 2009 and 2021 mean values. In fact, the max mean value of $0.27$ $mm$ relative to the 2016 is the only value higher than the max 2021 mean value, while the max mean value in 2009 is of about $0.06$ $mm$.\par Moreover, in general an increasing trend on peaks over years is quite visible, except for the peak in 2016. Actually, the trend is more noticeable than in figure \ref{fig:augusta}, thanks to the reduction of noise. This suggested that the averaged dataset could lead to interesting conclusions. 

\subsection{Annual view}
The following figures show the annual data from 2009 to 2021. They represent 13 different zooms of the full plot. The structure of the figures is the same as before, but this time colors represent different months.\par
\begin{figure}[h!]
\subfloat[Annual view - 2009 Augusta]{\includegraphics[width = 3.25in]{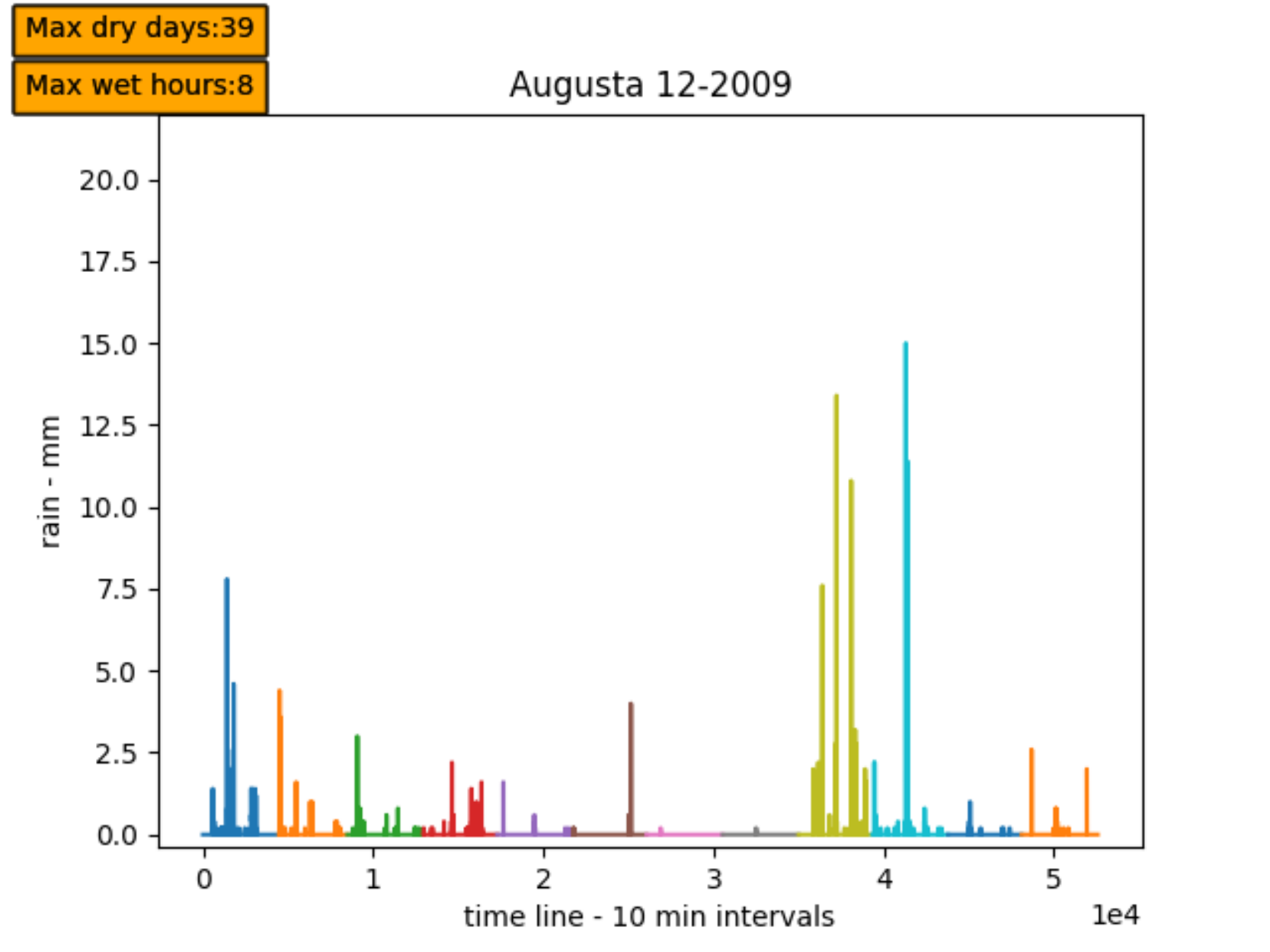}} 
\subfloat[Annual view - 2011 Augusta]{\includegraphics[width = 3.25in]{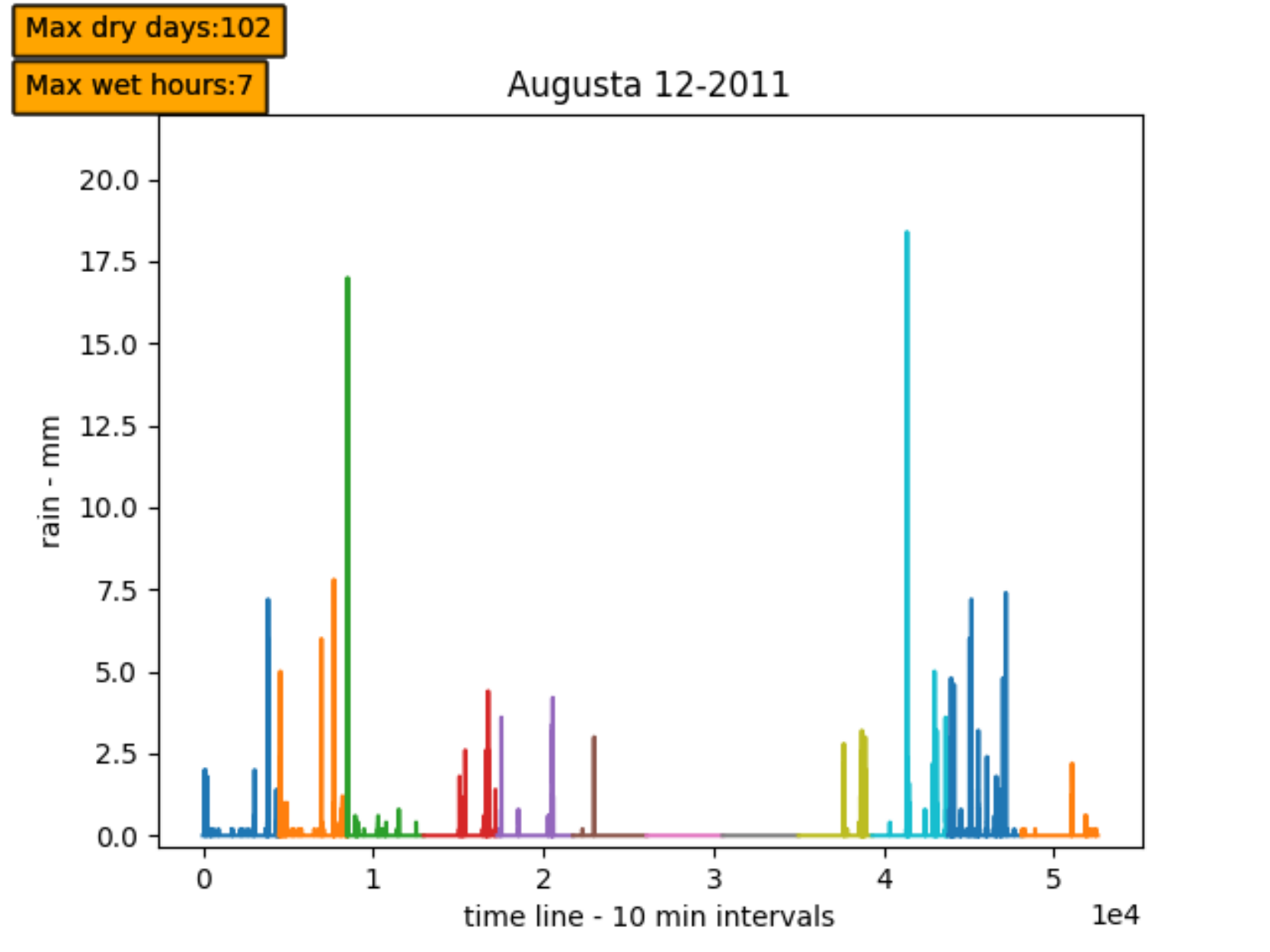}}\\
\subfloat[Annual view - 2012 Augusta]{\includegraphics[width = 3.25in]{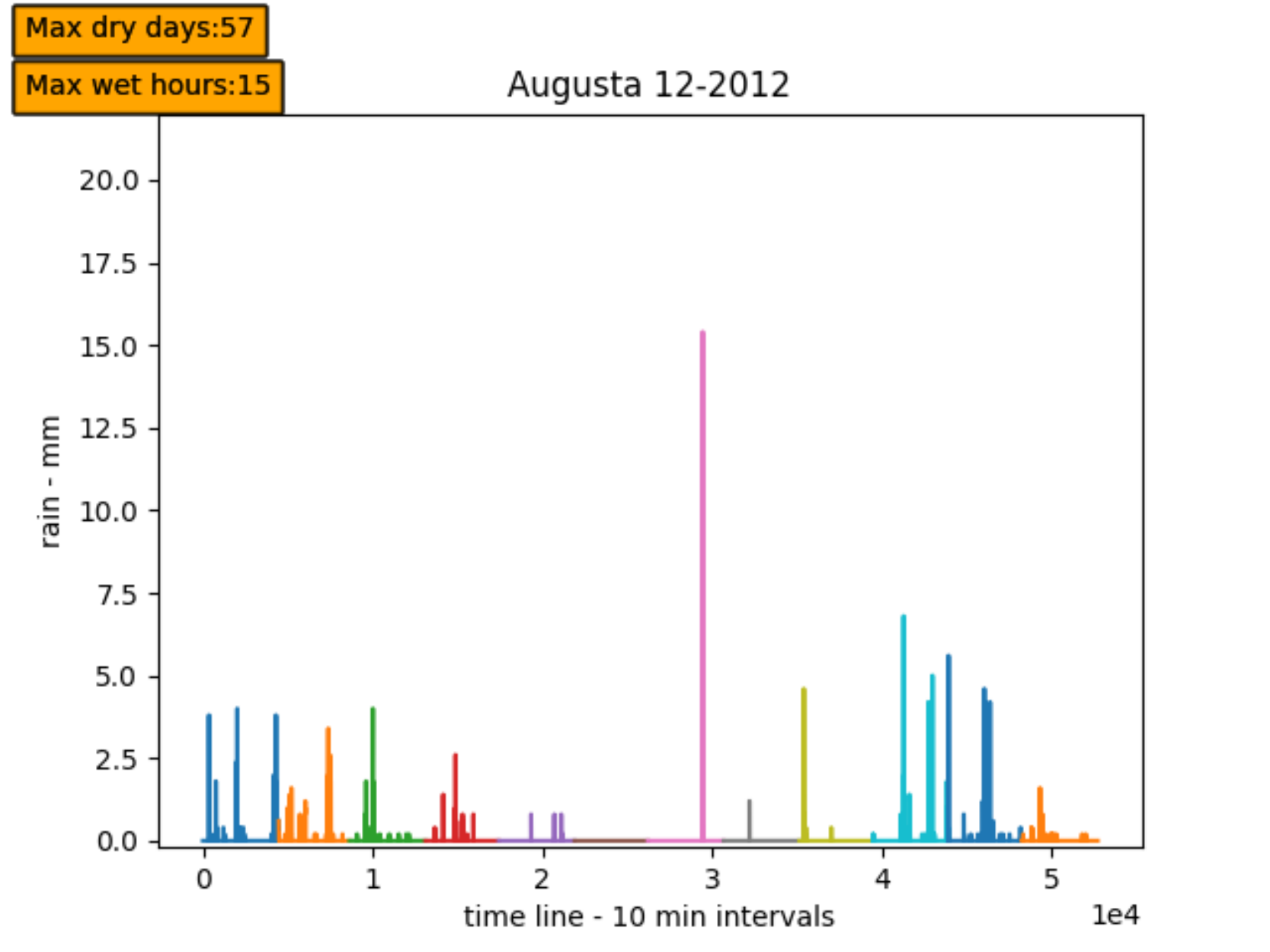}}
\subfloat[Annual view - 2013 Augusta]{\includegraphics[width = 3.25in]{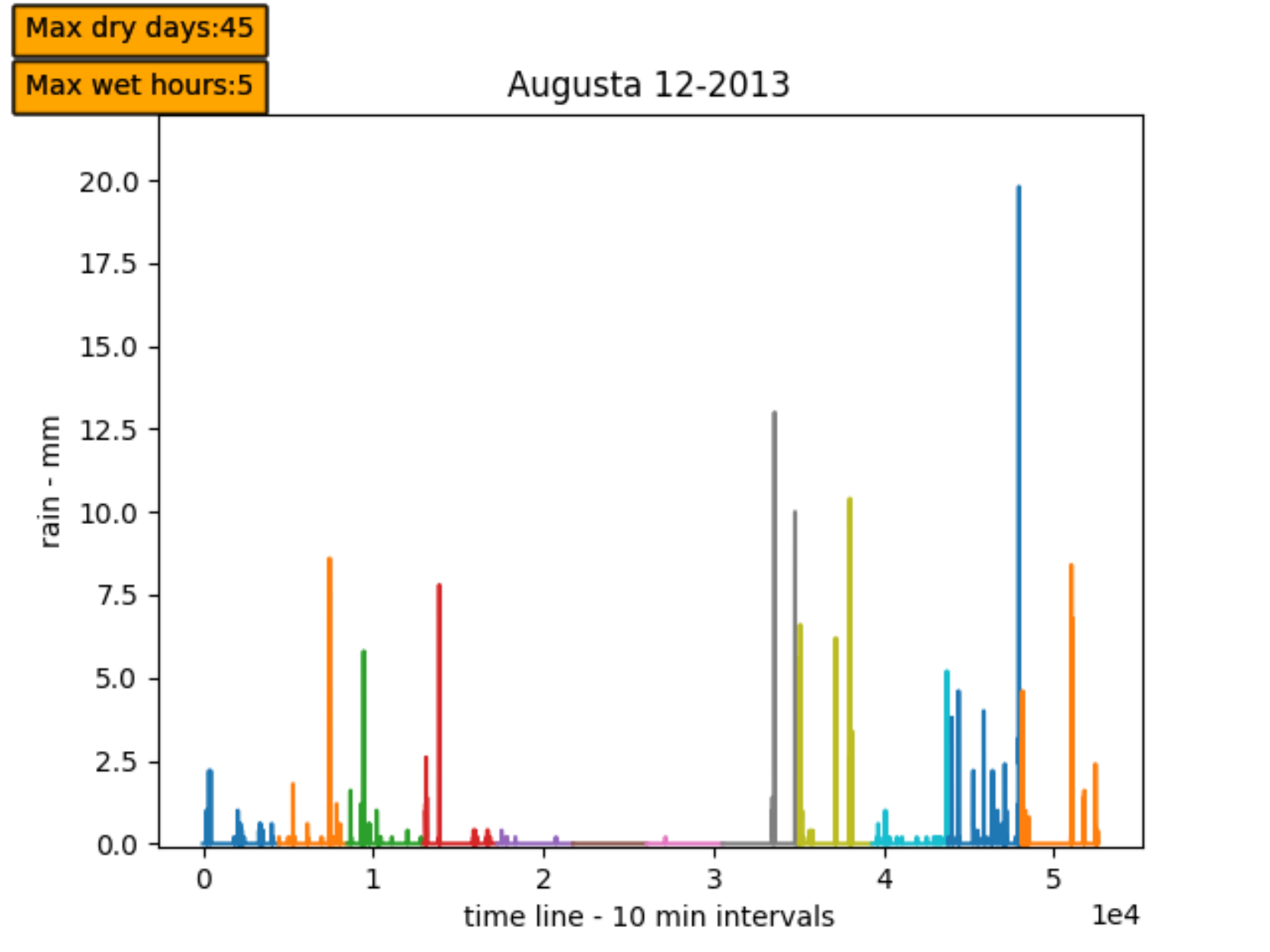}}
\label{fig:augusta1}
\caption{Annual view - Augusta}
\end{figure}
In 2009 we observed the \textit{Max dry days} value of $39$ achieved between July and August, and the \textit{Max wet hours} value of $8$ achieved in January. From this graphic we see seasonality, in fact the higher peaks lies in the autumn season, which is considered the rainy season for the Mediterranean area. There are some anomalies in November, due to the absence of peaks.\par Moreover, we observed a very similar situation in 2010, this is why we do not report its plot. In this case the \textit{Max dry days} value of $36$, achieved between July and August, is lower than the one in 2009 and the \textit{Max wet hours} value of $11$ achieved in March is higher than the one of 2009.\par 2011 is very different from the other years. Its \textit{Max dry days} value coincides with the global one, therefore it is the driest year. This behaviour is also associated with rainfall peaks higher than 2009 and 2010 in the rainy season.\par On the other hand, 2012 represents the wettest year - in the sense of continuous rain - since its \textit{Max wet hours} value achieved in February coincides with the global one. Moreover, 2012 reached the maximum value in July. This behaviour is anomalous and shows the increase of the rainstorms phenomenon.\par
\begin{figure}[h!]
\subfloat[Annual view - 2015 Augusta]{\includegraphics[width = 3.25in]{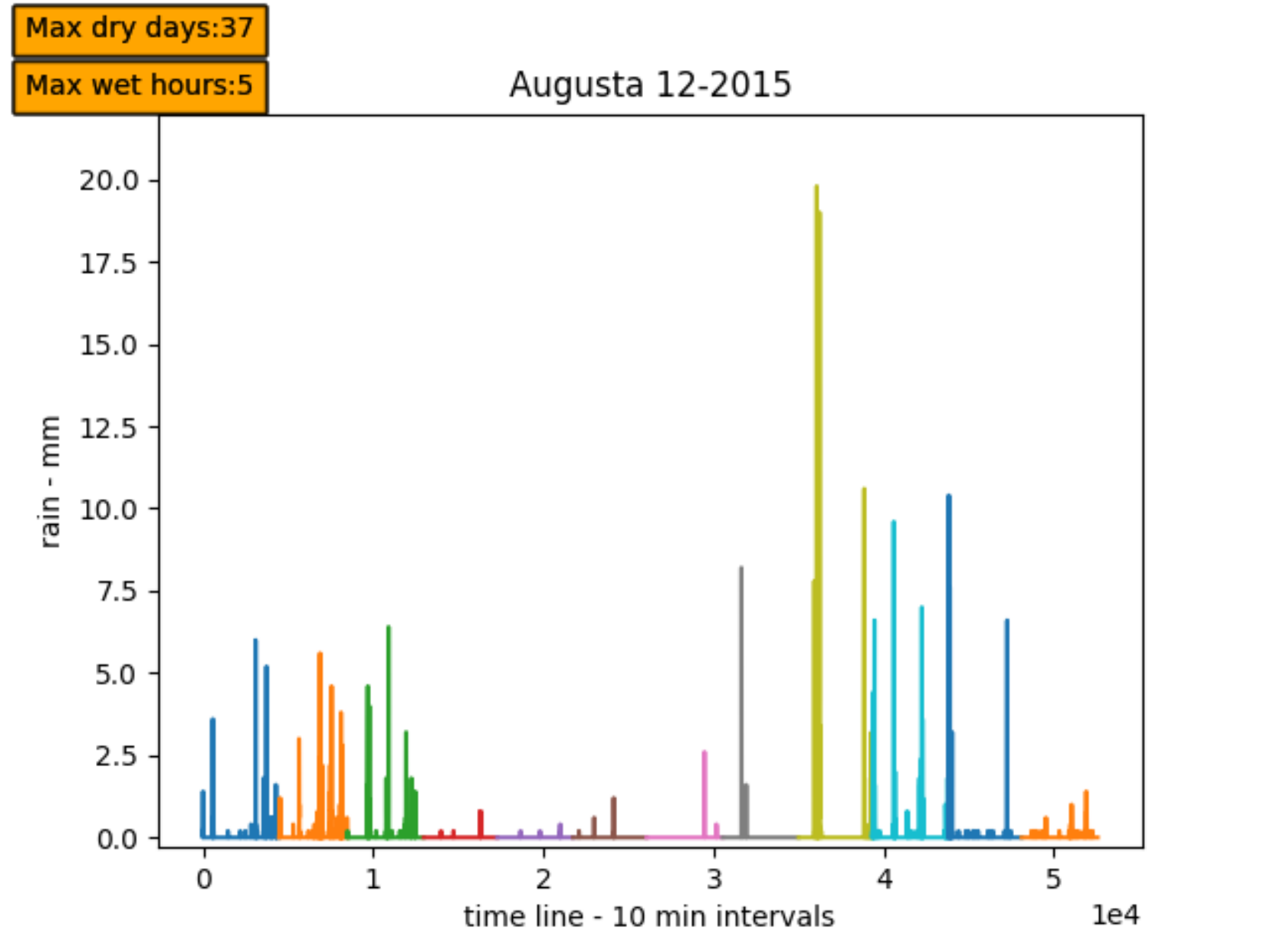}} 
\subfloat[Annual view - 2016 Augusta]{\includegraphics[width = 3.25in]{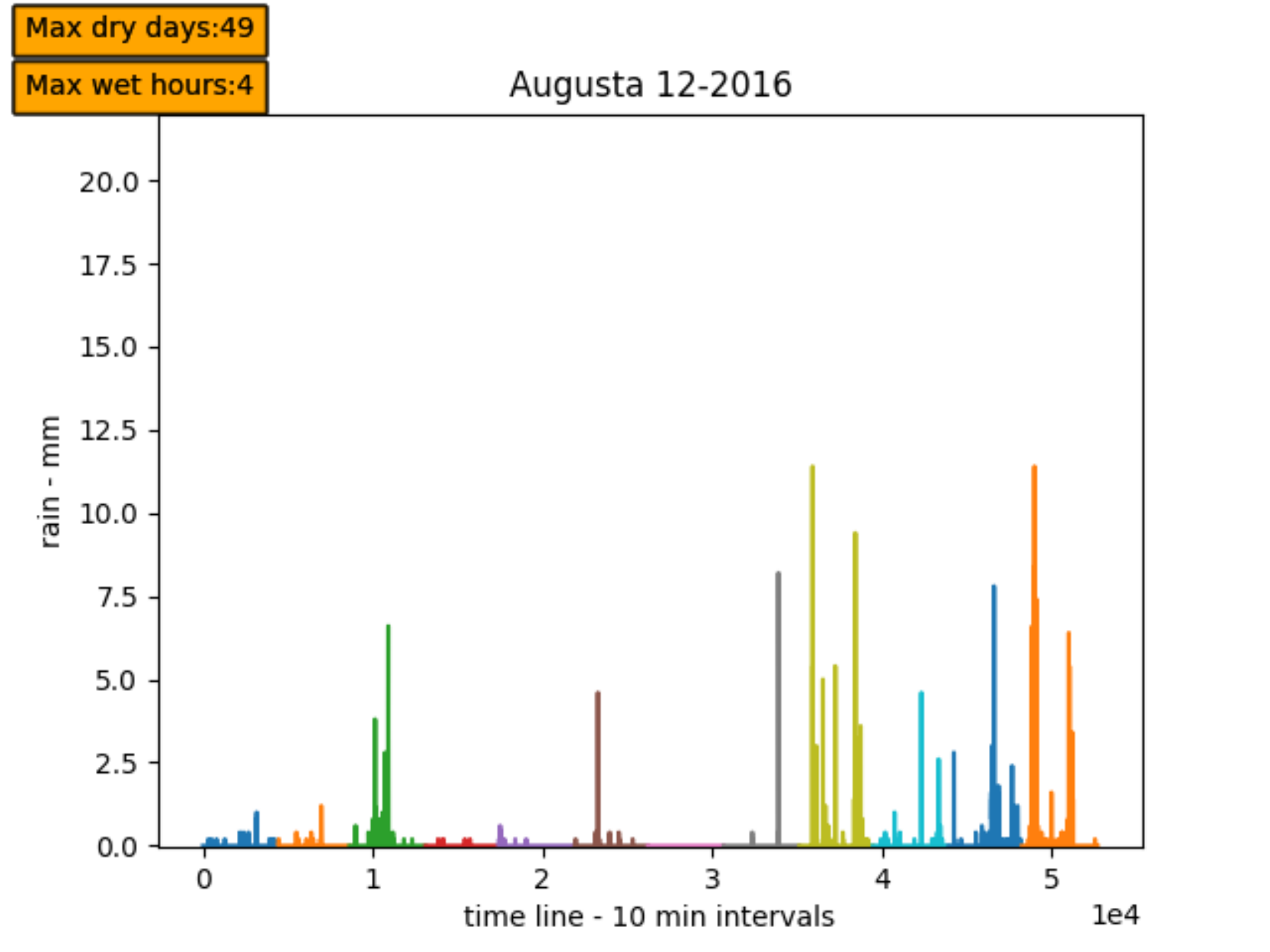}}\\
\subfloat[Annual view - 2017 Augusta]{\includegraphics[width = 3.25in]{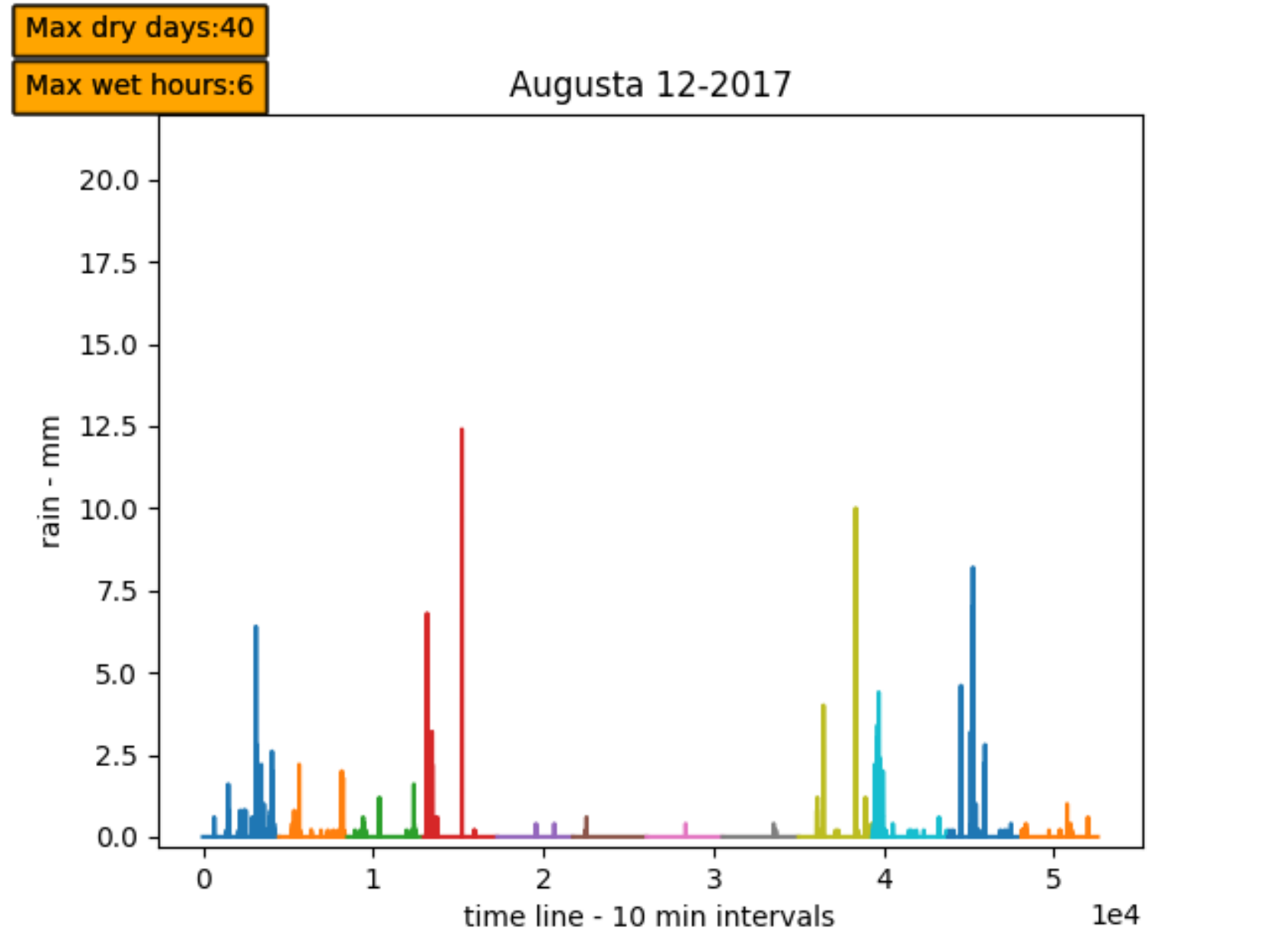}}
\subfloat[Annual view - 2018 Augusta]{\includegraphics[width = 3.25in]{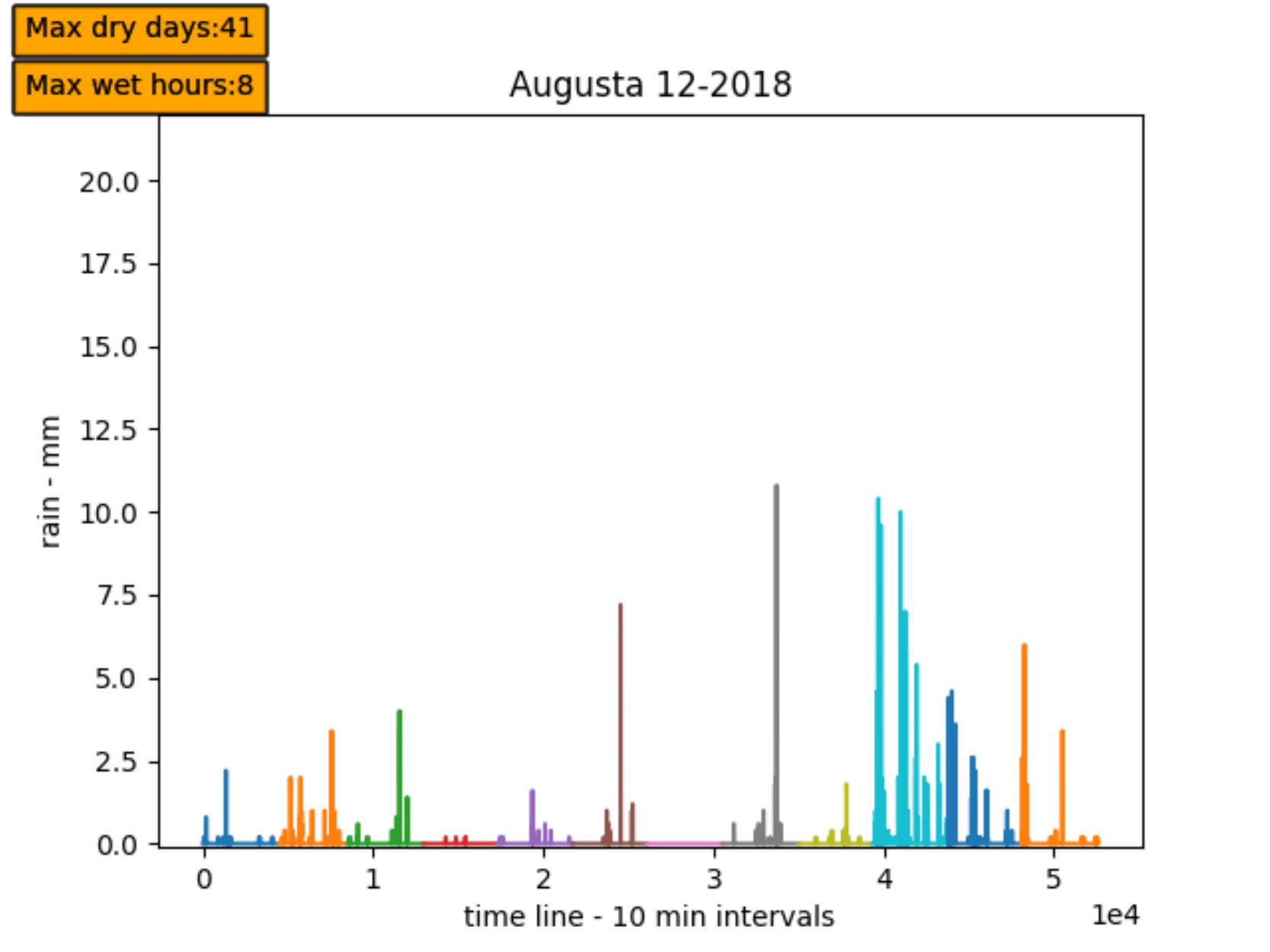}}
\label{fig:augusta2}
\caption{Annual view - Augusta}
\end{figure}
Moreover 2013 shows more or less a standard behaviour for Sicily, with dryness in the warmer months and wetness in the colder ones. We actually observe some anomalous peaks in August which refer to extreme unexpected rainfall events and a high annual peak of about 20 $mm$.\par
Furthermore 2014 did not highlight any remarkable event. This is why its plot is not reported here.\par 2015 has its peak in September, achieving again about $20$ $mm$. It is a very high value for a 10 minutes temporal step, in fact 2016, 2017 and 2018 never achieve that. \par 2017 unusually has its peak in April, whereas the rest of its behaviour seem standard. 2015, 2016 and 2018 show again peaks of about $10$ $mm$ in the summer months. 2019 shows a standard evolution, again with some peaks in July. For this reason we decided not to report it here.\par
\begin{figure}[h!]
\subfloat[Annual view - 2020 Augusta]{\includegraphics[width = 3.25in]{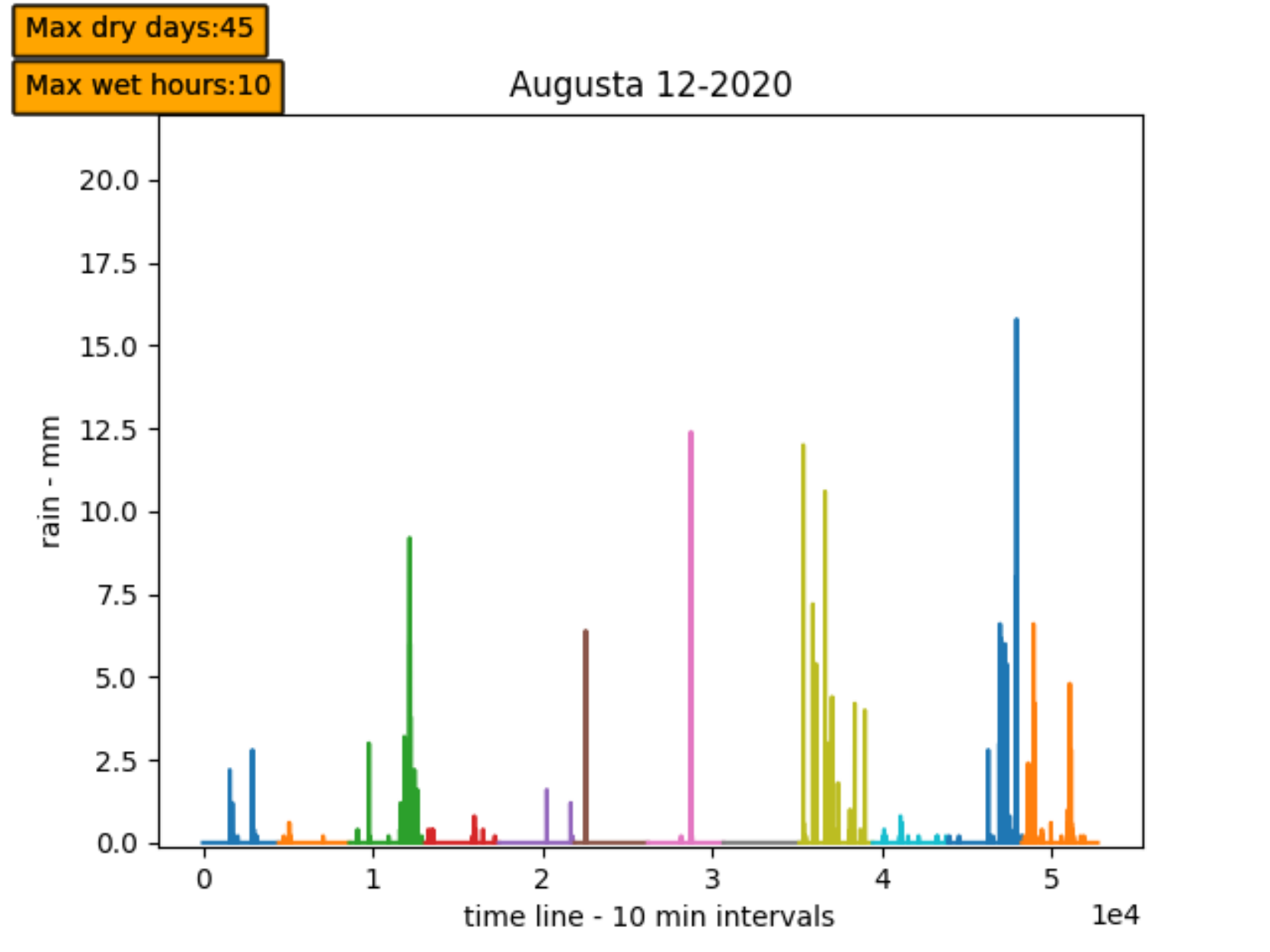}} 
\subfloat[Annual view - 2021 Augusta]{\includegraphics[width = 3.25in]{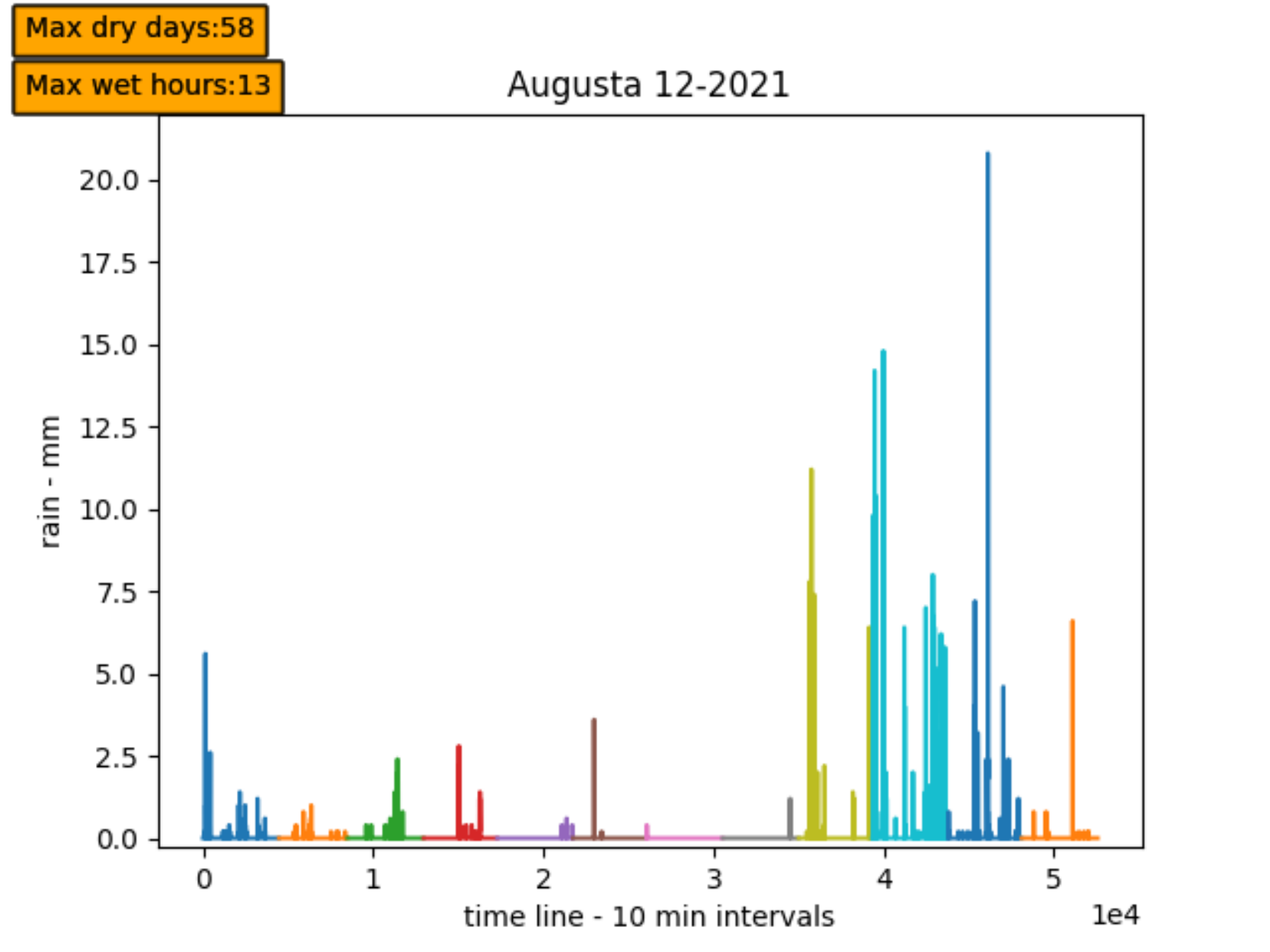}}\\
\label{fig:augusta3}
\caption{Annual view - Augusta}
\end{figure}
Eventually, 2020 showed an unusual October with no peaks, whereas 2021 reveals a tendency of very low rain during the year, except for September, October and November, where in contrast the peak coincide with the global one. This indicates a sort of redistribution of rainfall events over the years.\par The general trend passed from an annual balanced distribution, for example in 2011 or 2013, to an unequal one. In other words, from this annual observation we conclude that in Augusta it rains less frequently than before but harder and especially in unexpected periods.
\subsection{Monthly view}
We report here the November distributions over years. We take it as an example, but we could do the same for any month, obtaining different new observations.
\begin{figure}[h!]
\subfloat[2009]{\includegraphics[width = 3.1in]{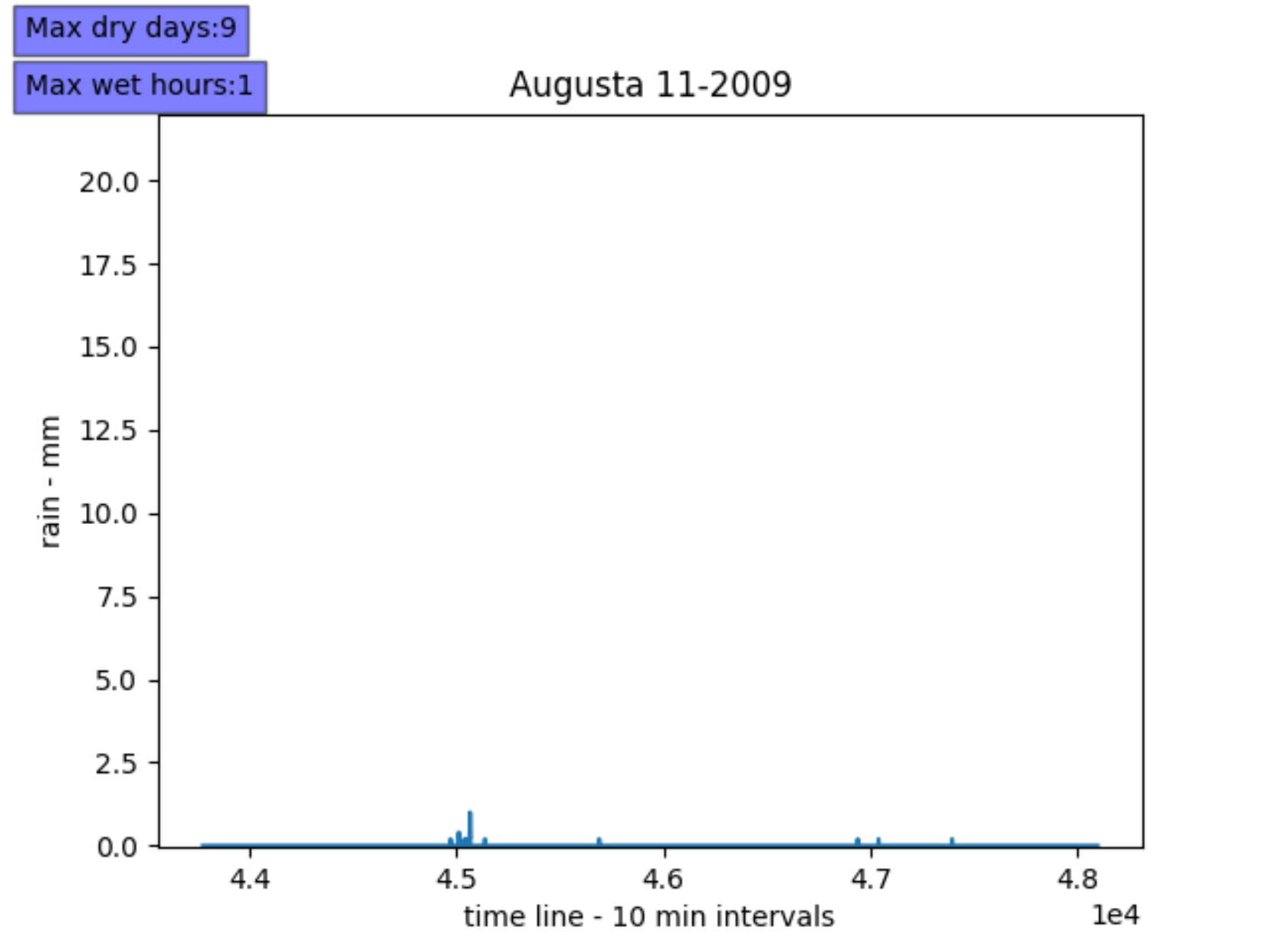}} 
\subfloat[2010]{\includegraphics[width = 3.1in]{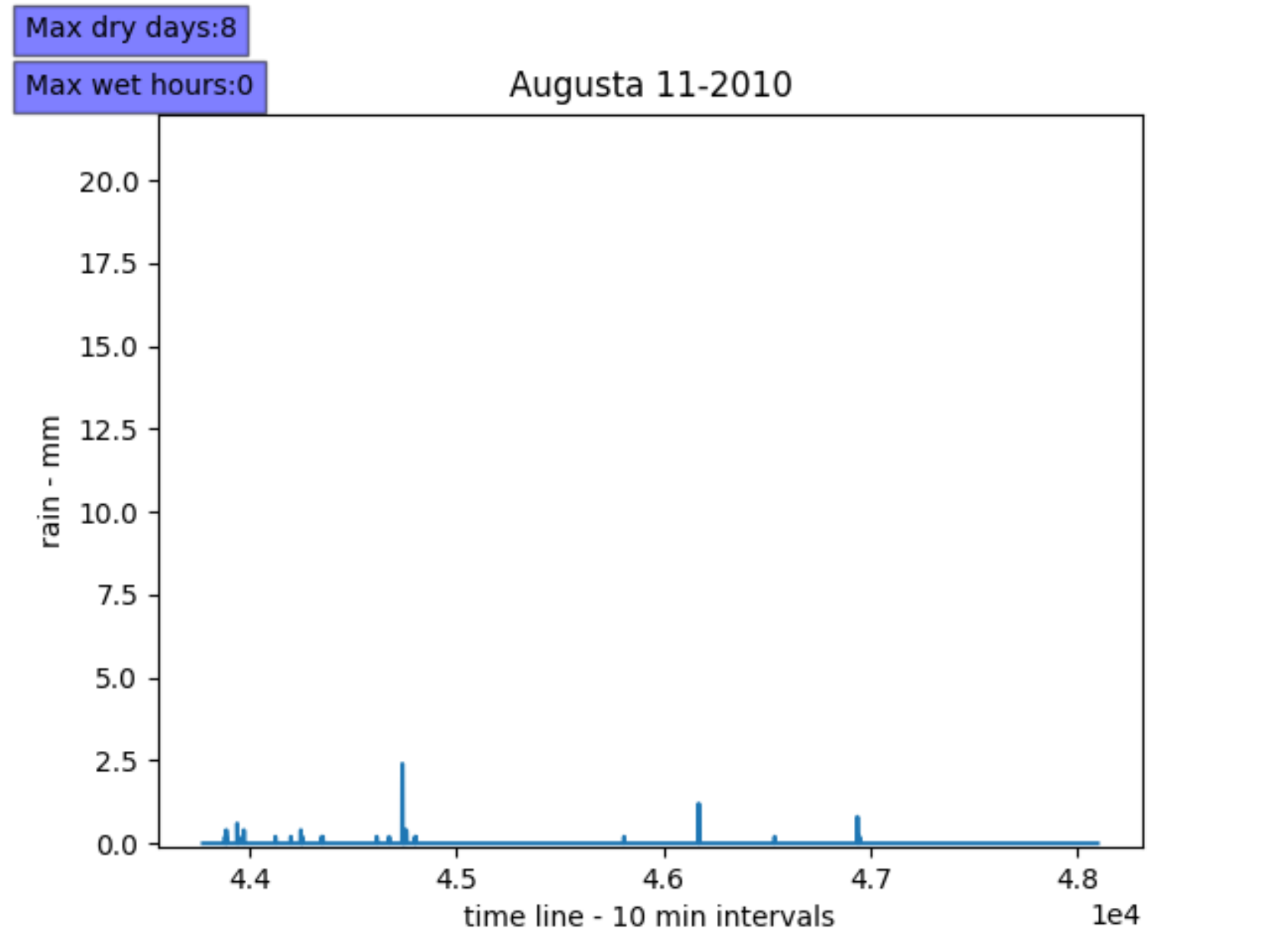}}\\
\subfloat[2011]{\includegraphics[width = 3.1in]{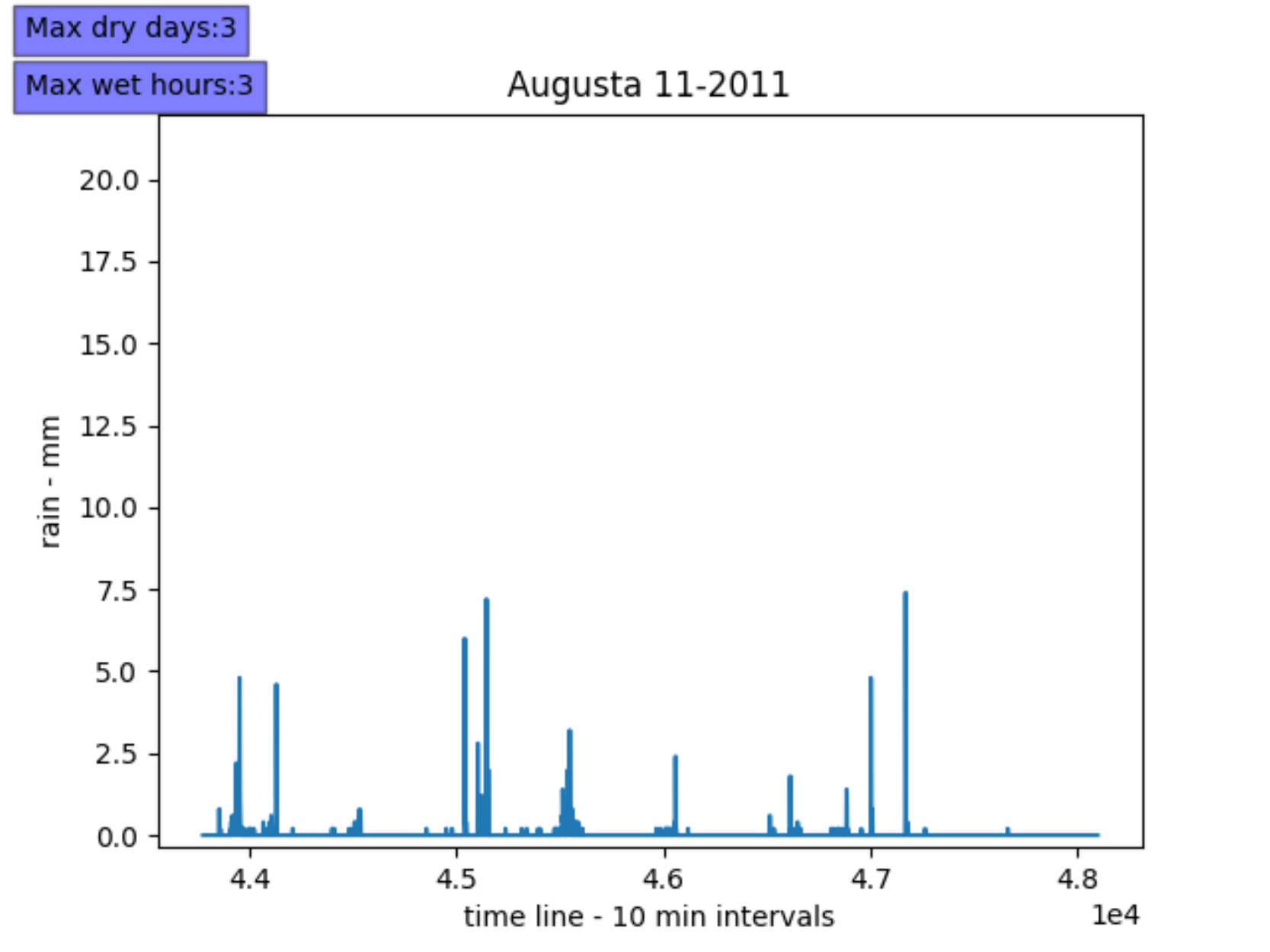}}
\subfloat[2012]{\includegraphics[width = 3.1in]{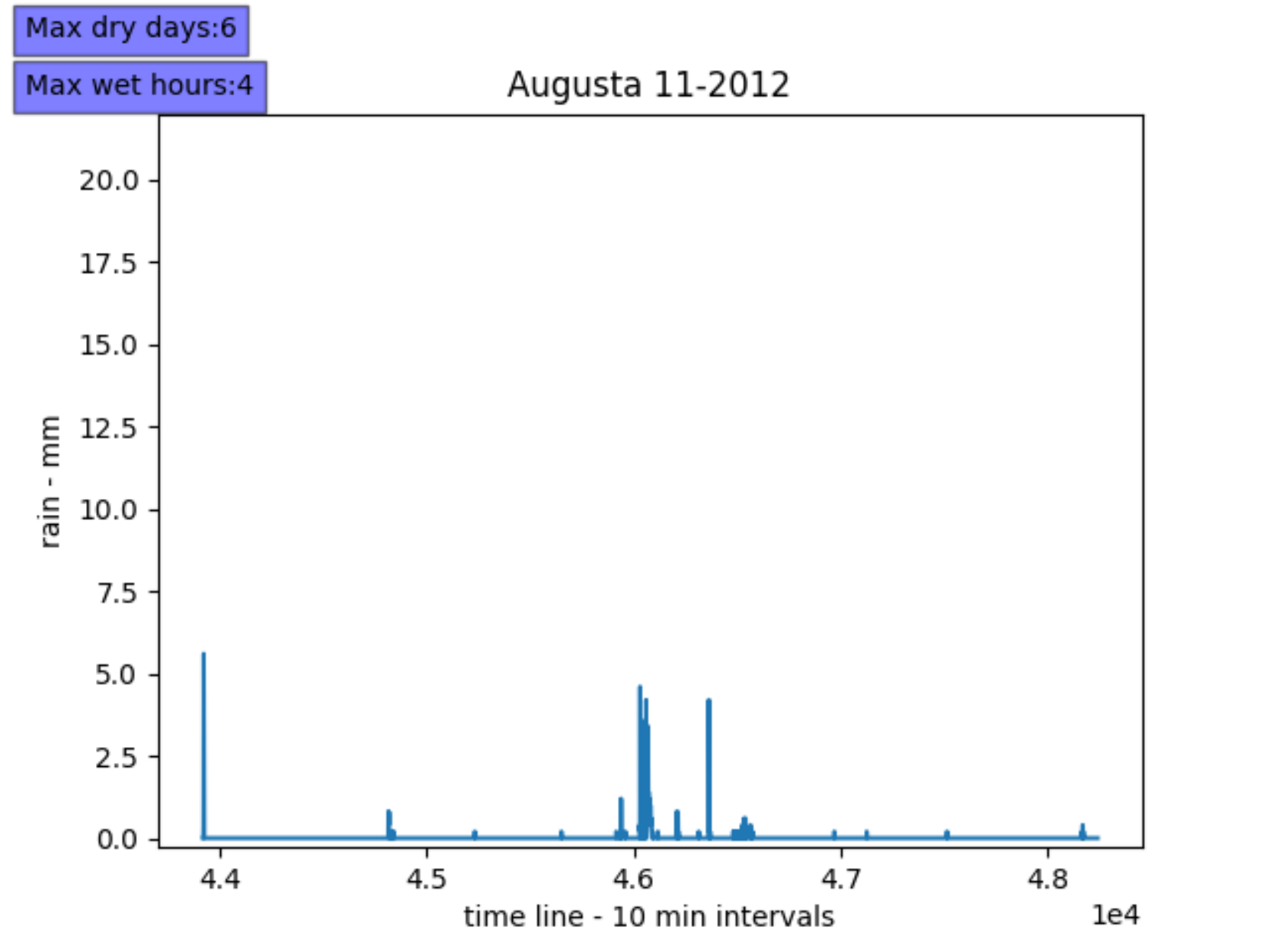}}\\
\subfloat[2013]{\includegraphics[width = 3.1in]{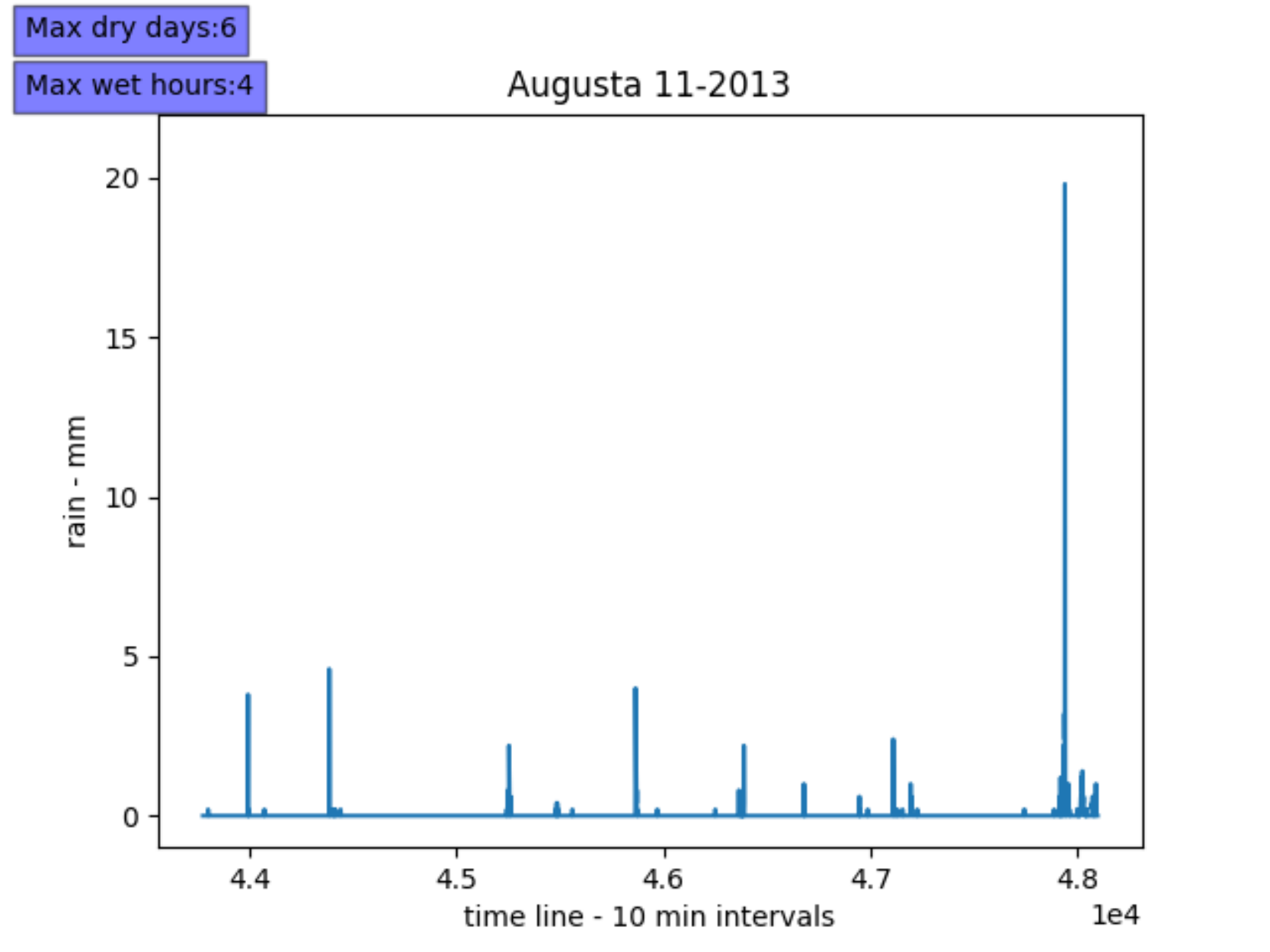}}
\subfloat[2014]{\includegraphics[width = 3.1in]{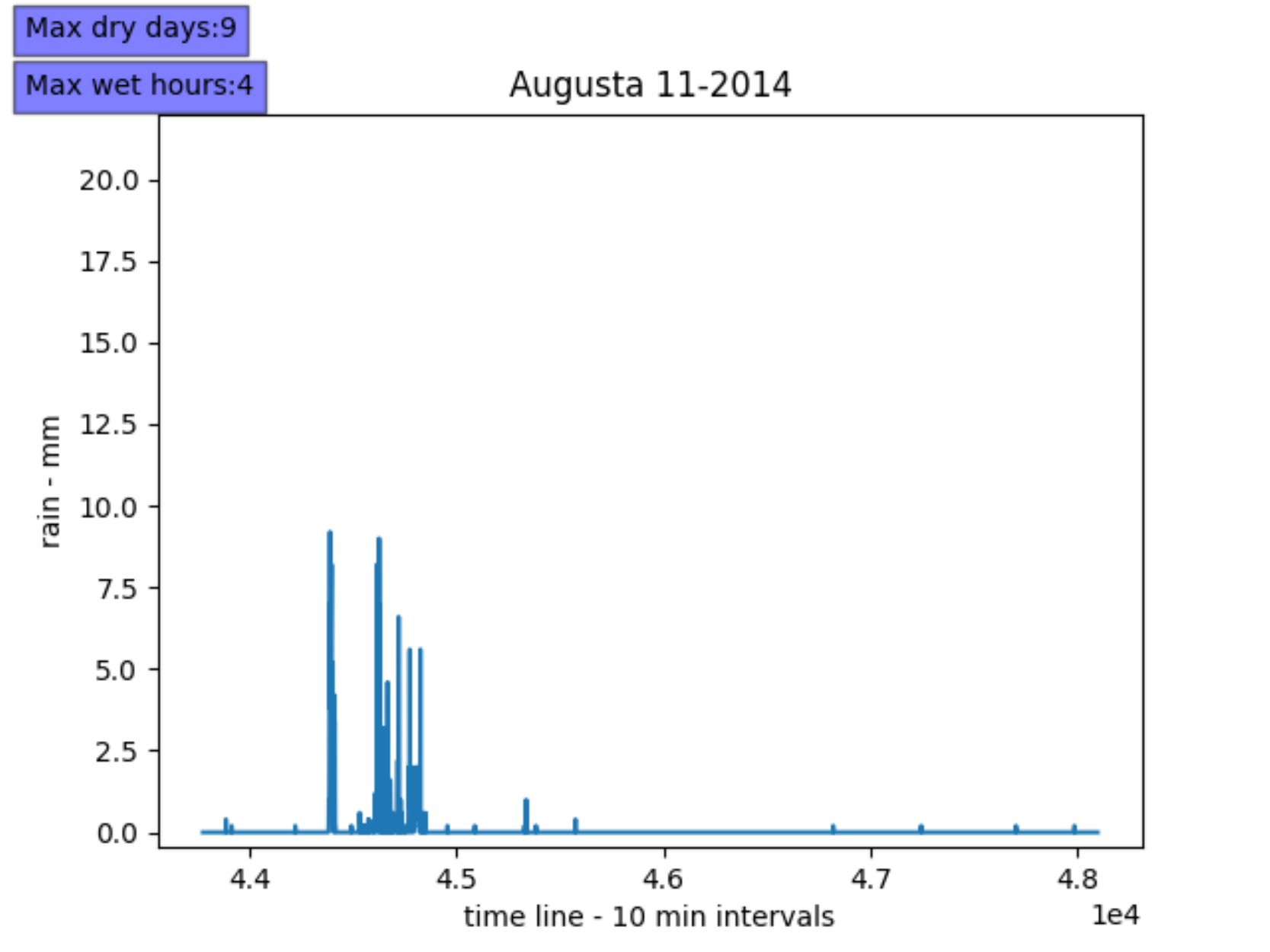}}\\
\caption{Monthly view - November - Augusta}
\label{fig:augusta4}
\end{figure}
Figure \ref{fig:augusta4} shows November in the first 6 years of our investigation, from 2009 to 2014. We see that November rainfall events are very rare in 2009 and 2010, whereas they are  common in 2011. The other three years show several rainfall peaks, notably the one of about $20$ $mm$ in 2013.
\begin{figure}[h!]
\subfloat[2015]{\includegraphics[width = 3.1in]{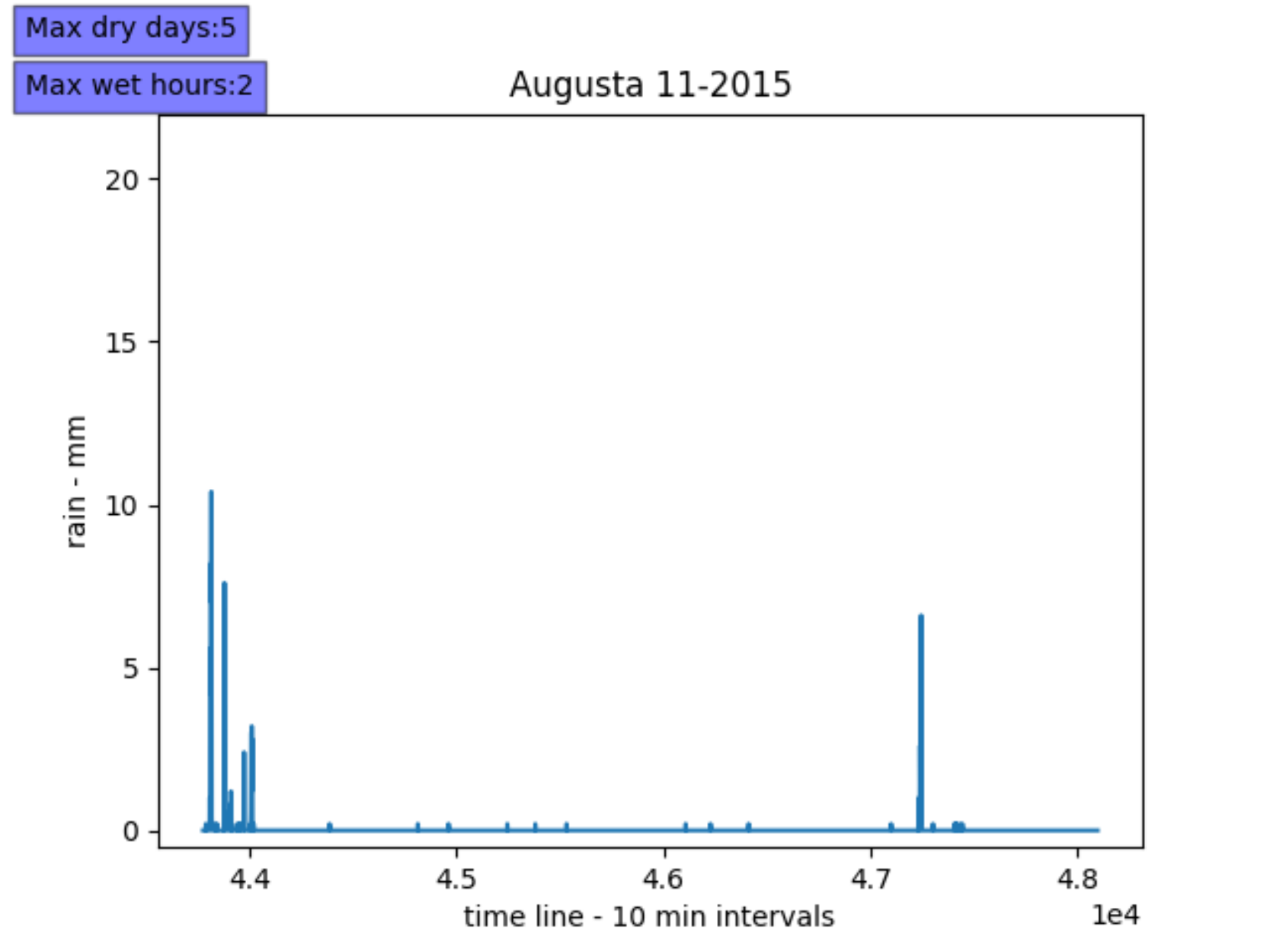}}
\subfloat[2016]{\includegraphics[width = 3.1in]{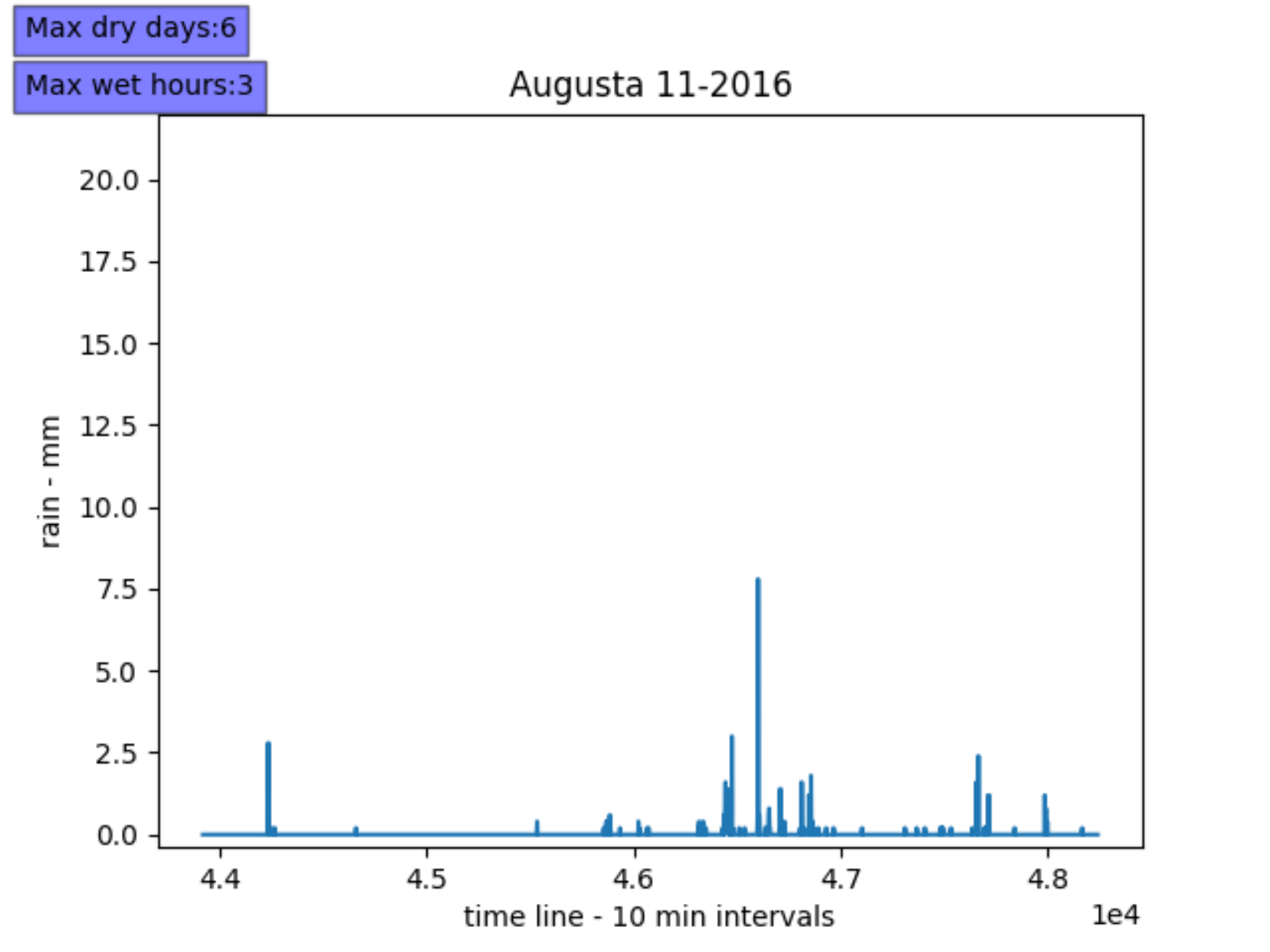}}\\
\subfloat[2017]{\includegraphics[width = 3.1in]{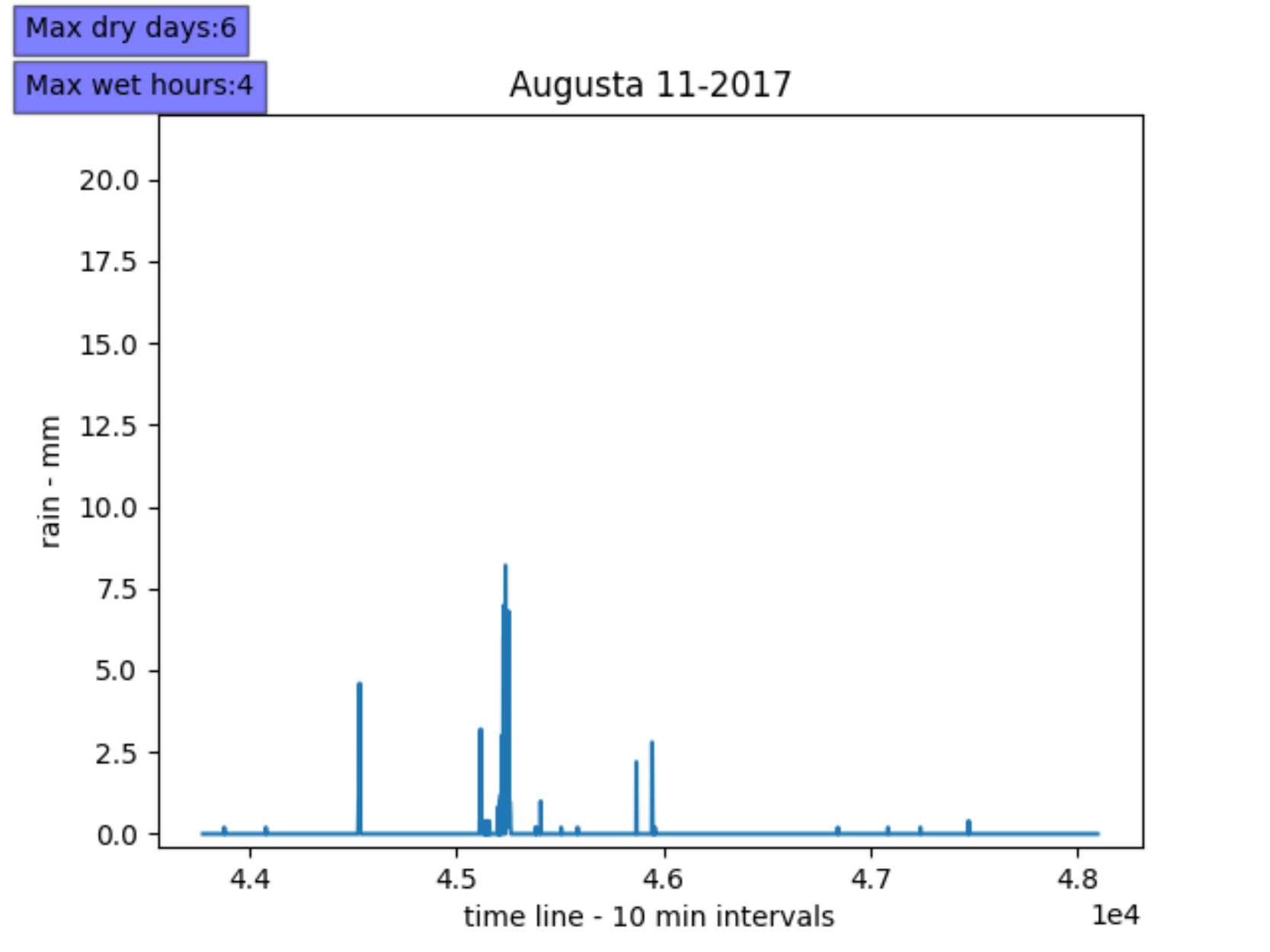}}
\subfloat[2018]{\includegraphics[width = 3.1in]{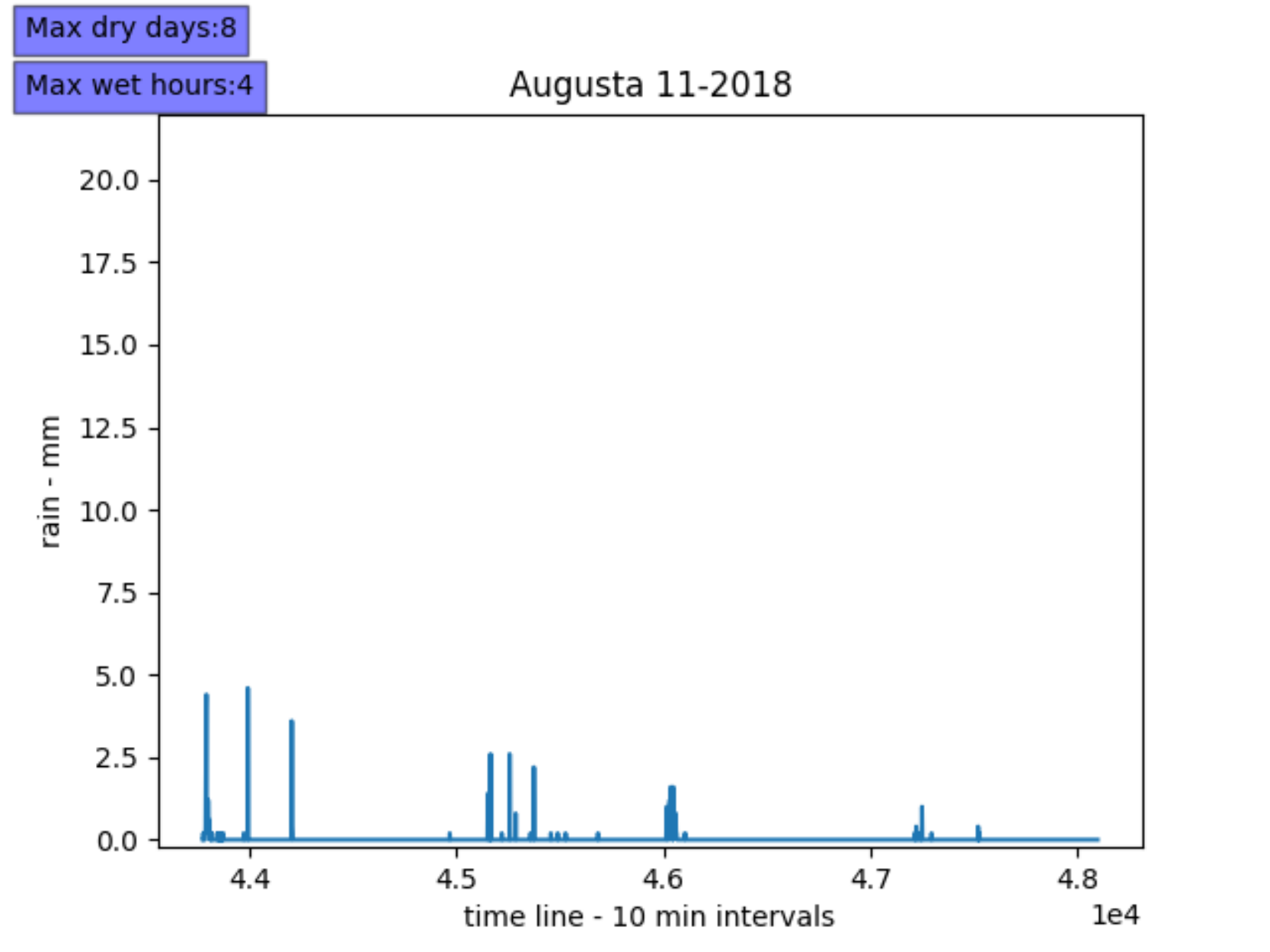}}\\
\subfloat[2019]{\includegraphics[width = 3.1in]{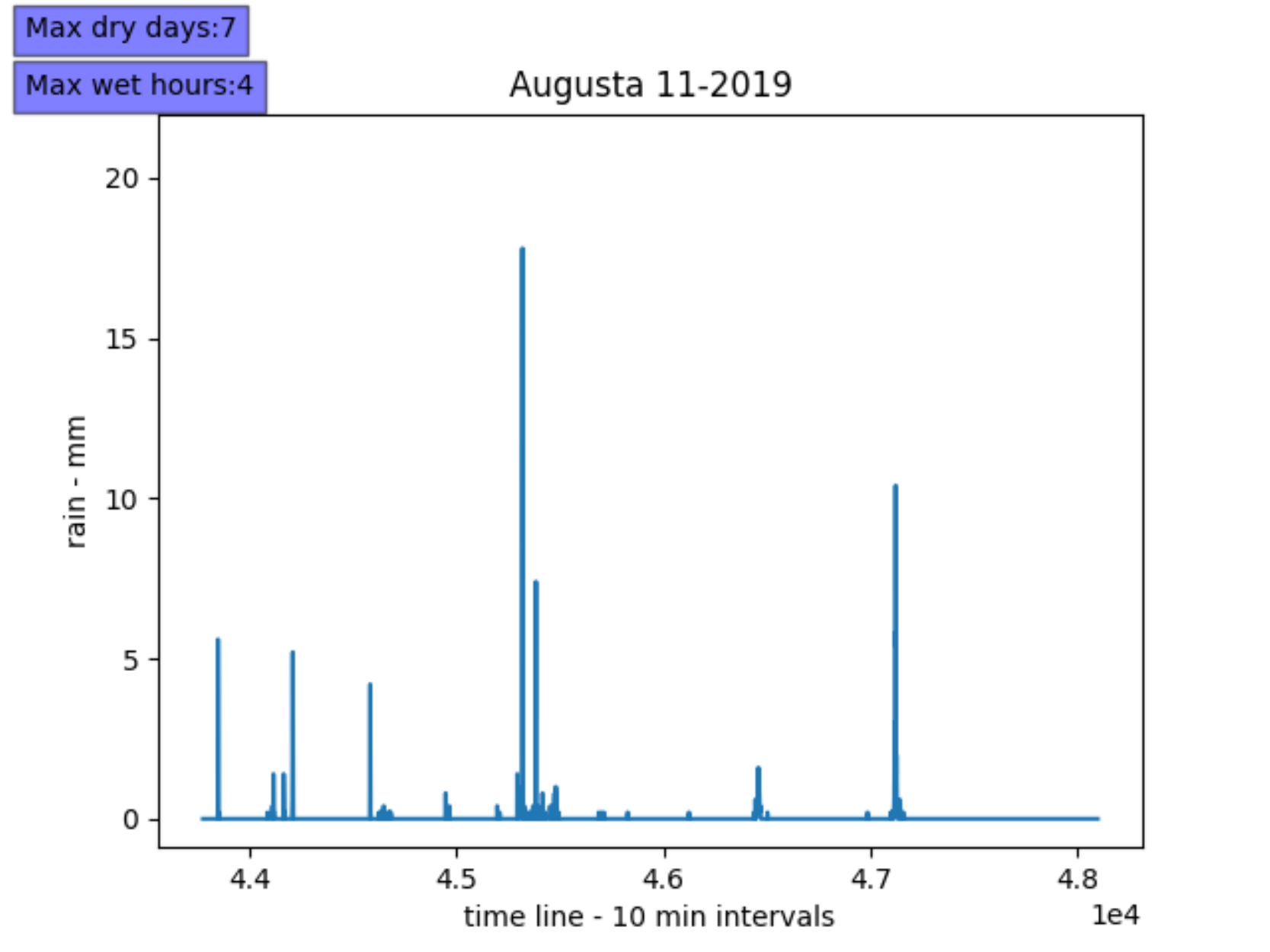}}
\subfloat[2020]{\includegraphics[width = 3.1in]{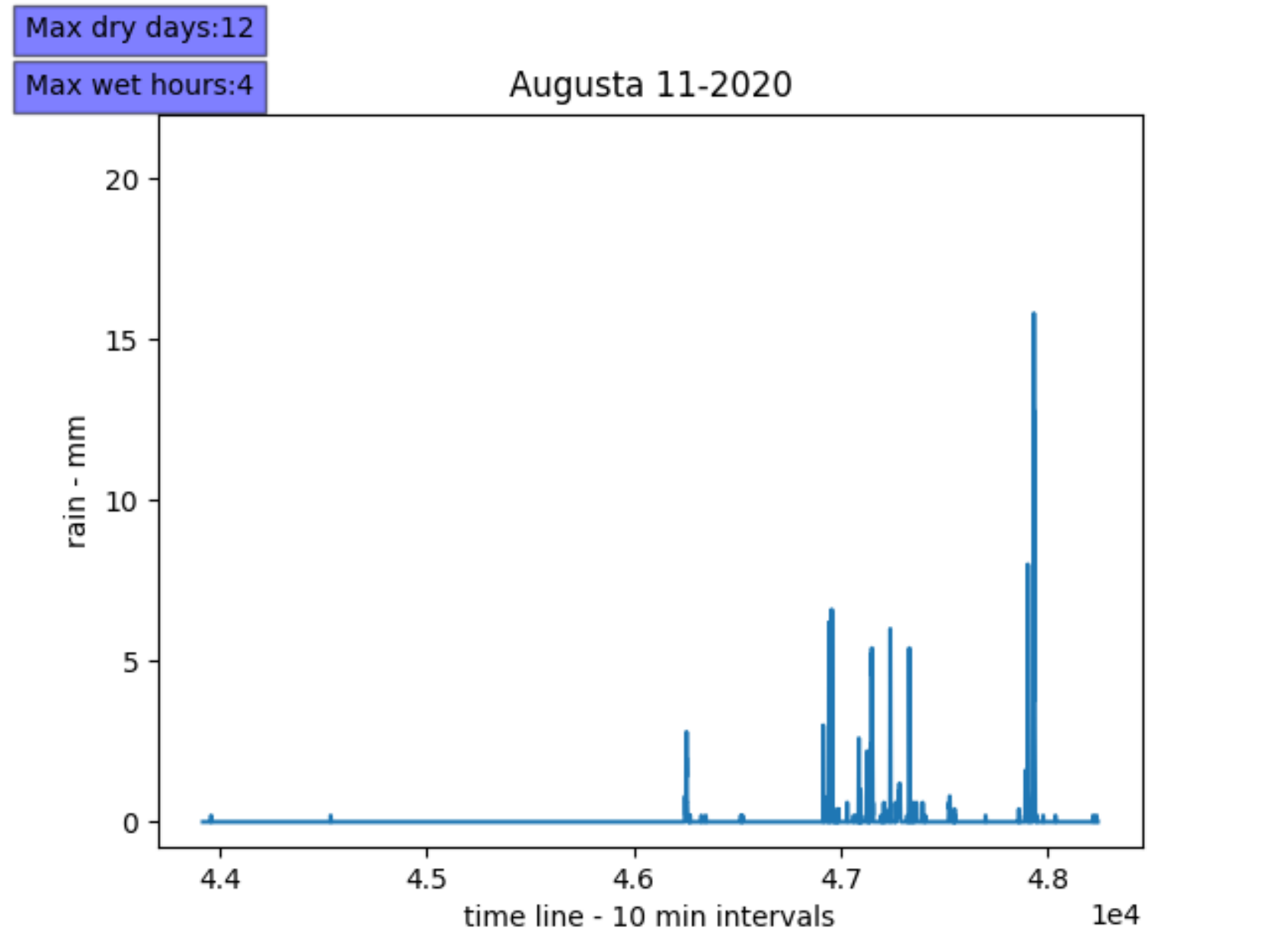}}\\
\caption{Monthly view - November - Augusta}
\label{fig:augusta5}
\end{figure}
\clearpage
\begin{figure}[h!]
\subfloat[2021]{\includegraphics[width = 3.1in]{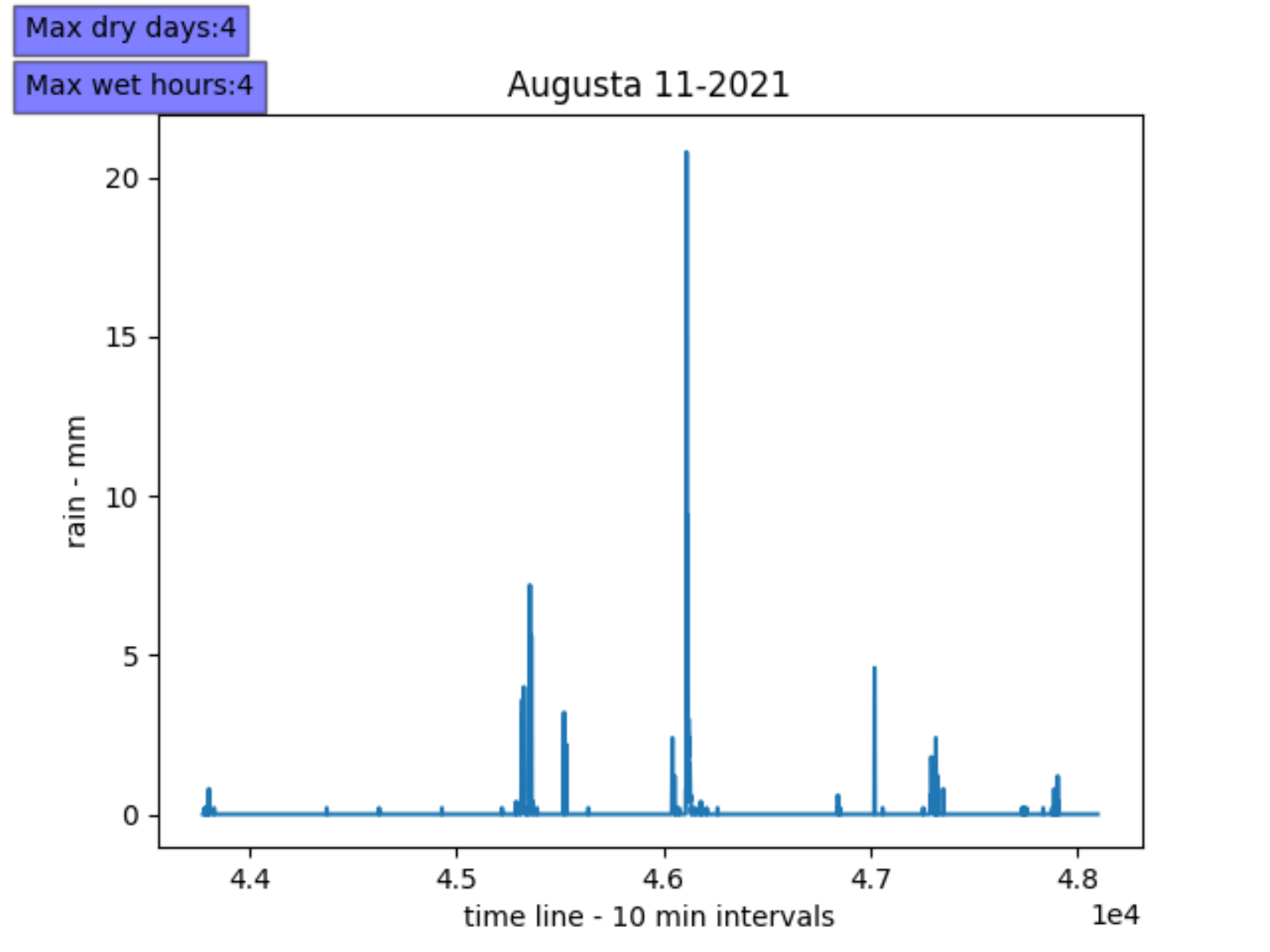}}
\caption{Monthly view - November - Augusta}
\label{fig:augusta6}
\end{figure}
Moreover figures \ref{fig:augusta5} and \ref{fig:augusta6} show November in the last 7 years of our investigation, from 2015 to 2021. Again we observe a general trend to storms with peaks reaching more or less the global maximum of Augusta, for instance in 2019 and in 2021. This is an alarming observation that confirms a change in the climate of Augusta. Data suggest that those huge peaks will be more and more frequent, becoming a threat to the environment and the population.
\section{Annual results}\label{sec.c}
Here we report the annual clustering results in all the considered settings. In particular, in the case of Euclidean metrics and C.$A$ (see the main document for the Collections descriptions), the results consist in a principal cluster and some exceptions, mainly in the East side of the island. This suggests the vulnerability to extreme events of the East side of Sicily respect to the West. In the case of C.$B$, there are many principal clusters and some exceptions, again mostly in the East side.
Differently, in the cases of Correlation metrics with both C.$A$ and C.$B$, most of the time results consist in two clusters splitting Sicily in half.
It follows that Euclidean metrics let to better detect outliers respect to the Correlation metrics.
Consequently, the Euclidean metrics seems to be more suitable and precise than the Correlation metrics.
\begin{figure}[h!]
\begin{minipage}{.5\linewidth}
\centering
\subfloat[Euclidean metrics and  C.$A$ ]{\label{2009ea}\includegraphics[width = 3.2in]{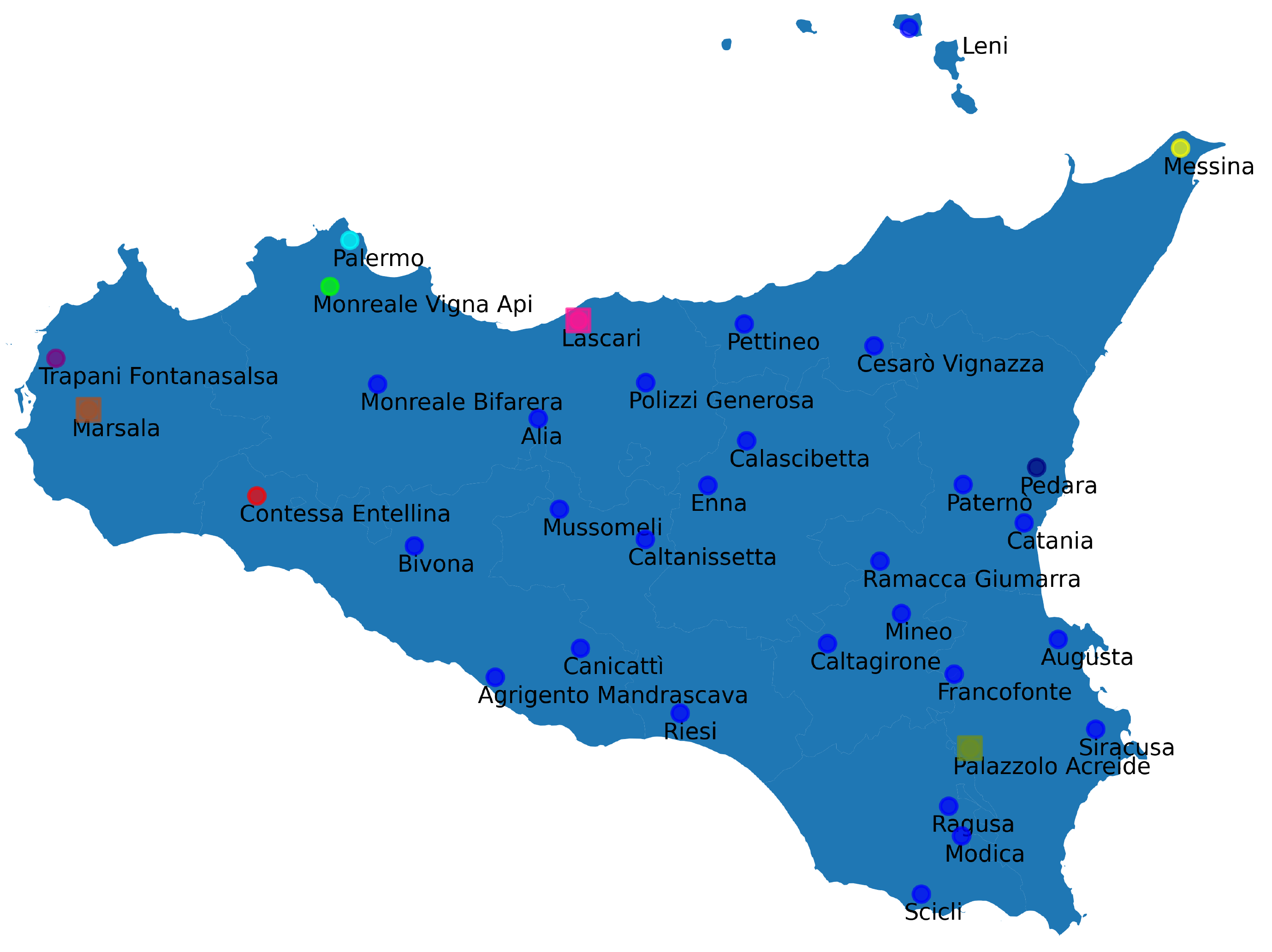}}
\end{minipage}%
\begin{minipage}{.5\linewidth}
\centering
\subfloat[Correlation metrics and C.$A$]{\label{2009ca}\includegraphics[width = 3.2in]{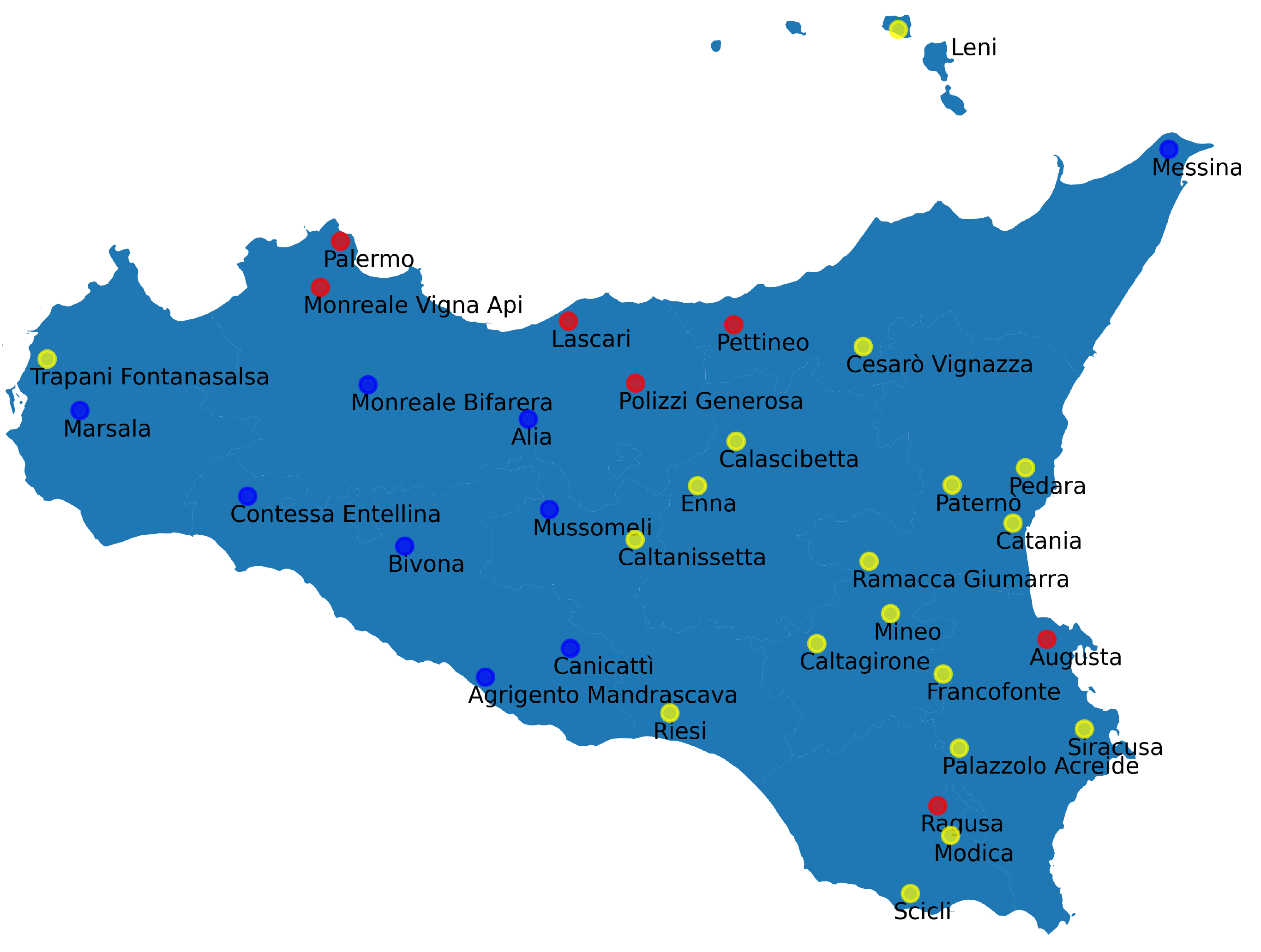}}
\end{minipage}\par\medskip
\begin{minipage}{.5\linewidth}
\centering
\subfloat[Euclidean metrics and C.$B$ ]{\label{2009eb}\includegraphics[width = 3.2in]{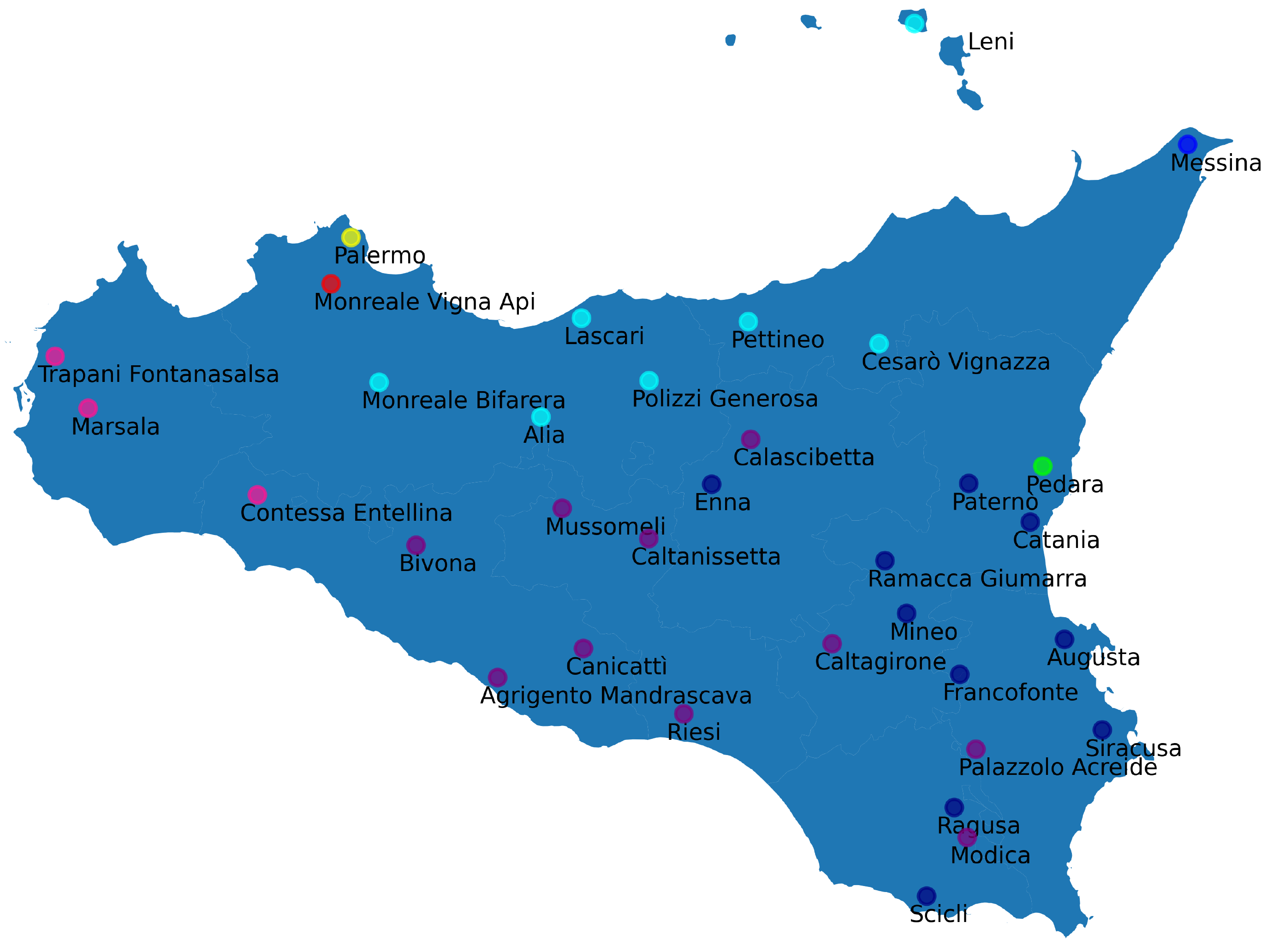}}
\end{minipage}%
\begin{minipage}{.5\linewidth}
\centering
\subfloat[Correlation metrics and C.$B$]{\label{2009cb}\includegraphics[width = 3.2in]{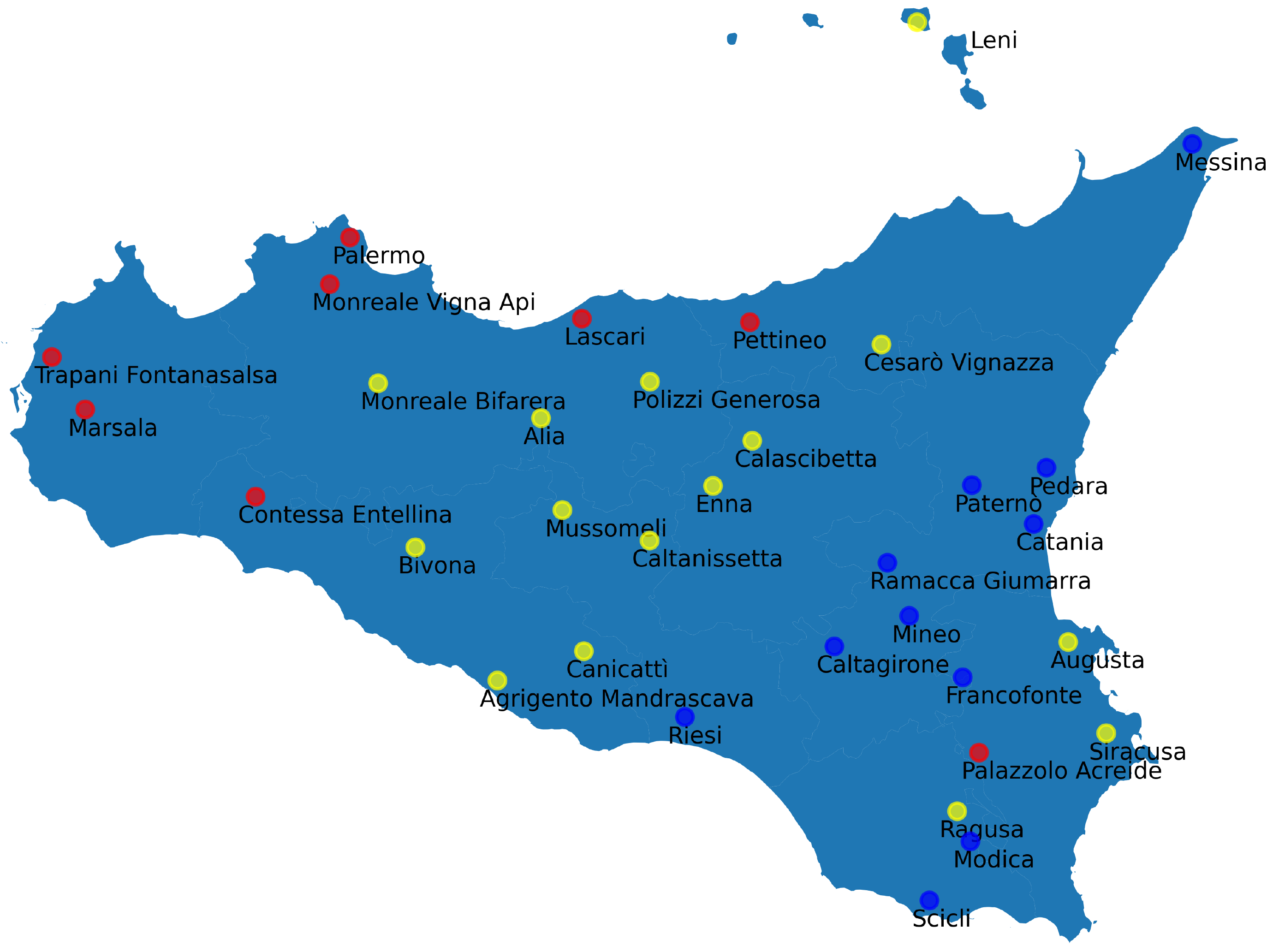}}
\end{minipage}\par\medskip
\caption{Annual case - 2009\\ \textbf{Panel a}: The principal cluster is reported in blue. \textbf{Panel b}: The three clusters are reported in blue, red and yellow. \textbf{Panel c}: The four principal clusters are reported in light blue, dark blue, purple and pink. \textbf{Panel d}: The three clusters are reported in blue, red and yellow.}
\label{annual2009}
\end{figure}

\begin{figure}[h!]
\begin{minipage}{.5\linewidth}
\centering
\subfloat[Euclidean metrics and C.$A$]{\label{2010ea}\includegraphics[width = 3.2in]{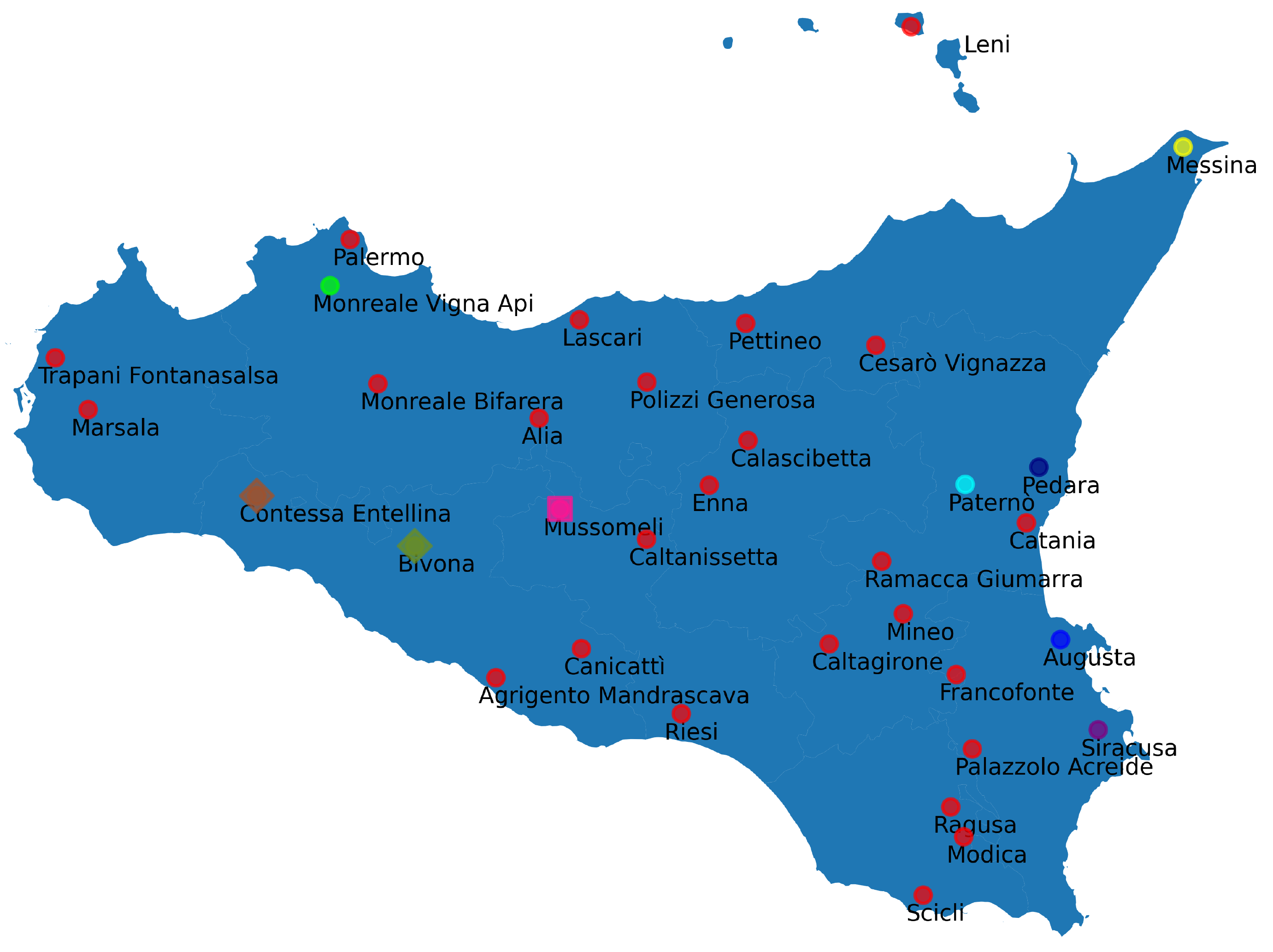}}
\end{minipage}%
\begin{minipage}{.5\linewidth}
\centering
\subfloat[Correlation metrics and C.$A$]{\label{2010ca}\includegraphics[width = 3.2in]{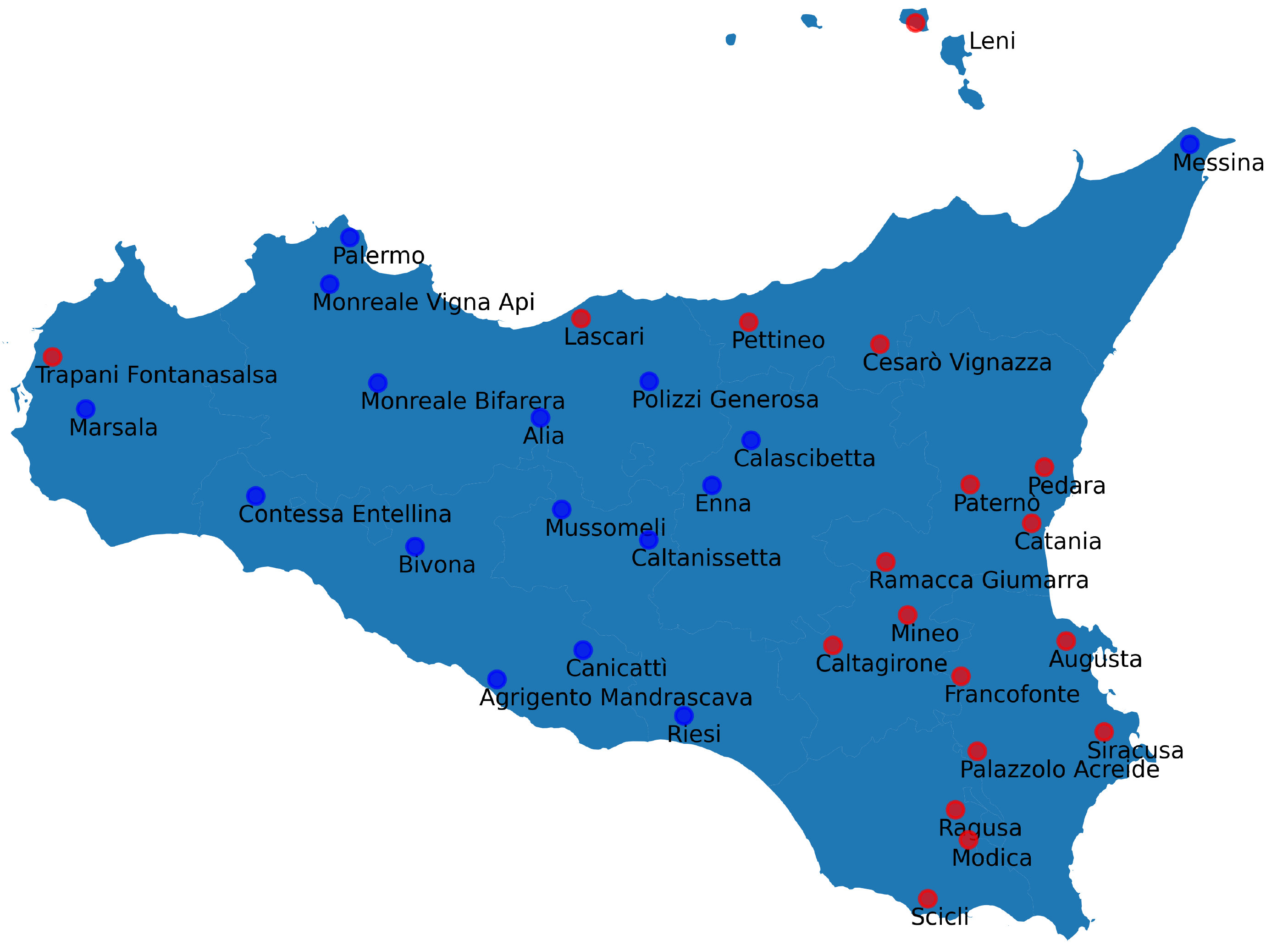}}
\end{minipage}\par\medskip
\begin{minipage}{.5\linewidth}
\centering
\subfloat[Euclidean metrics and C.$B$]{\label{2010eb}\includegraphics[width = 3.2in]{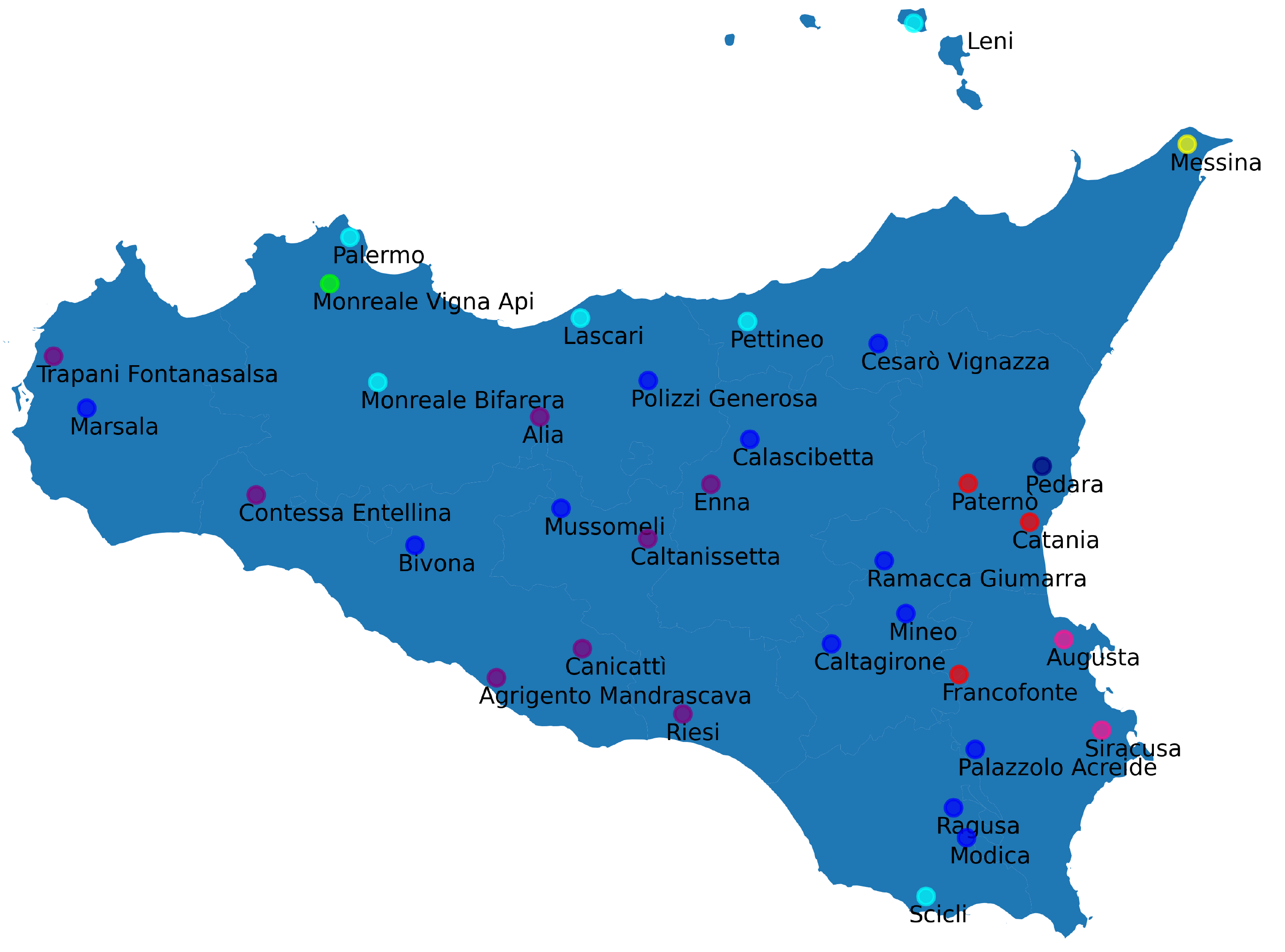}}
\end{minipage}%
\begin{minipage}{.5\linewidth}
\centering
\subfloat[Correlation metrics and C.$B$]{\label{2010cb}\includegraphics[width = 3.2in]{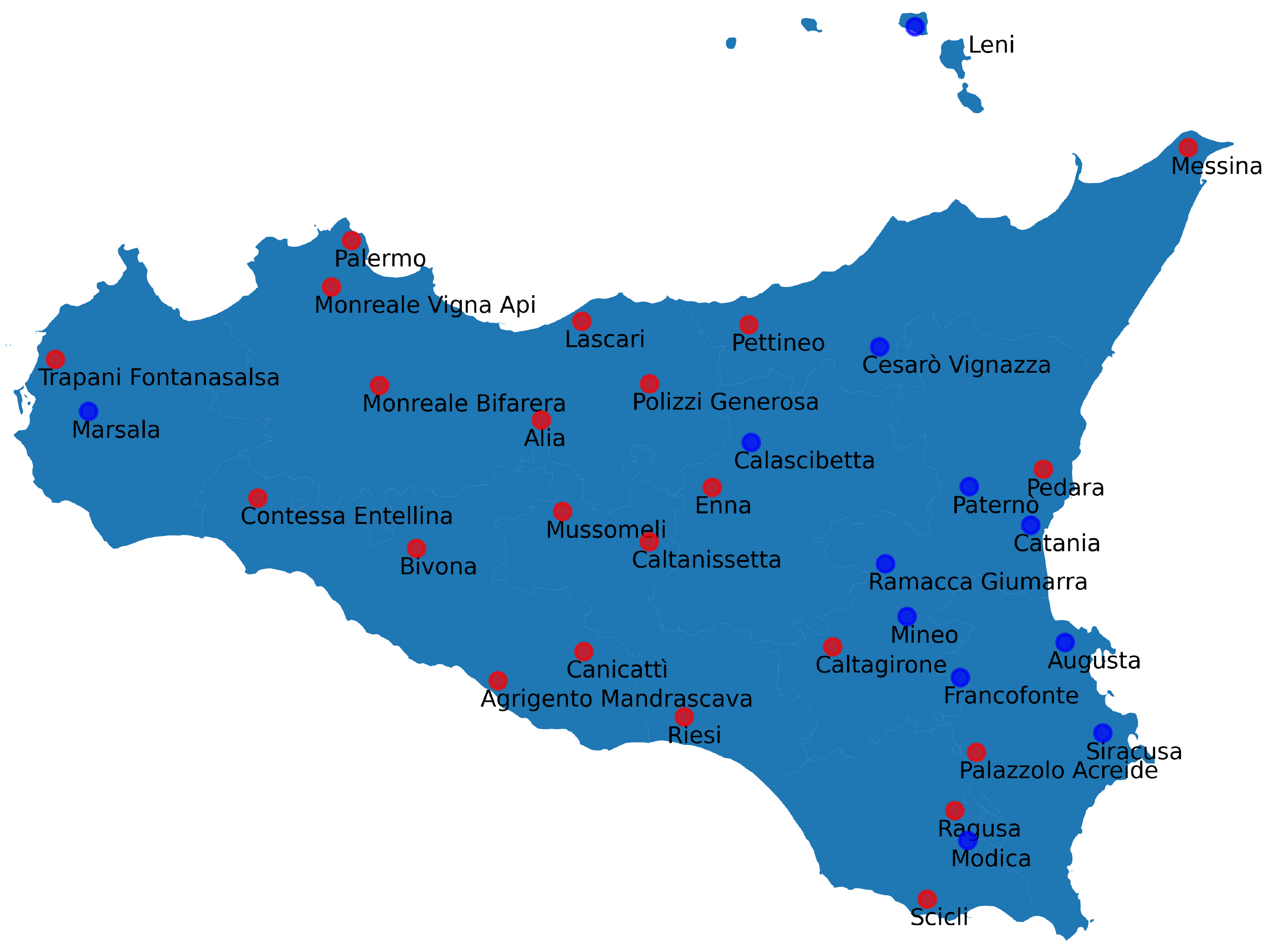}}
\end{minipage}\par\medskip
\caption{Annual case - 2010\\ \textbf{Panel a}: The principal cluster is reported in red. \textbf{Panel b}: The two clusters are reported in red and blue. \textbf{Panel c}: The five principal clusters are reported in blue, red, light blue, pink and purple. \textbf{Panel d}: The two clusters are reported in red and blue.}
\label{annual2010}
\end{figure}

\begin{figure}[h!]
\begin{minipage}{.5\linewidth}
\centering
\subfloat[Euclidean metrics and C.$A$]{\label{2011ea}\includegraphics[width = 3.2in]{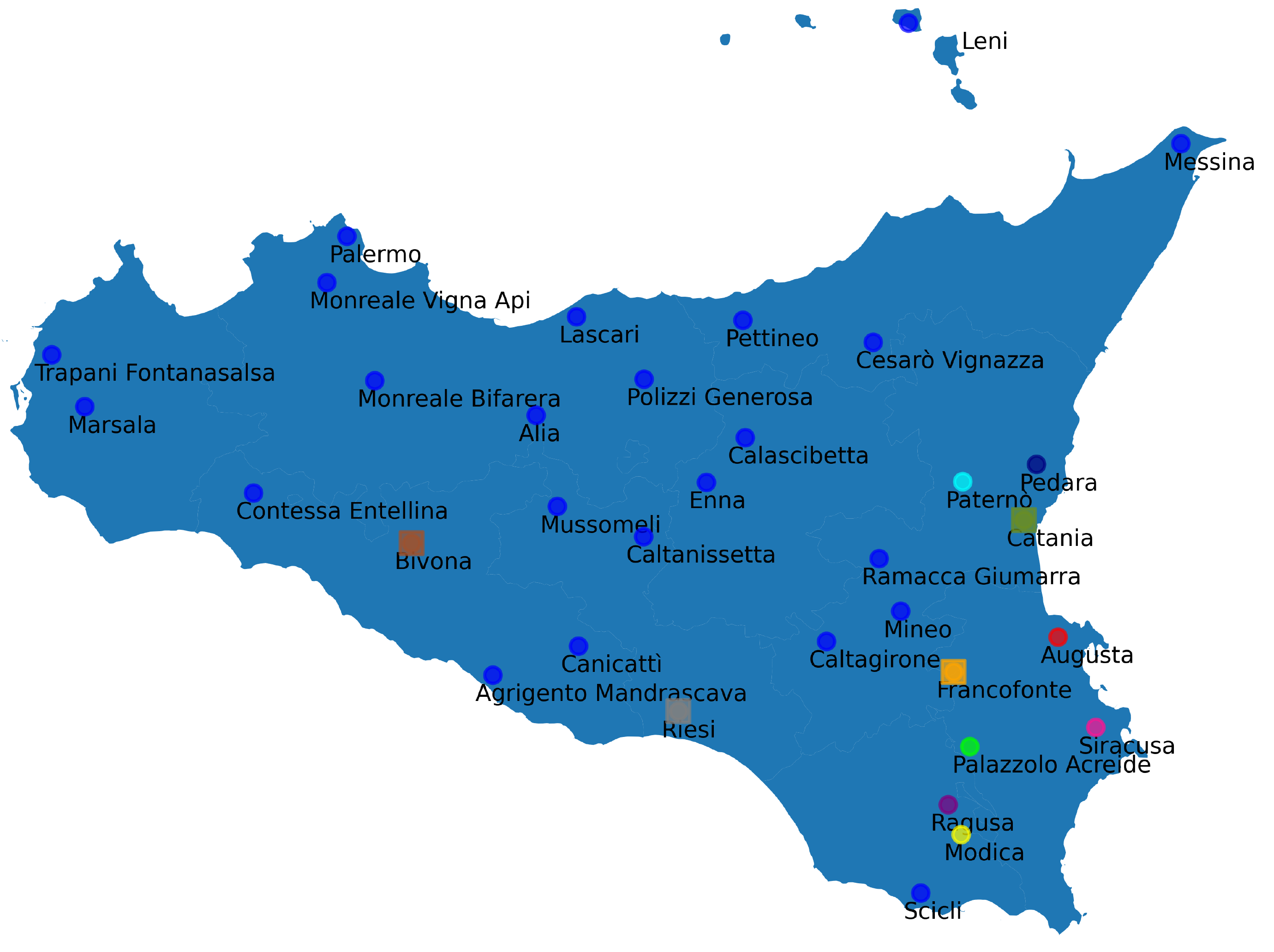}}
\end{minipage}%
\begin{minipage}{.5\linewidth}
\centering
\subfloat[Correlation metrics and C.$A$]{\label{2011ca}\includegraphics[width = 3.2in]{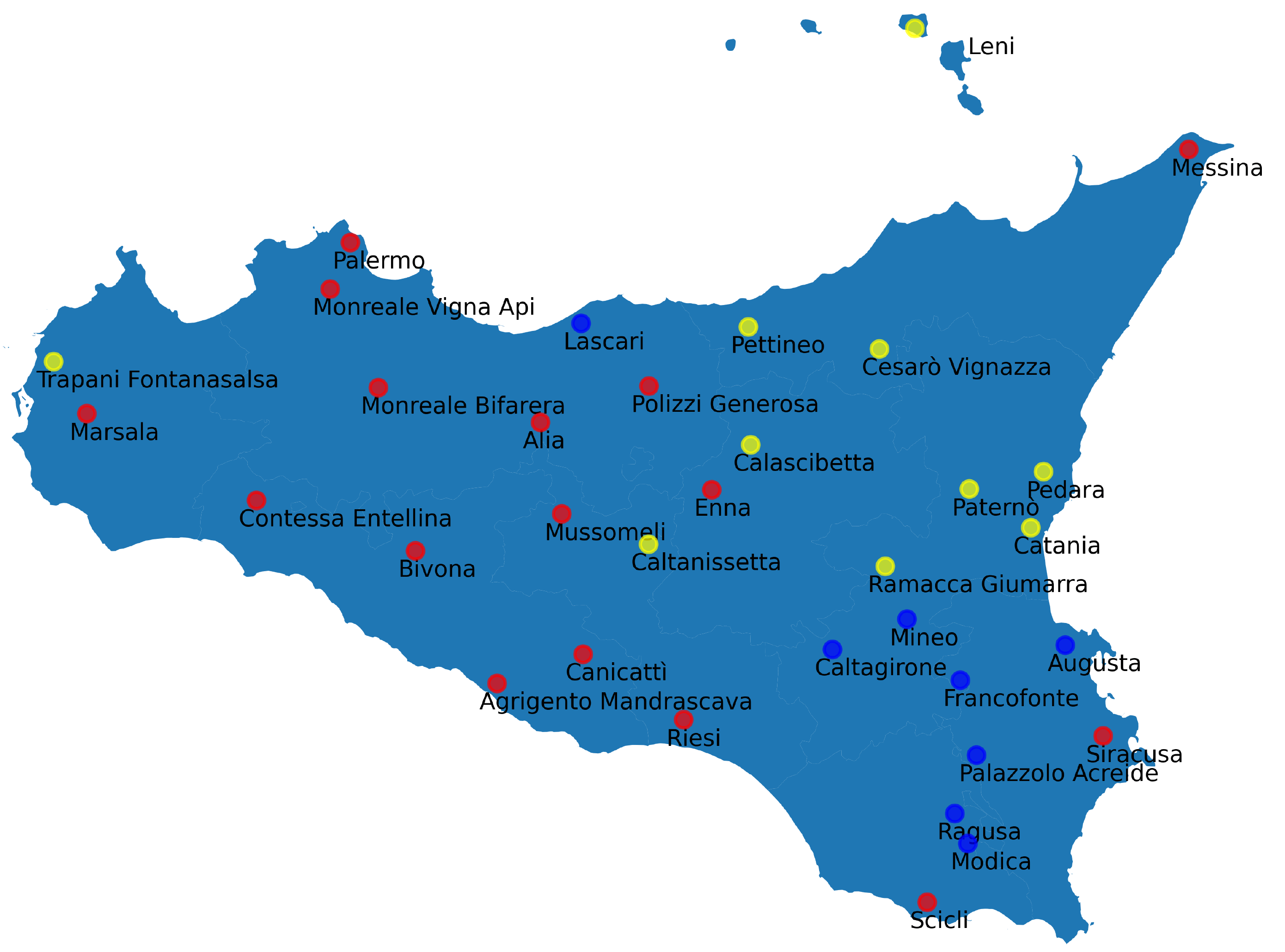}}
\end{minipage}\par\medskip
\begin{minipage}{.5\linewidth}
\centering
\subfloat[Euclidean metrics and C.$B$]{\label{2011eb}\includegraphics[width = 3.2in]{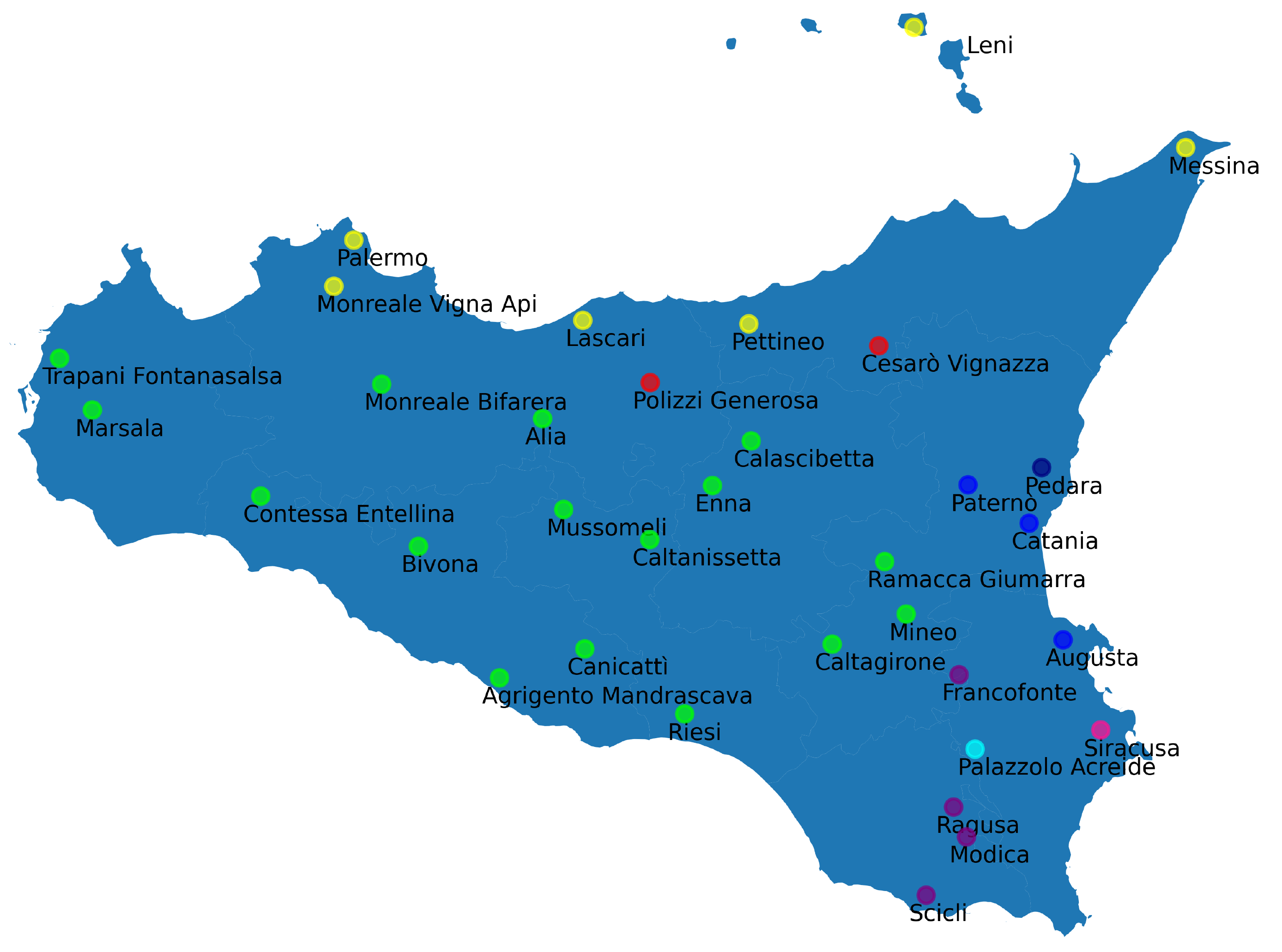}}
\end{minipage}%
\begin{minipage}{.5\linewidth}
\centering
\subfloat[Correlation metrics and C.$B$]{\label{2011cb}\includegraphics[width = 3.2in]{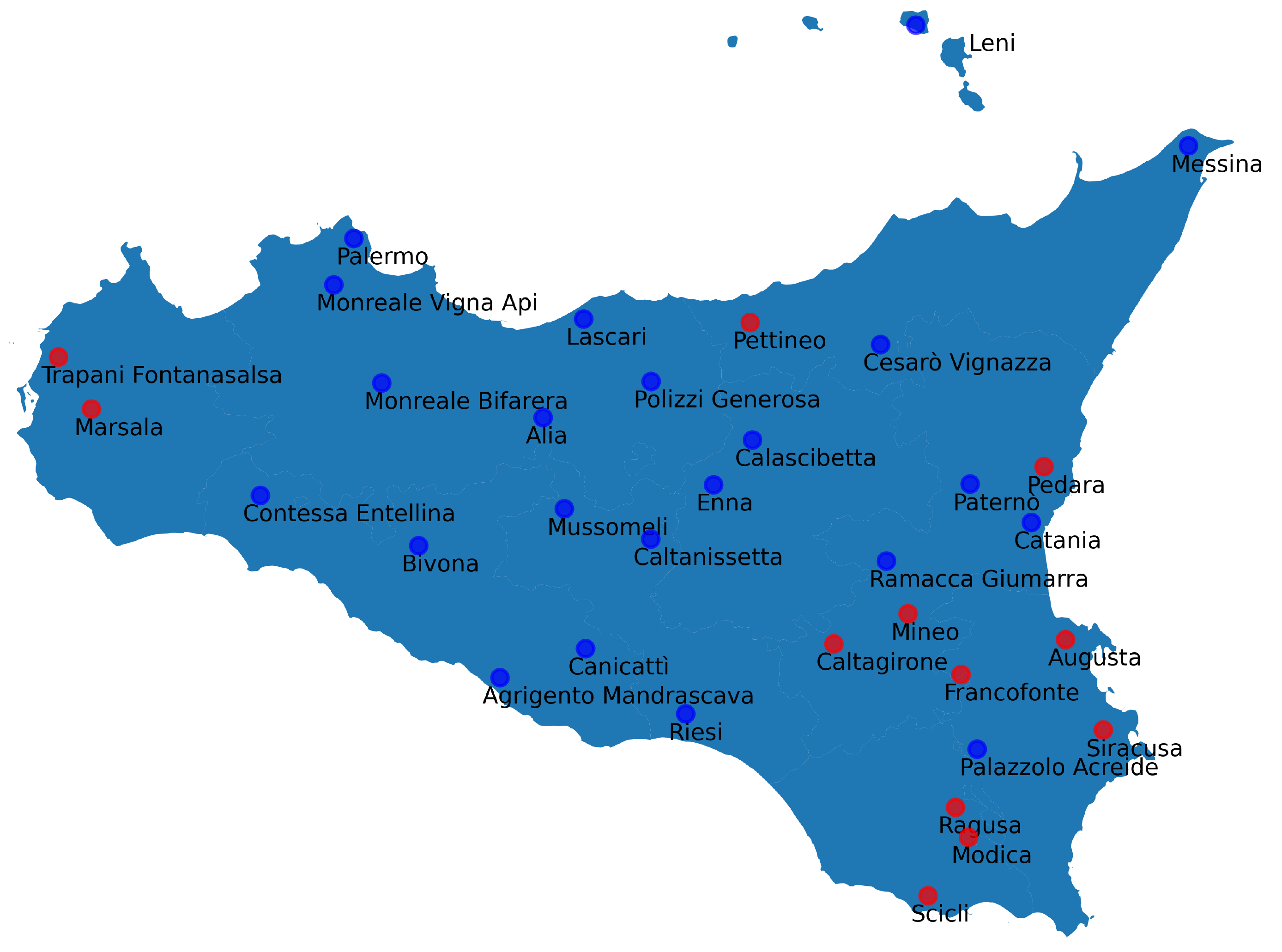}}
\end{minipage}\par\medskip
\caption{Annual case - 2011\\ \textbf{Panel a}: The principal cluster is reported in blue. \textbf{Panel b}: The three clusters are reported in red, blue and yellow. \textbf{Panel c}: The five principal clusters are reported in green, purple, blue, red and yellow. \textbf{Panel d}: The two clusters are reported in red and blue.}
\label{annual2011}
\end{figure}

\begin{figure}[h!]
\begin{minipage}{.5\linewidth}
\centering
\subfloat[Euclidean metrics and C.$A$]{\label{2012ea}\includegraphics[width = 3.2in]{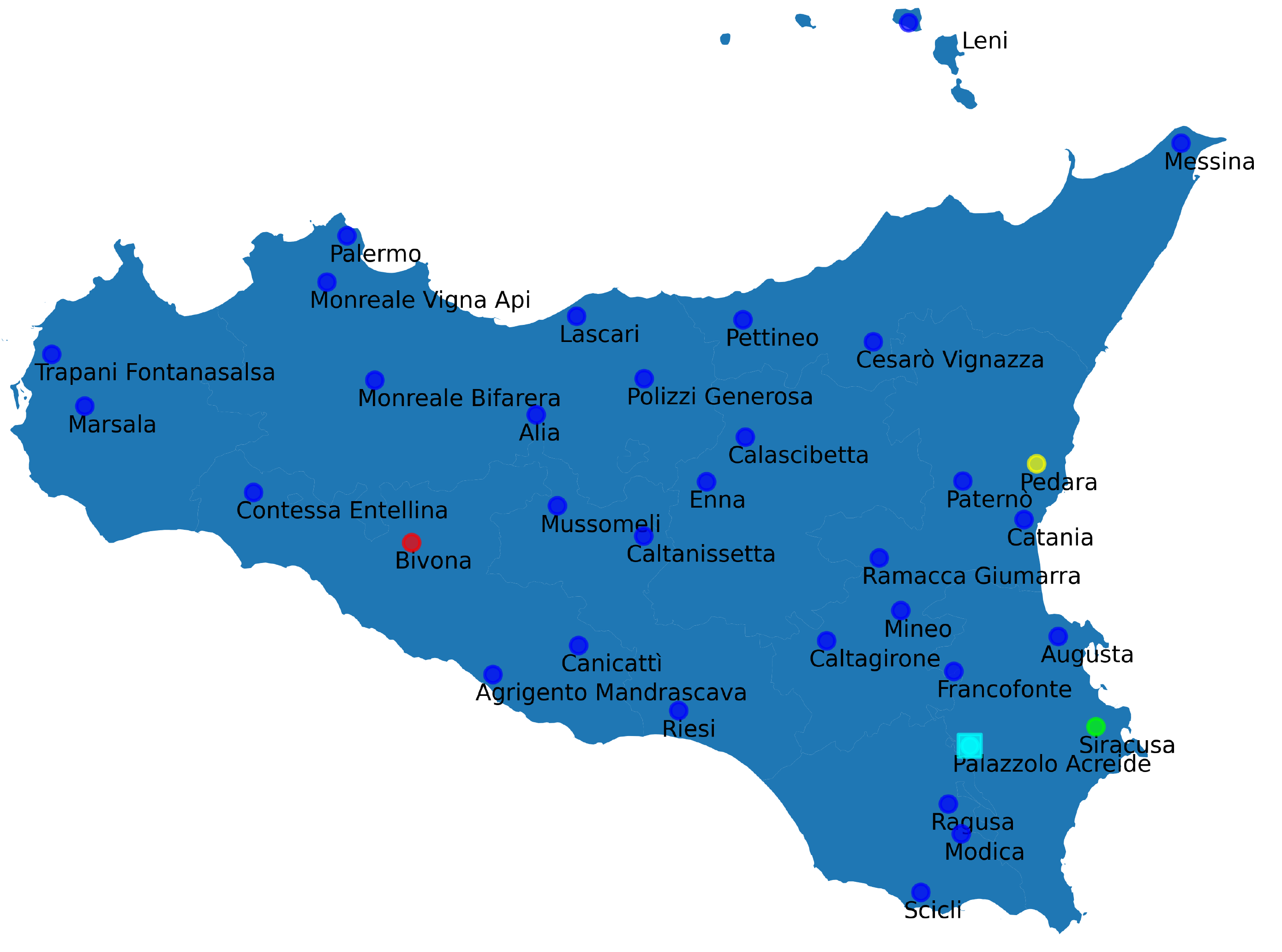}}
\end{minipage}%
\begin{minipage}{.5\linewidth}
\centering
\subfloat[Correlation metrics and C.$A$]{\label{2012ca}\includegraphics[width = 3.2in]{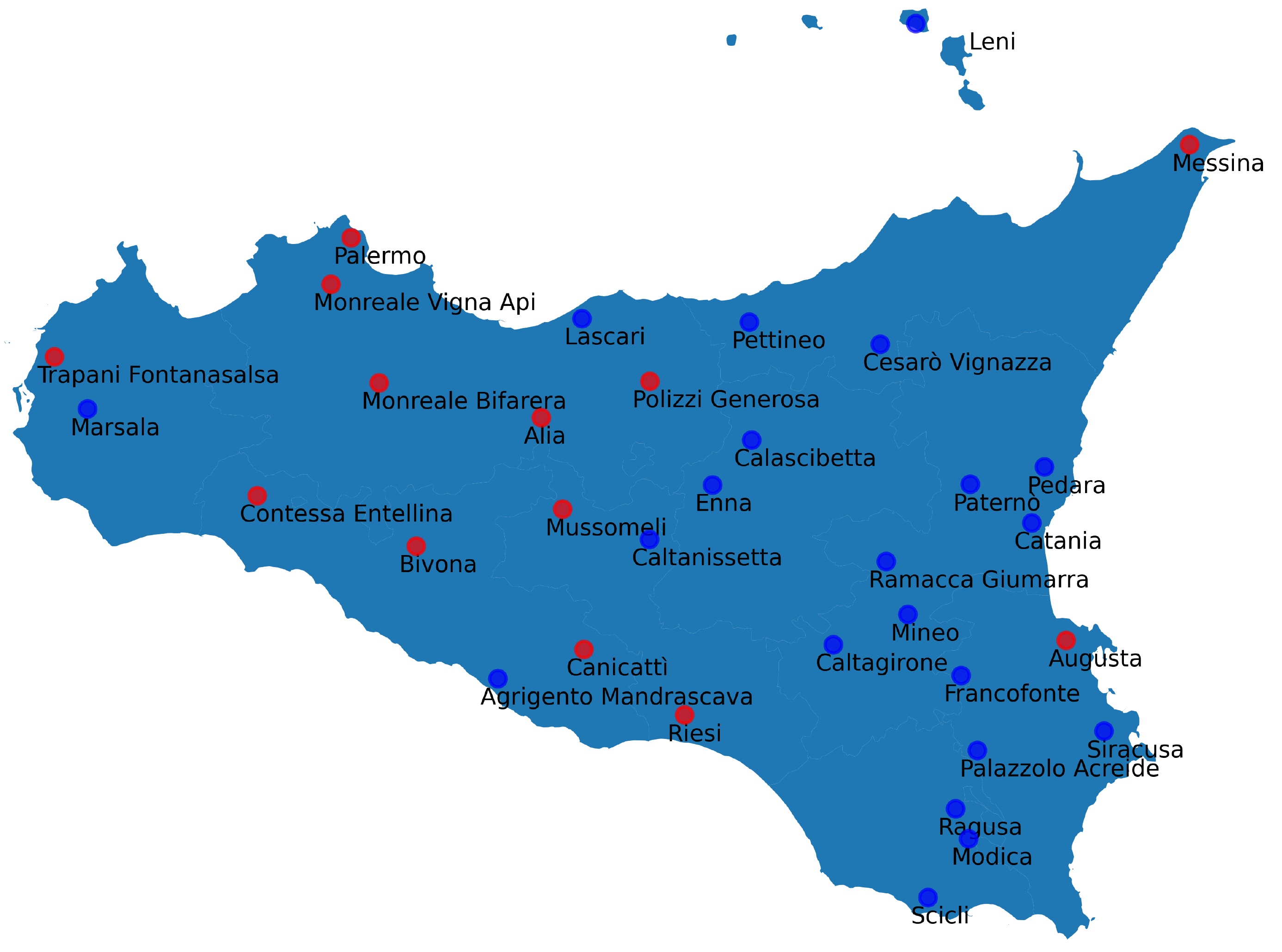}}
\end{minipage}\par\medskip
\begin{minipage}{.5\linewidth}
\centering
\subfloat[Euclidean metrics and C.$B$]{\label{2012eb}\includegraphics[width = 3.2in]{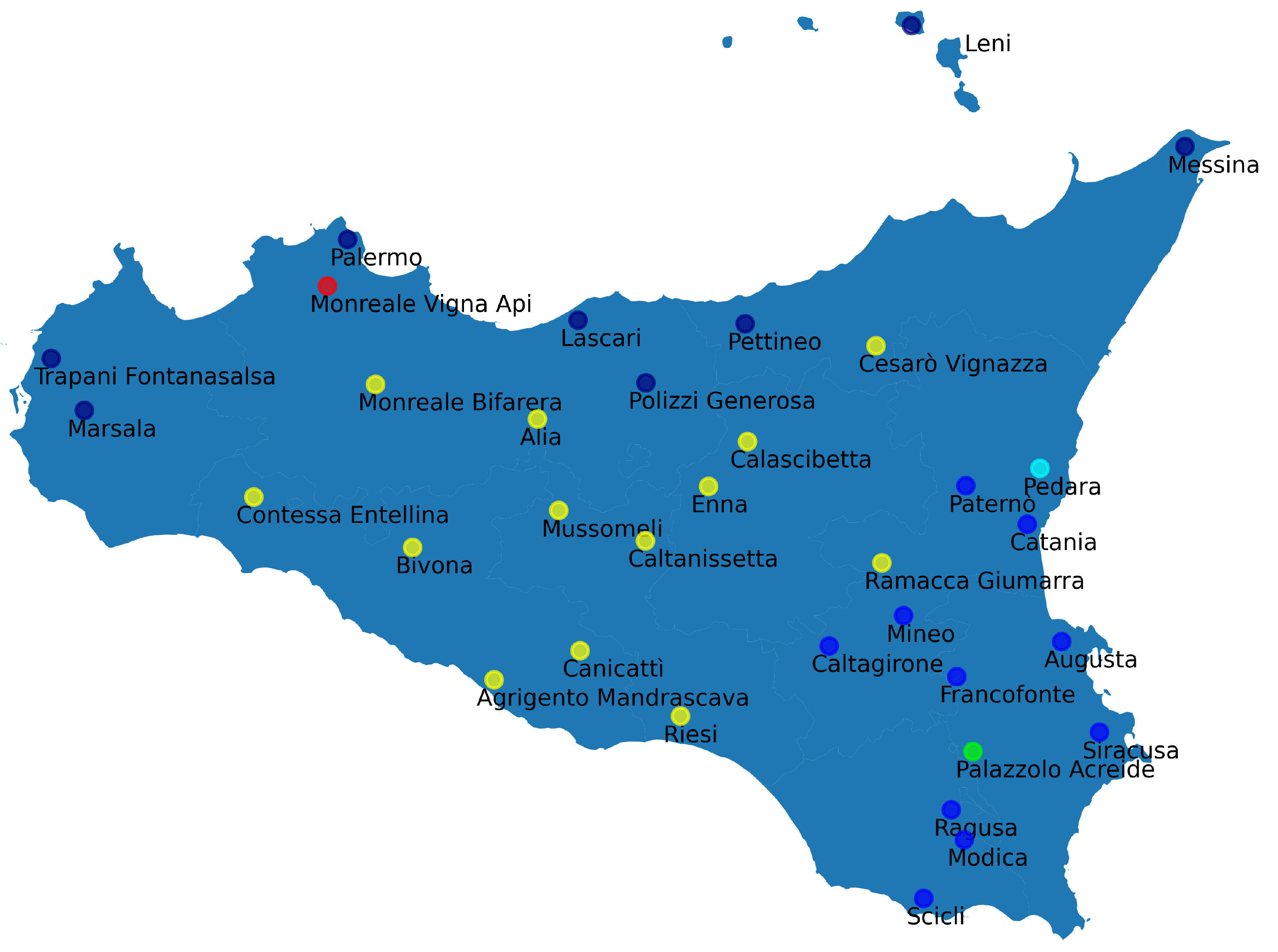}}
\end{minipage}%
\begin{minipage}{.5\linewidth}
\centering
\subfloat[Correlation metrics and C.$B$]{\label{2012cb}\includegraphics[width = 3.2in]{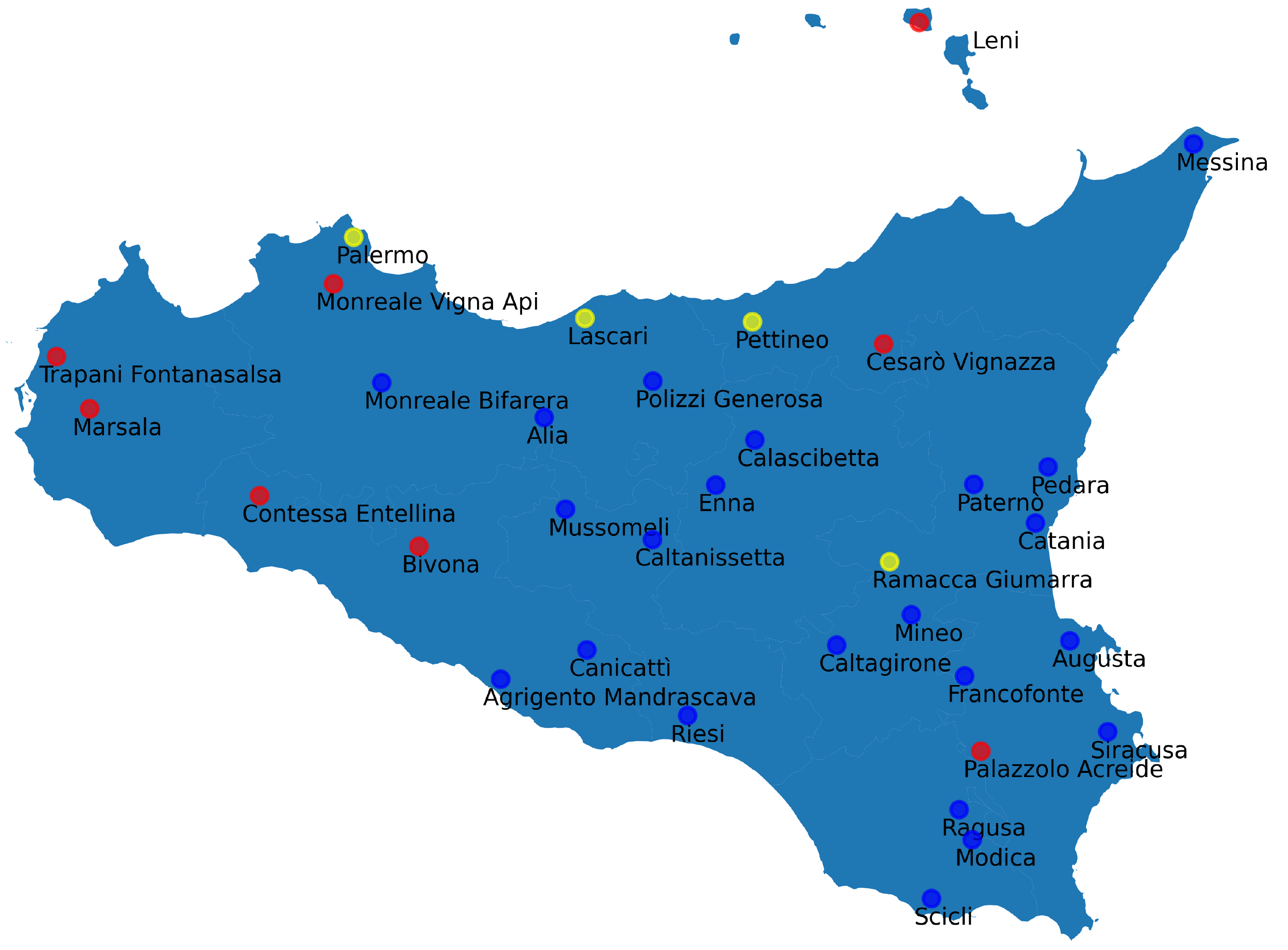}}
\end{minipage}\par\medskip
\caption{Annual case - 2012\\ \textbf{Panel a}: The principal cluster is reported in blue. \textbf{Panel b}: The two clusters are reported in red and blue. \textbf{Panel c}: The three principal clusters are reported in blue, yellow and dark blue. \textbf{Panel d}: The three clusters are reported in blue, red and yellow}.
\label{annual2012}
\end{figure}

\begin{figure}[h!]
\begin{minipage}{.5\linewidth}
\centering
\subfloat[Euclidean metrics and C.$A$]{\label{2013ea}\includegraphics[width = 3.2in]{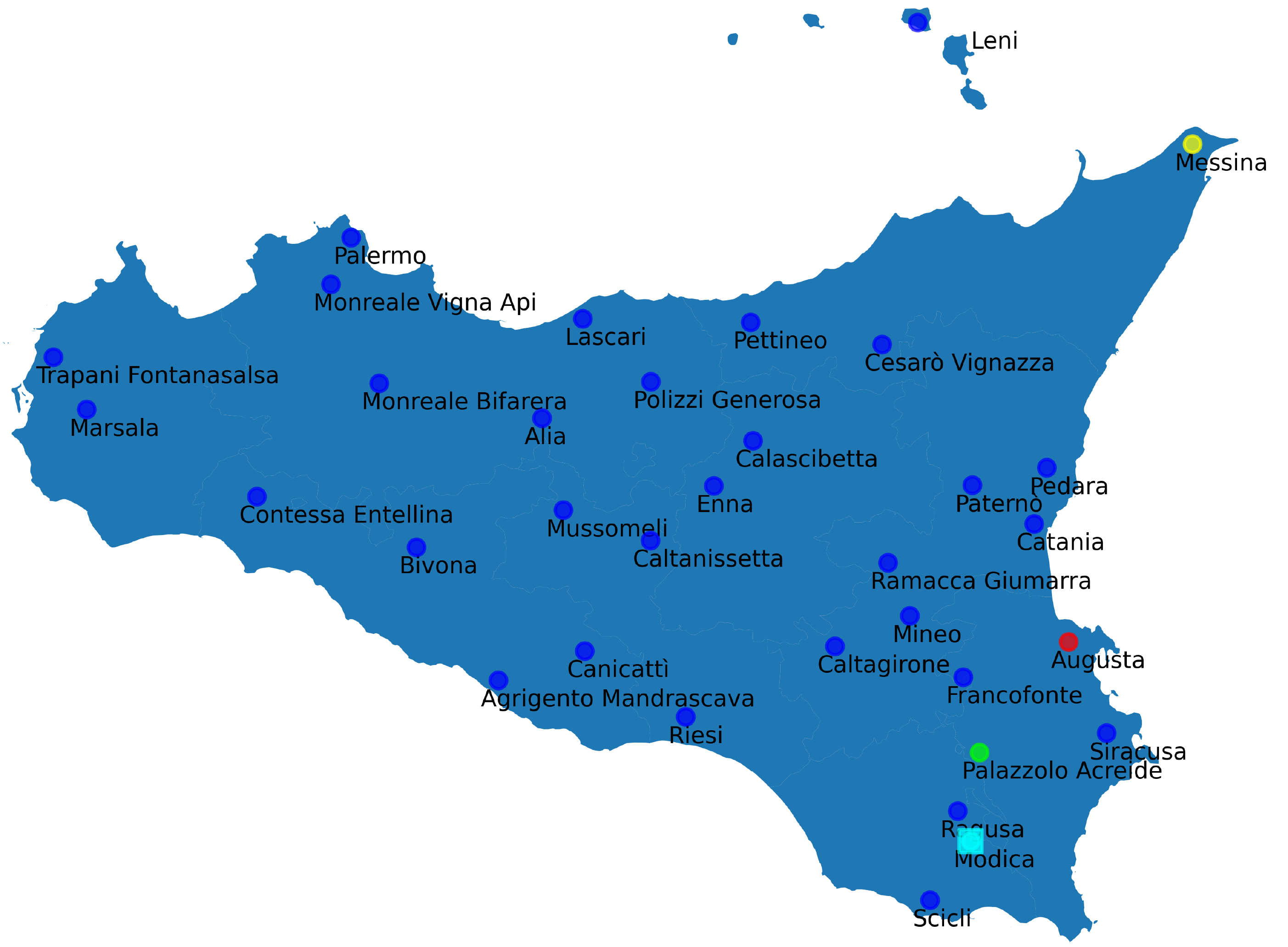}}
\end{minipage}%
\begin{minipage}{.5\linewidth}
\centering
\subfloat[Correlation metrics and C.$A$]{\label{2013ca}\includegraphics[width = 3.2in]{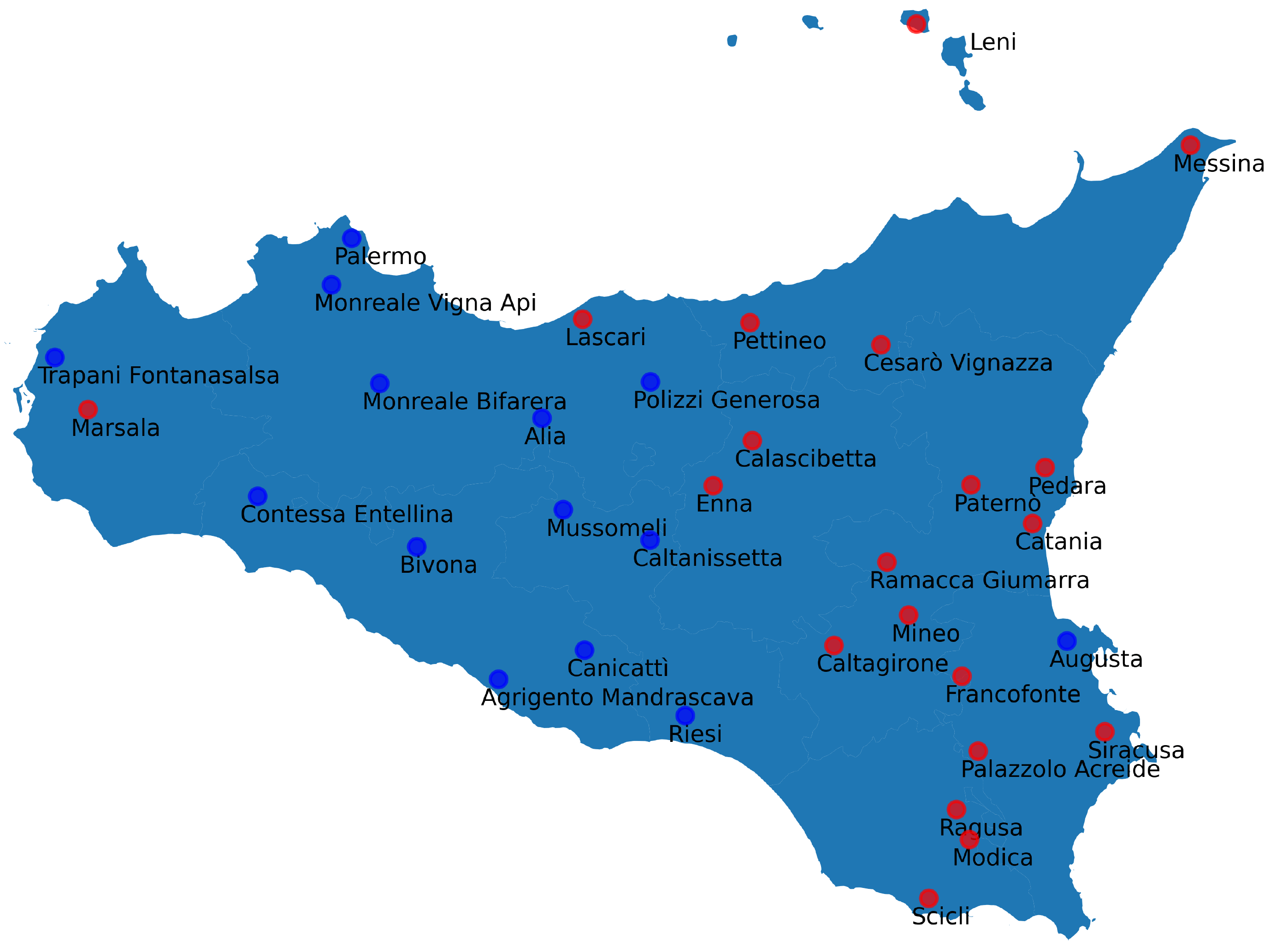}}
\end{minipage}\par\medskip
\begin{minipage}{.5\linewidth}
\centering
\subfloat[Euclidean metrics and C.$B$]{\label{2013eb}\includegraphics[width = 3.2in]{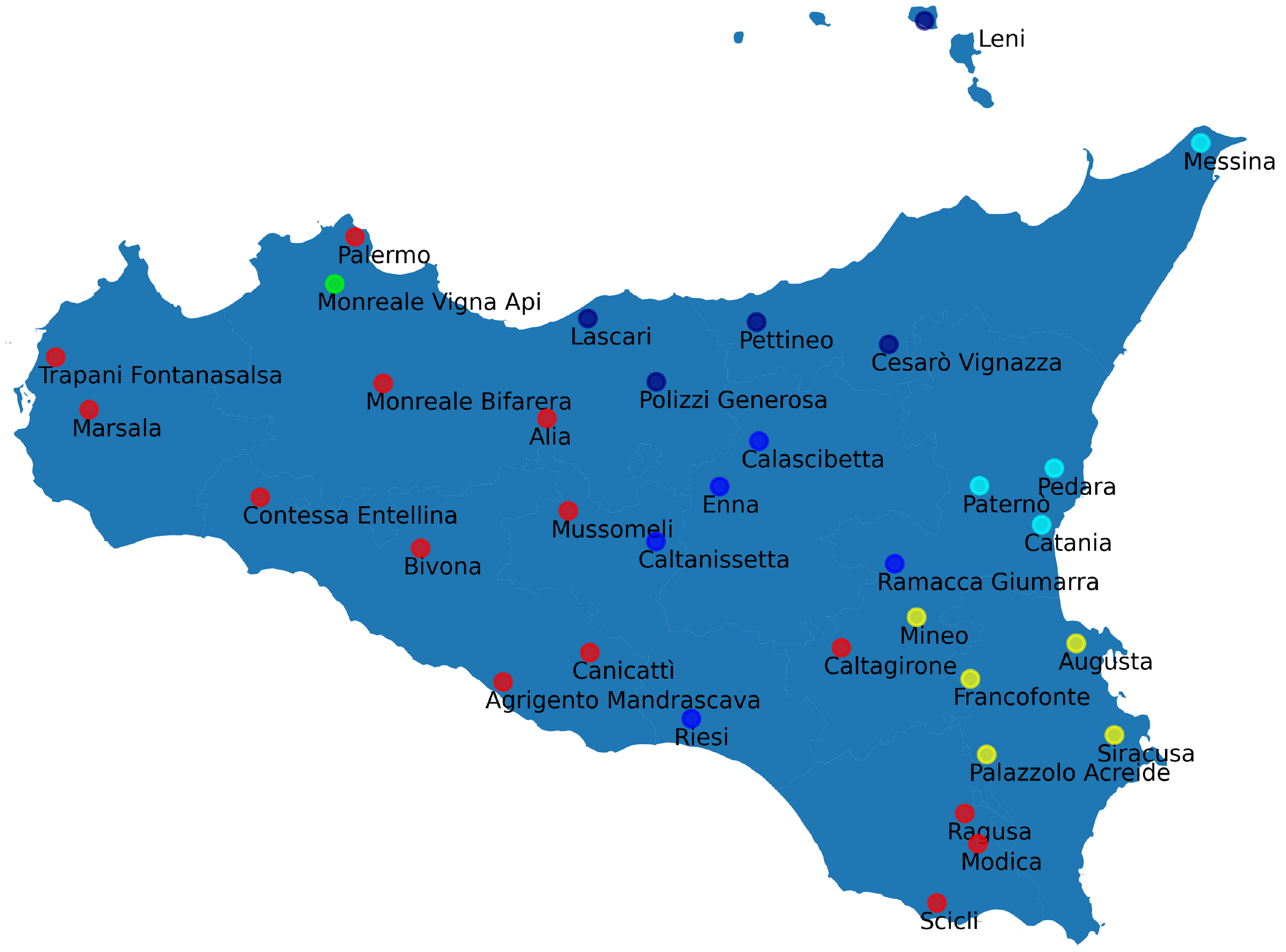}}
\end{minipage}%
\begin{minipage}{.5\linewidth}
\centering
\subfloat[Correlation metrics and C.$B$]{\label{2013cb}\includegraphics[width = 3.2in]{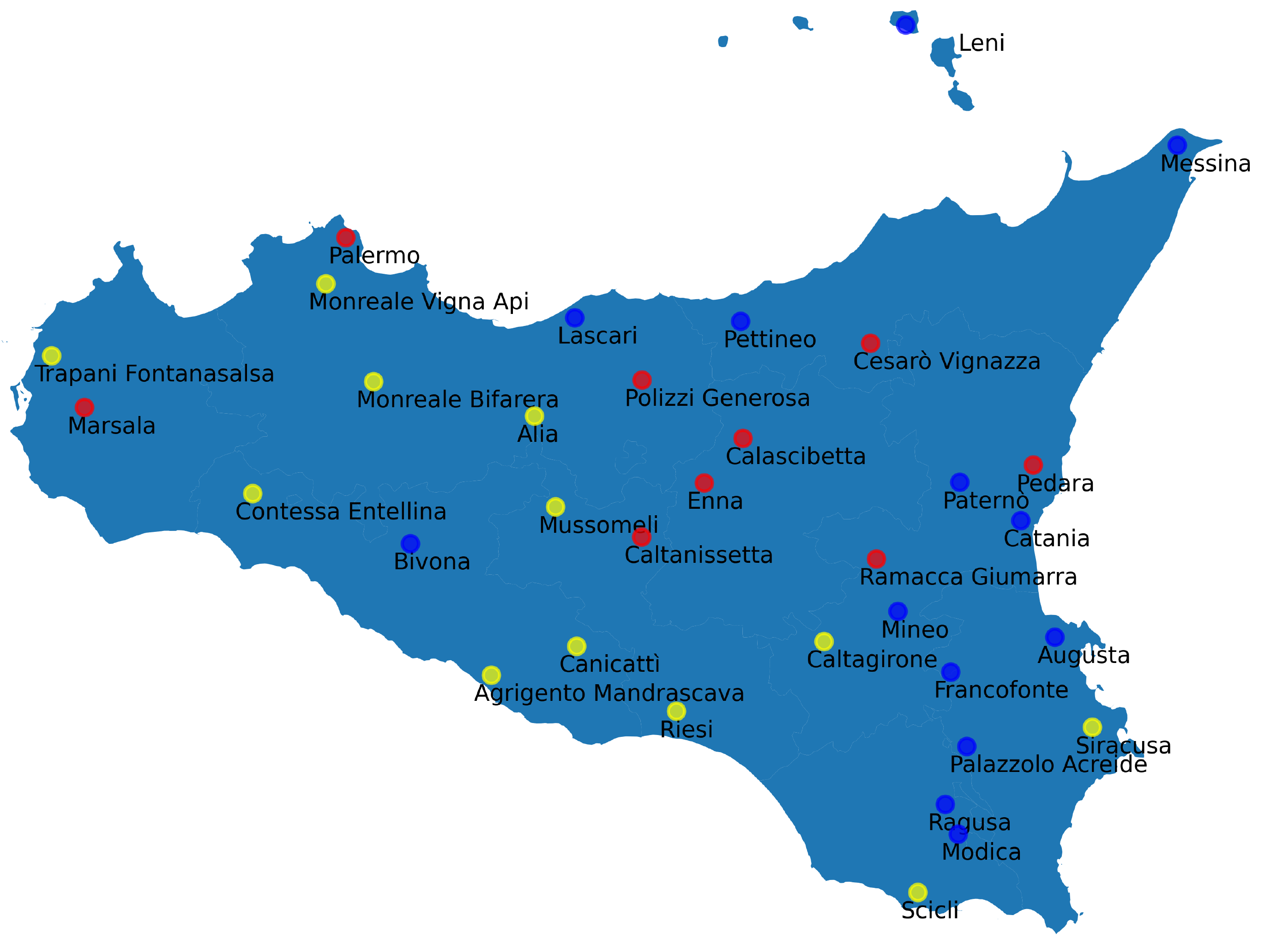}}
\end{minipage}\par\medskip
\caption{Annual case - 2013\\ \textbf{Panel a}: The principal cluster is reported in blue. \textbf{Panel b}: The two clusters are reported in red and blue\textbf{Panel c}: The five principal clusters are reported in red, blue, dark blue, light blue and yellow.\textbf{Panel d}: The three clusters are reported in red, blue and yellow.}
\label{annual2013}
\end{figure}

\begin{figure}[h!]
\begin{minipage}{.5\linewidth}
\centering
\subfloat[Euclidean metrics and C.$A$]{\label{2014ea}\includegraphics[width = 3.2in]{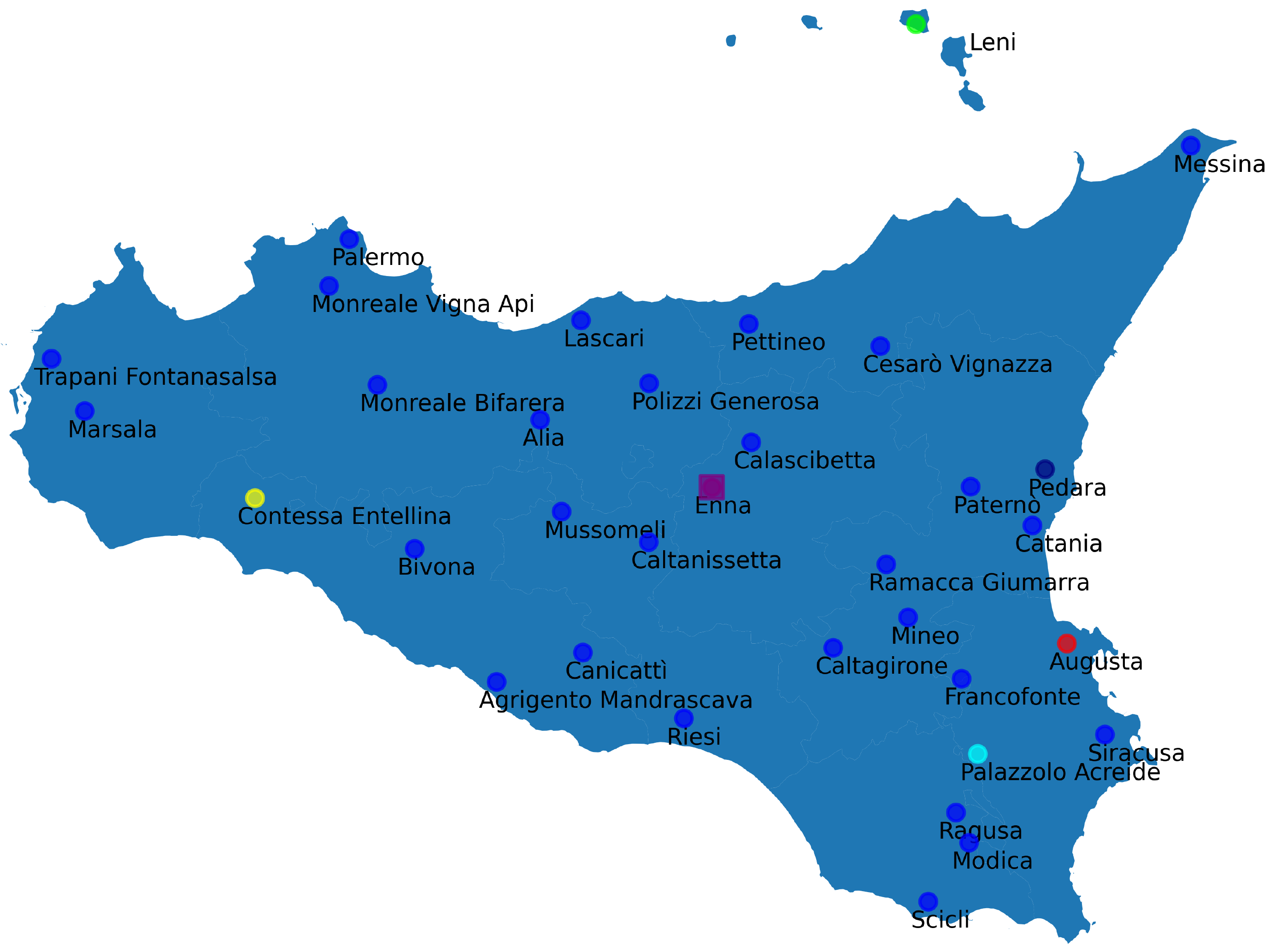}}
\end{minipage}%
\begin{minipage}{.5\linewidth}
\centering
\subfloat[Correlation metrics and C.$A$]{\label{2014ca}\includegraphics[width = 3.2in]{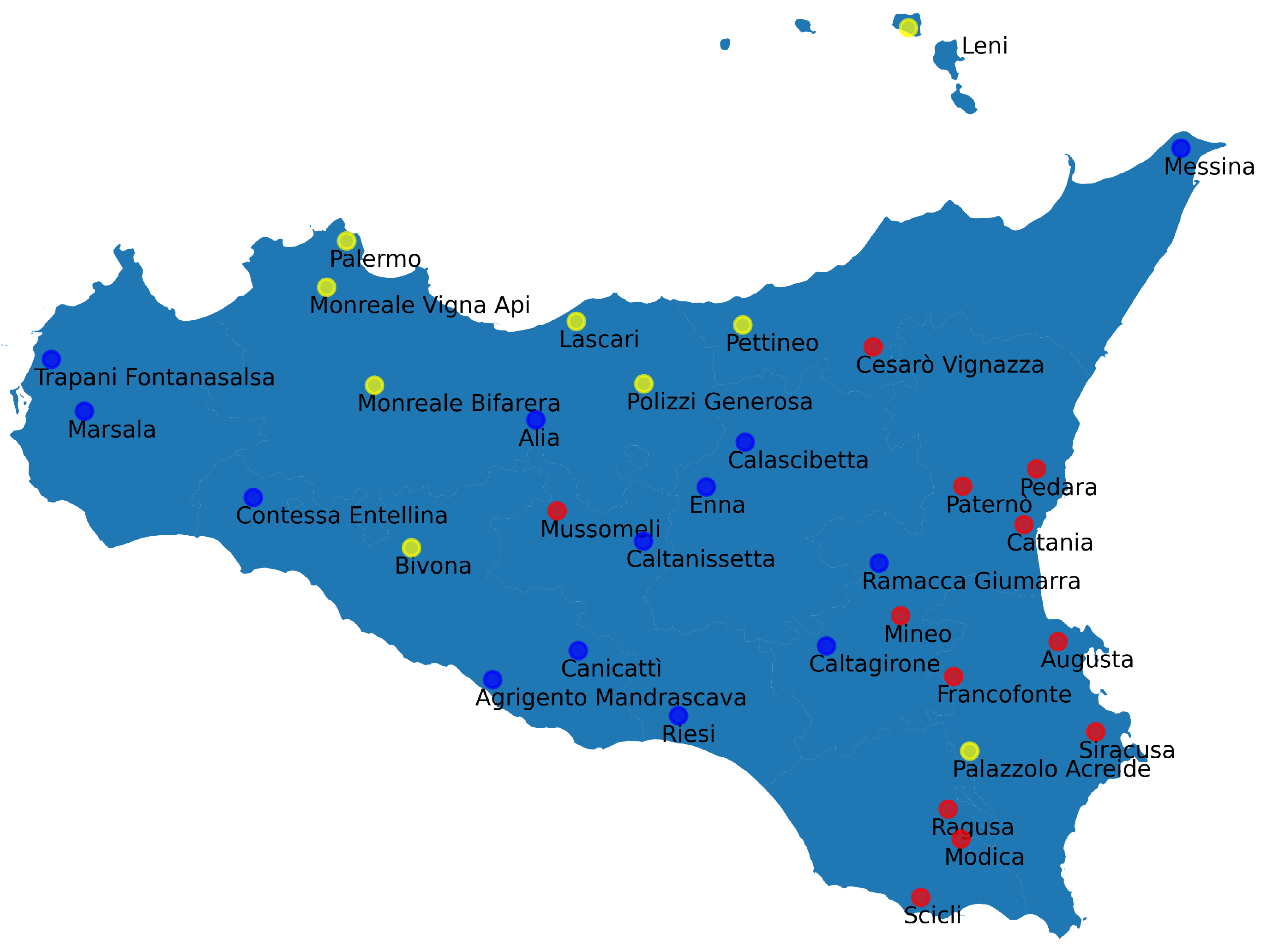}}
\end{minipage}\par\medskip
\begin{minipage}{.5\linewidth}
\centering
\subfloat[Euclidean metrics and C.$B$]{\label{2014eb}\includegraphics[width = 3.2in]{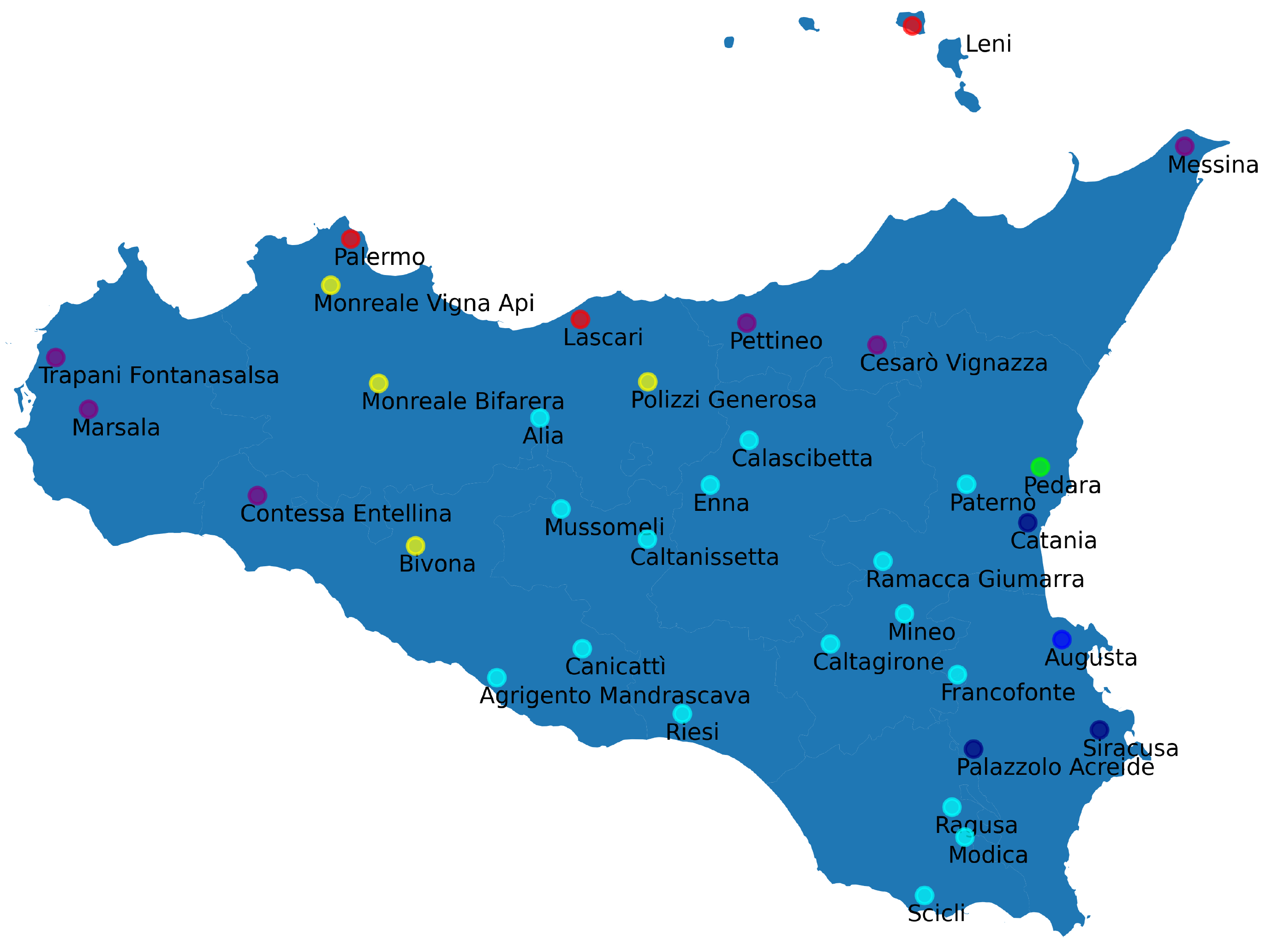}}
\end{minipage}%
\begin{minipage}{.5\linewidth}
\centering
\subfloat[Correlation metrics and C.$B$]{\label{2014cb}\includegraphics[width = 3.2in]{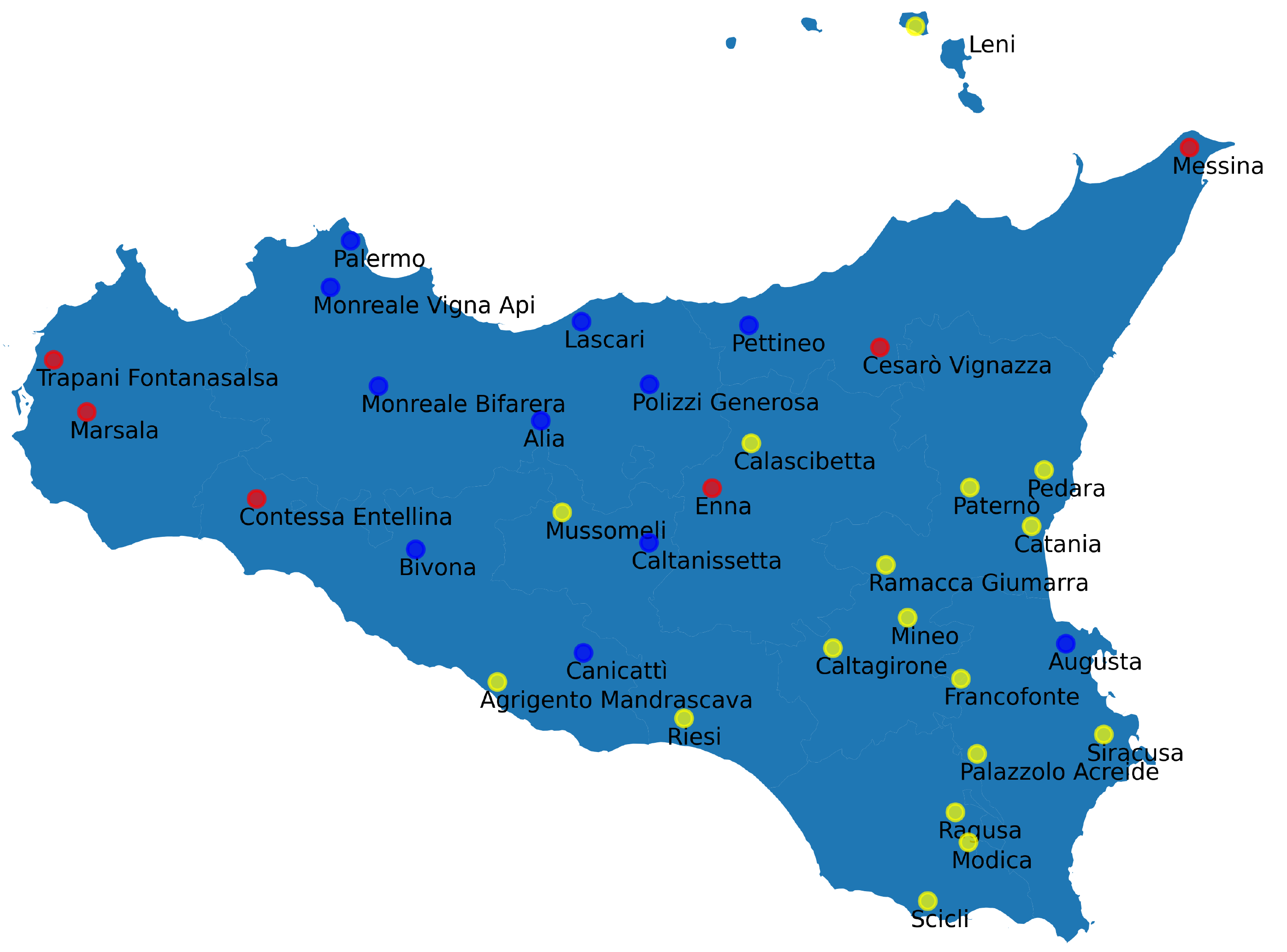}}
\end{minipage}\par\medskip
\caption{Annual case - 2014\\ \textbf{Panel a}: The principal cluster is reported in blue. \textbf{Panel b}: The three clusters are reported in red, blue and yellow. \textbf{Panel c}: The five principal clusters are reported in light blue, yellow, purple, blue and red. \textbf{Panel d}: The three clusters are reported in red, blue and yellow.}
\label{annual2014}
\end{figure}

\begin{figure}[h!]
\begin{minipage}{.5\linewidth}
\centering
\subfloat[Euclidean metrics and C.$A$]{\label{2015ea}\includegraphics[width = 3.2in]{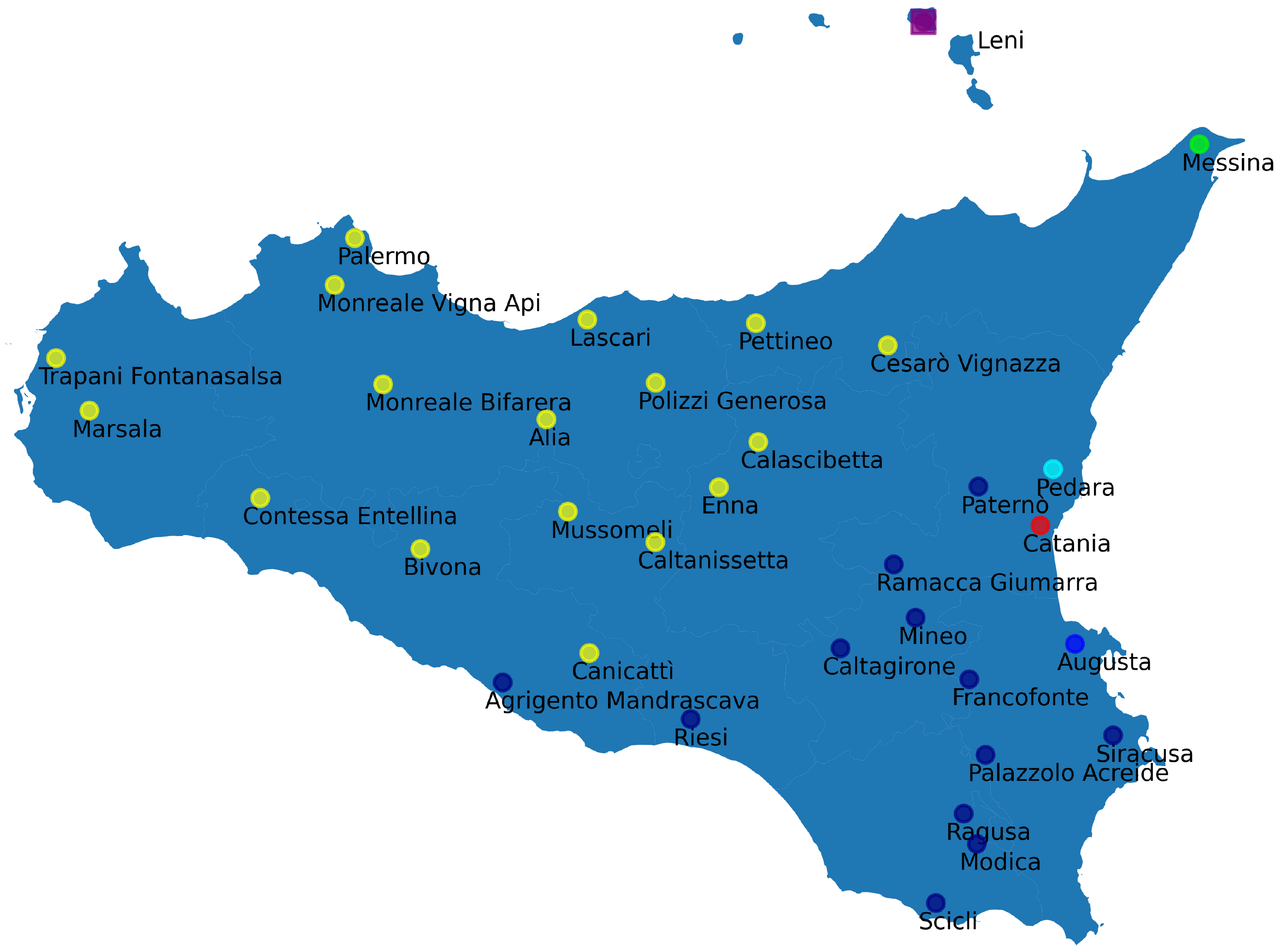}}
\end{minipage}%
\begin{minipage}{.5\linewidth}
\centering
\subfloat[Correlation metrics and C.$A$]{\label{2015ca}\includegraphics[width = 3.2in]{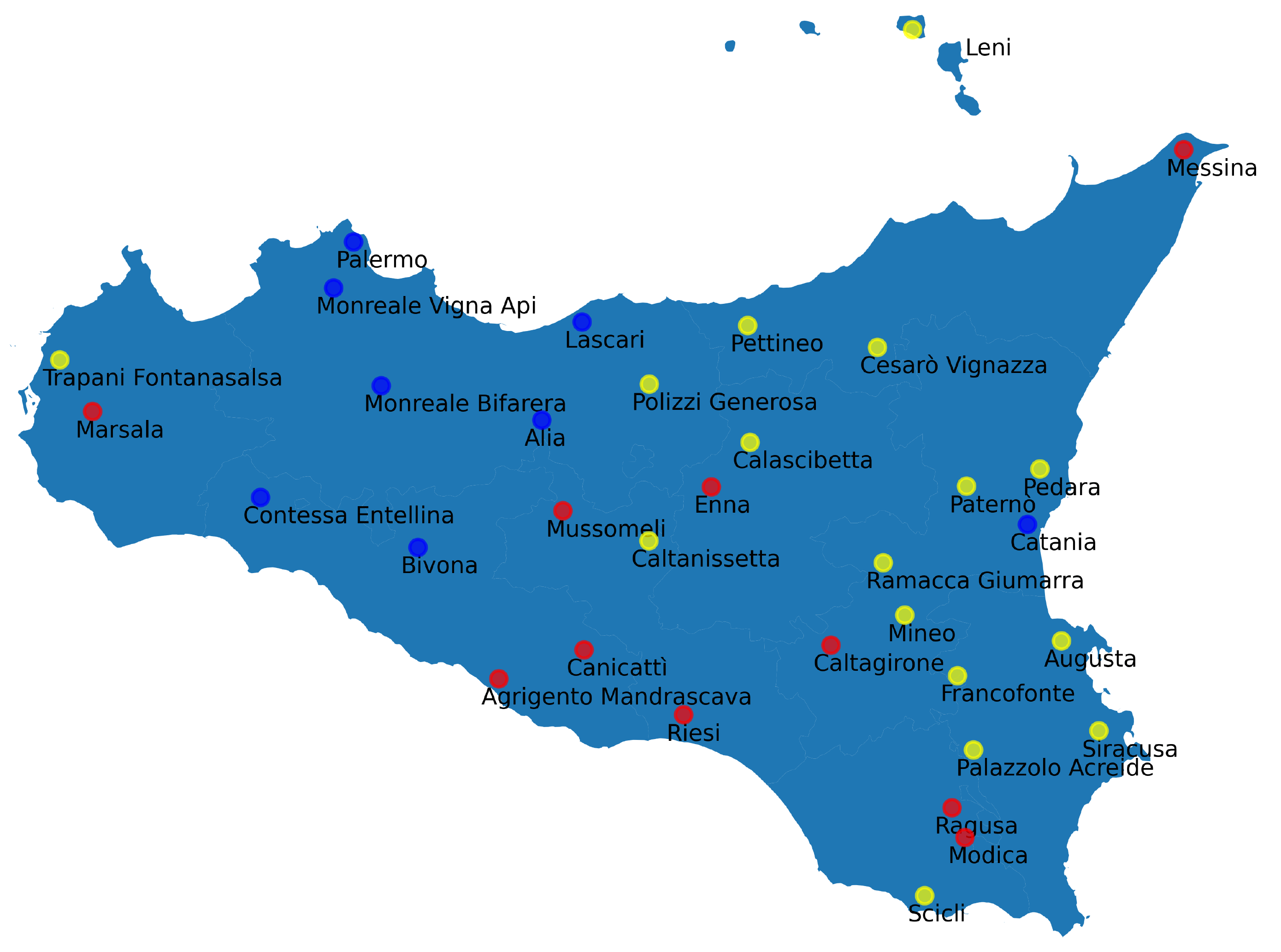}}
\end{minipage}\par\medskip
\begin{minipage}{.5\linewidth}
\centering
\subfloat[Euclidean metrics and C.$B$]{\label{2015eb}\includegraphics[width = 3.2in]{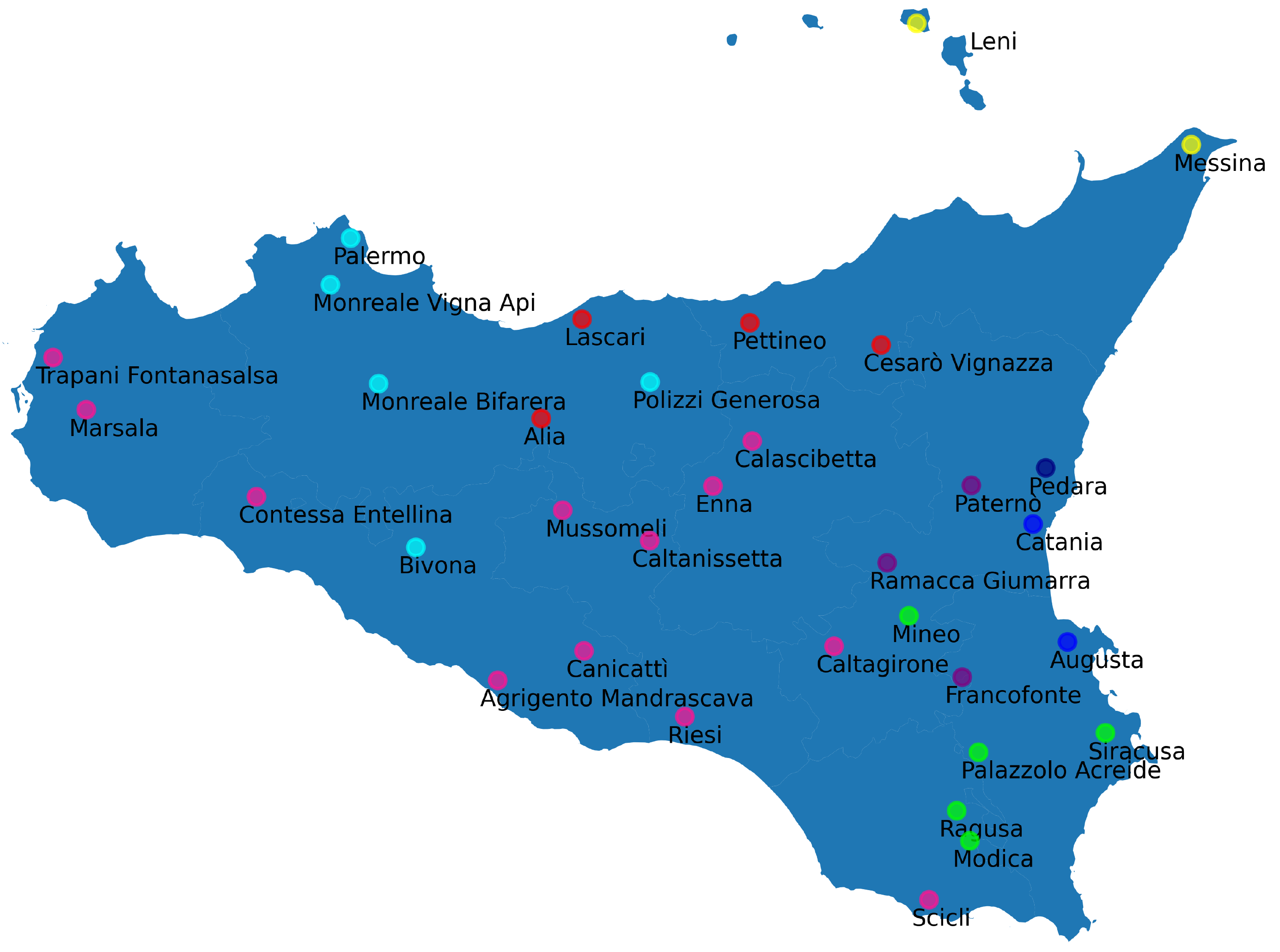}}
\end{minipage}%
\begin{minipage}{.5\linewidth}
\centering
\subfloat[Correlation metrics and C.$B$]{\label{2015cb}\includegraphics[width = 3.2in]{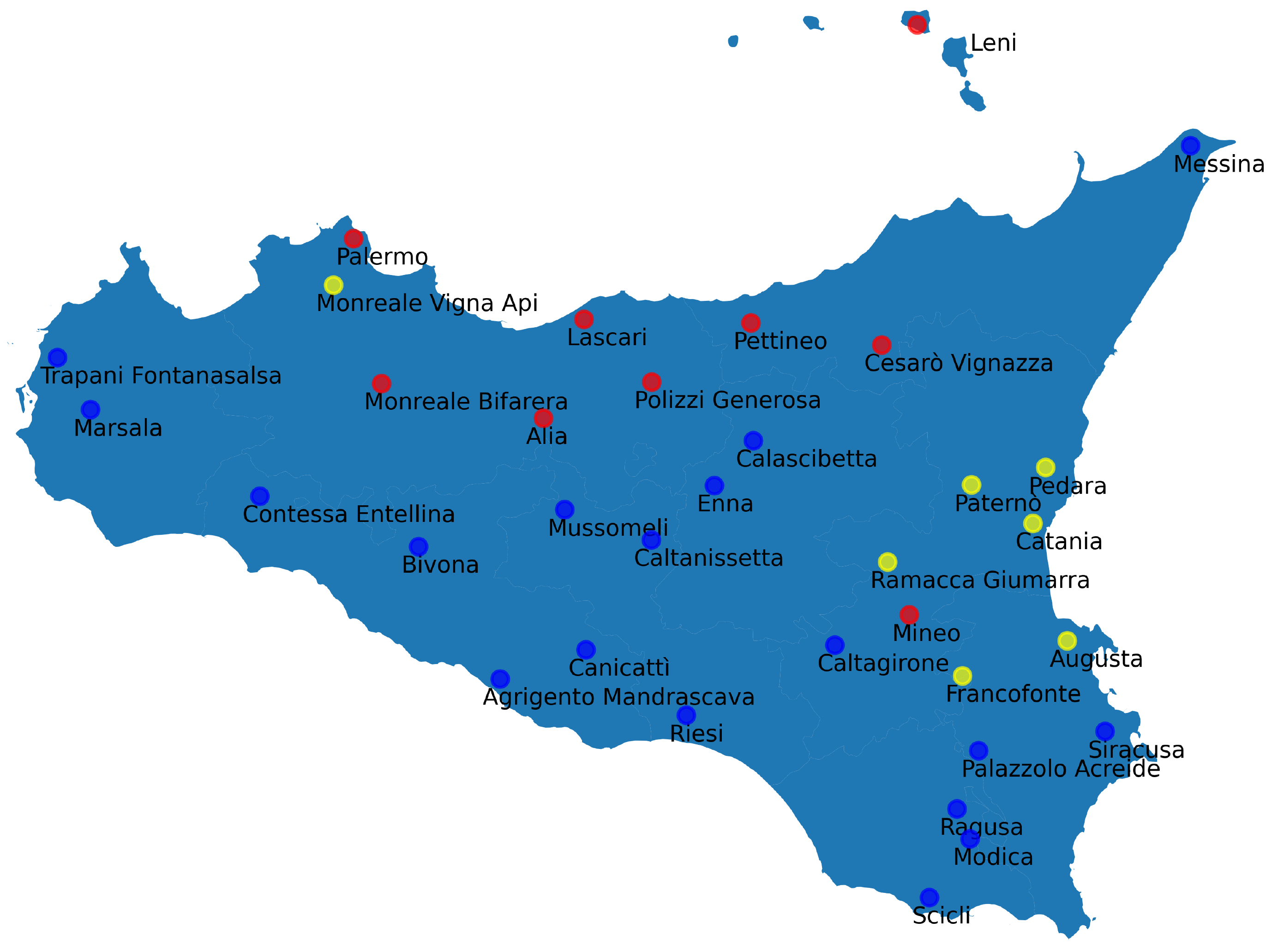}}
\end{minipage}\par\medskip
\caption{Annual case - 2015\\ \textbf{Panel a}: The two main clusters are reported in yellow and dark blue. \textbf{Panel b}: The three clusters are reported in red, blue and yellow. \textbf{Panel c}: The seven principal clusters are reported in pink, purple, green, blue, light blue, red and yellow \textbf{Panel d}: The three clusters are reported in red, blue and yellow.}
\label{annual2015}
\end{figure}

\begin{figure}[h!]
\begin{minipage}{.5\linewidth}
\centering
\subfloat[Euclidean metrics and C.$A$]{\label{2016ea}\includegraphics[width = 3.2in]{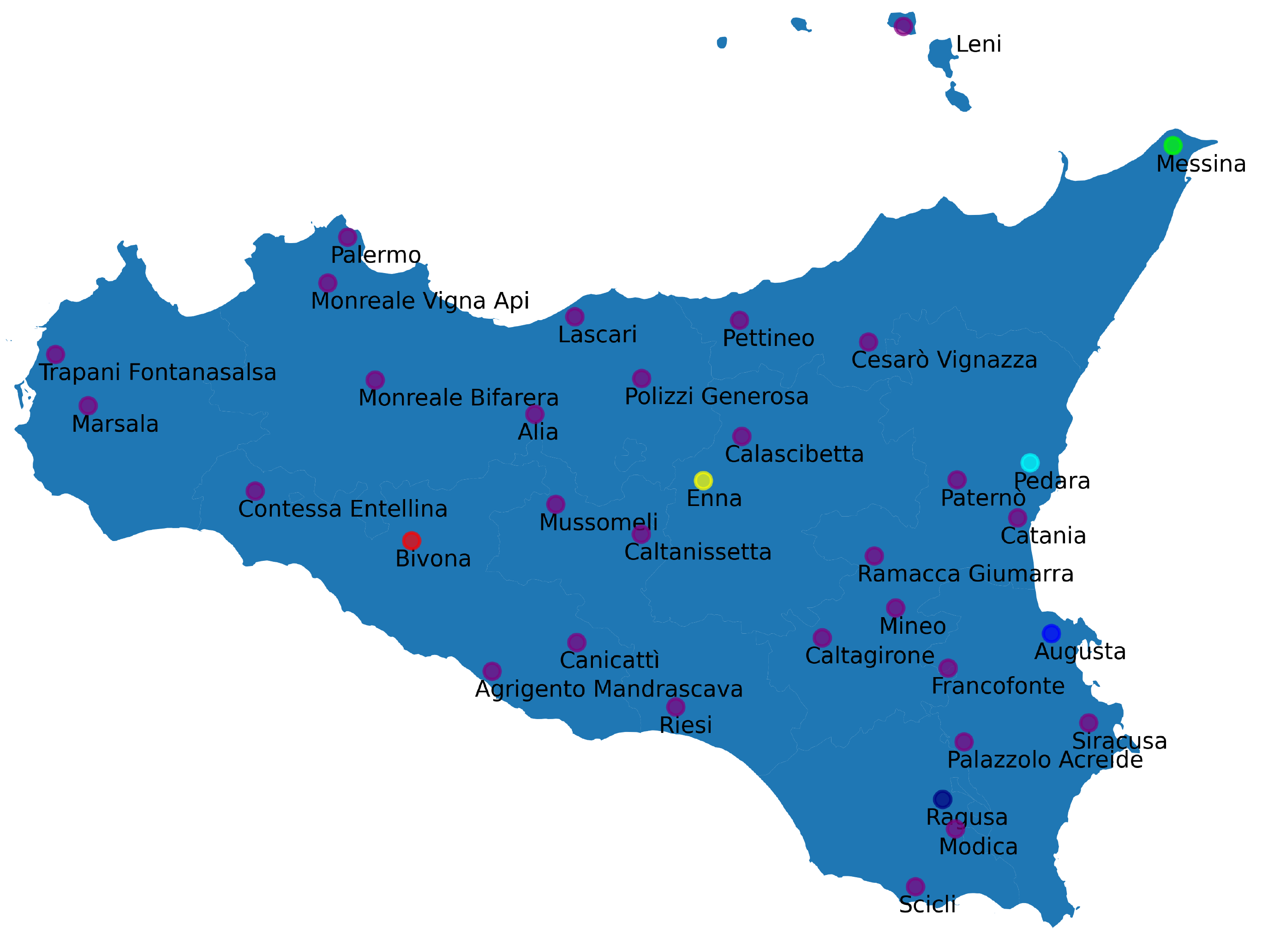}}
\end{minipage}%
\begin{minipage}{.5\linewidth}
\centering
\subfloat[Correlation metrics and C.$A$]{\label{2016ca}\includegraphics[width = 3.2in]{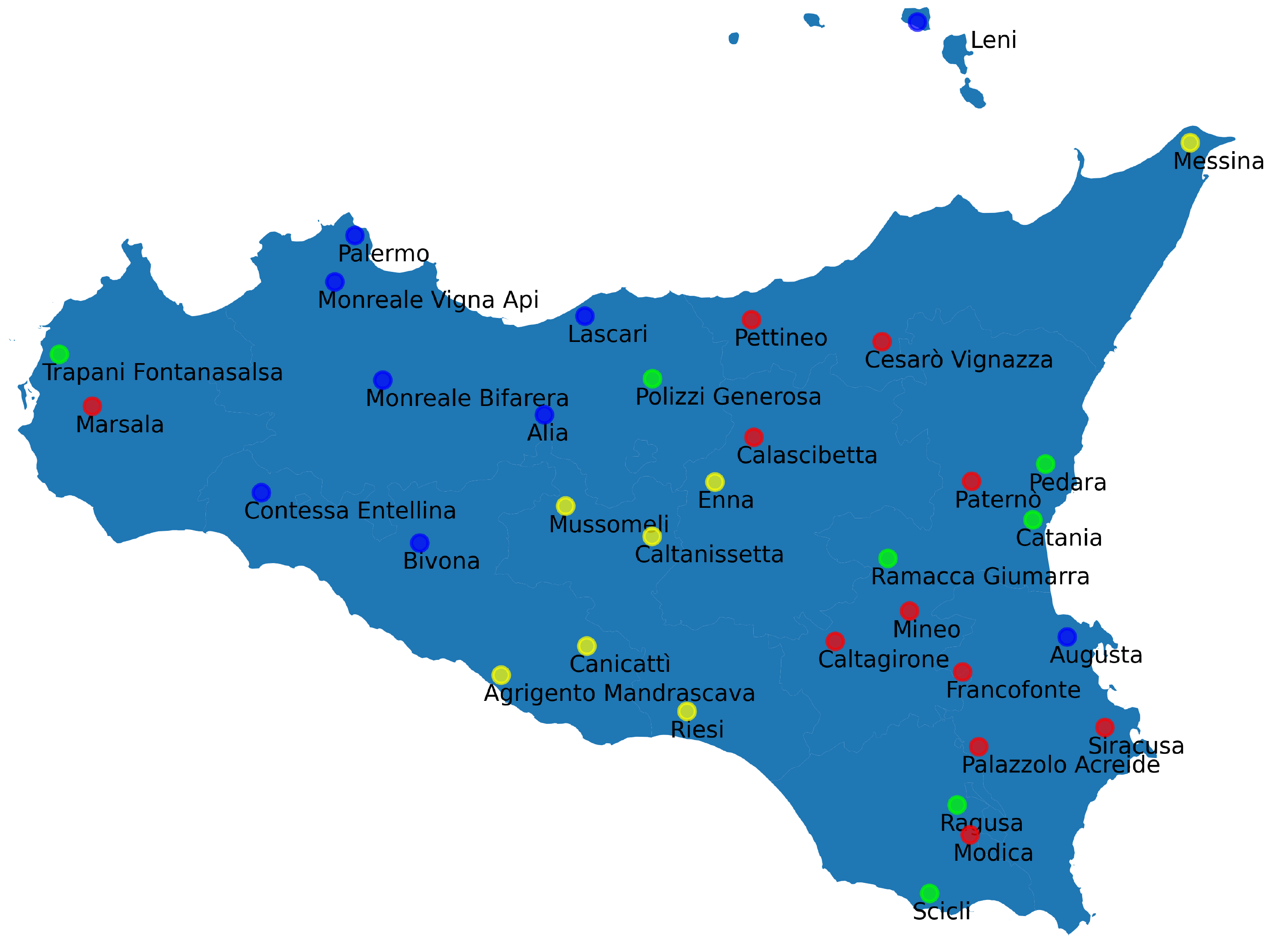}}
\end{minipage}\par\medskip
\begin{minipage}{.5\linewidth}
\centering
\subfloat[Euclidean metrics and C.$B$]{\label{2016eb}\includegraphics[width = 3.2in]{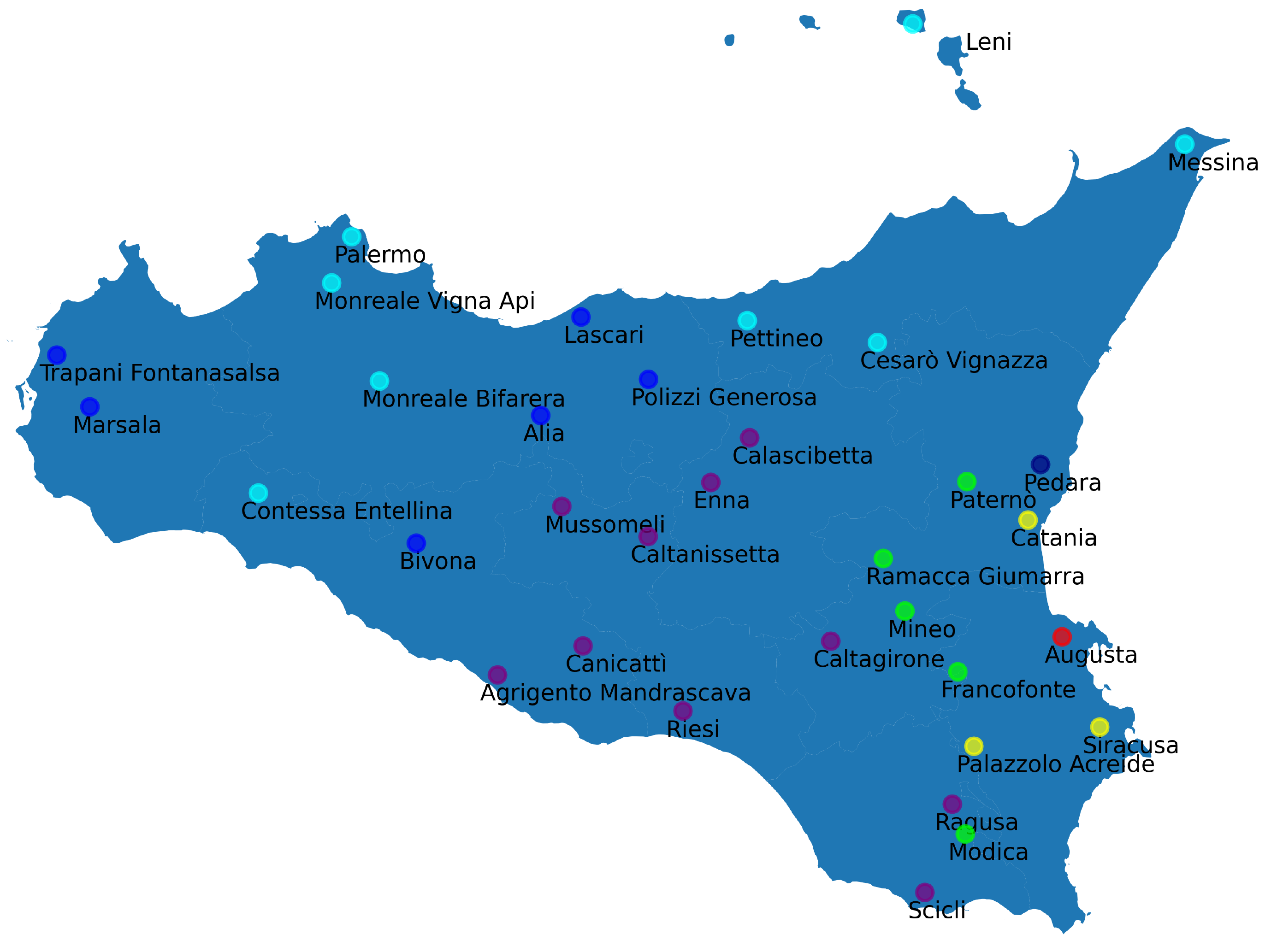}}
\end{minipage}%
\begin{minipage}{.5\linewidth}
\centering
\subfloat[Correlation metrics and C.$B$]{\label{2016cb}\includegraphics[width = 3.2in]{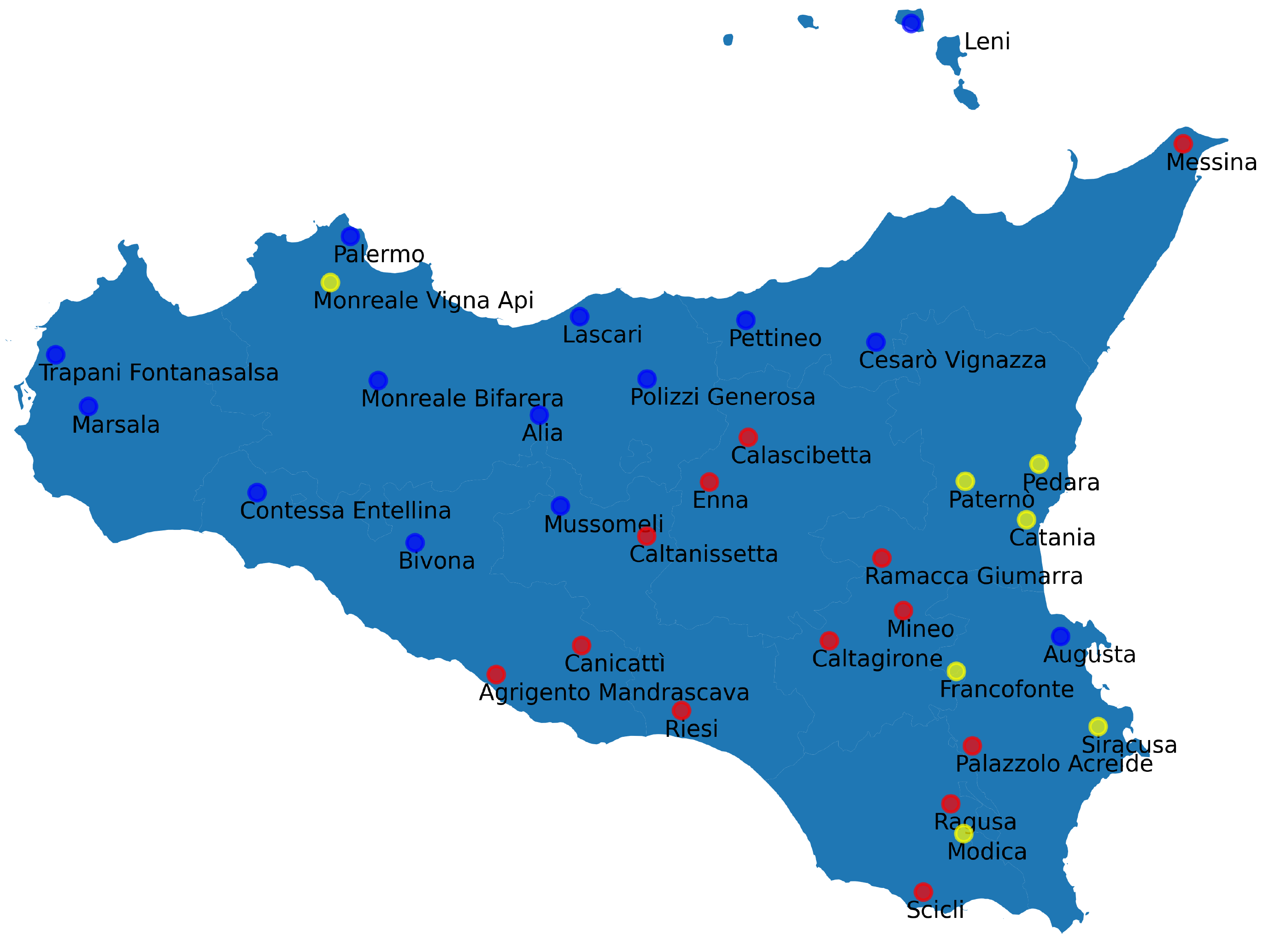}}
\end{minipage}\par\medskip
\caption{Annual case - 2016\\ \textbf{Panel a}: The principal cluster is reported in purple. \textbf{Panel b}: The four clusters are reported in blue, red, yellow and green. \textbf{Panel c}: The five principal clusters are reported in purple, blue, light blue, green and yellow. \textbf{Panel d}: The three clusters are reported in blue, red and yellow.}
\label{annual2016}
\end{figure}

\begin{figure}[h!]
\begin{minipage}{.5\linewidth}
\centering
\subfloat[Euclidean metrics and C.$A$]{\label{2017ea}\includegraphics[width = 3.2in]{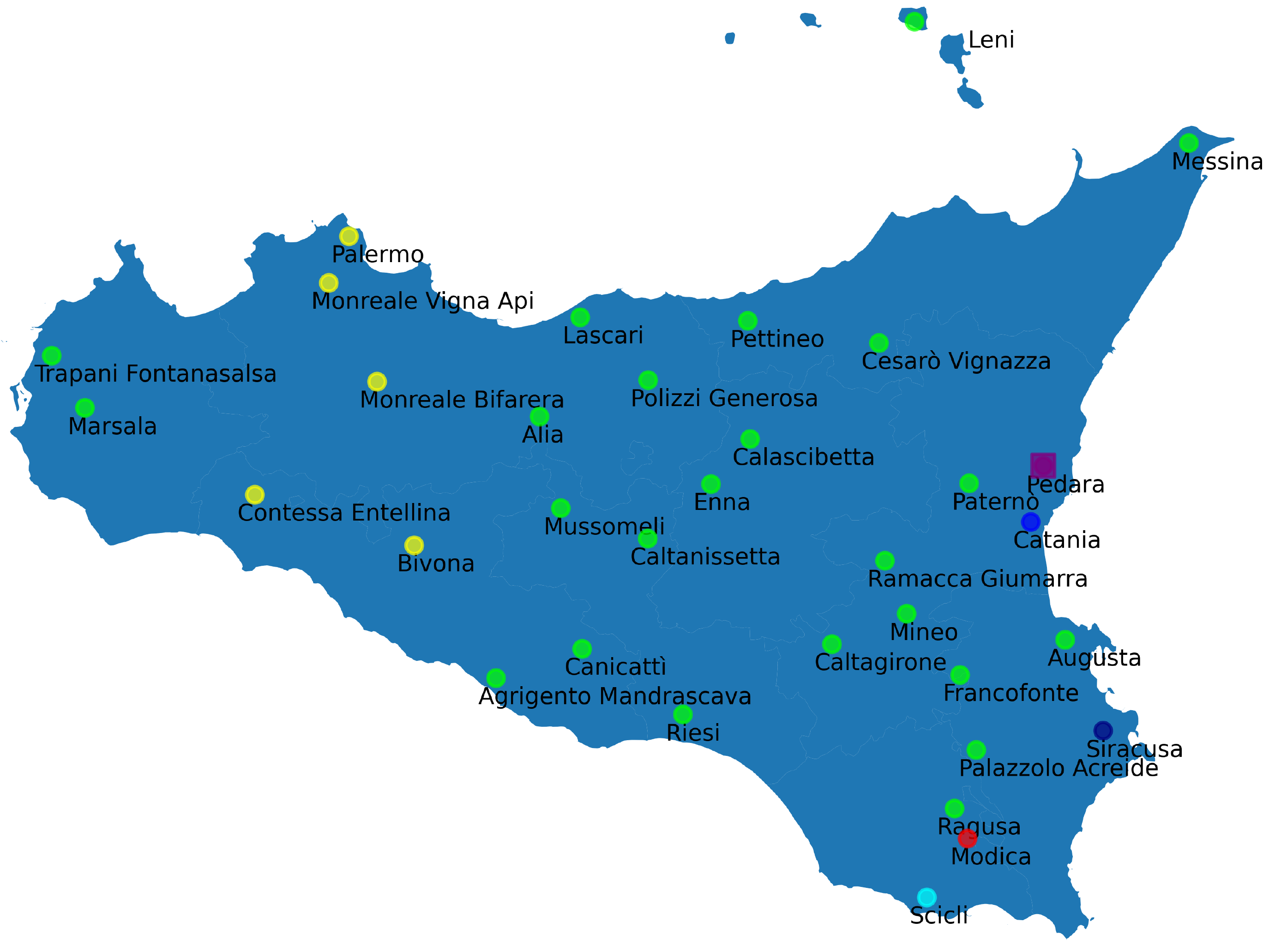}}
\end{minipage}%
\begin{minipage}{.5\linewidth}
\centering
\subfloat[Correlation metrics and C.$A$]{\label{2017ca}\includegraphics[width = 3.2in]{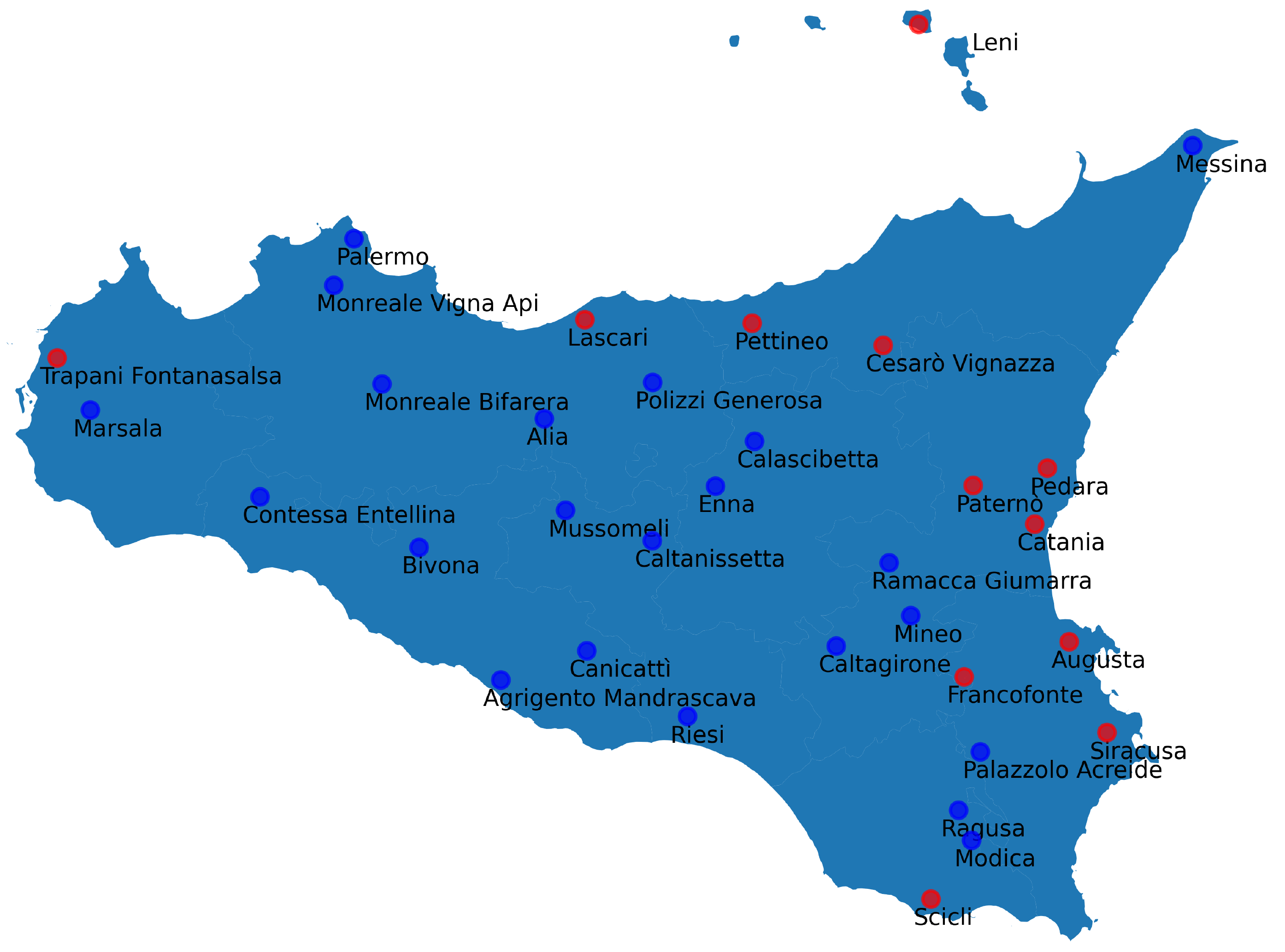}}
\end{minipage}\par\medskip
\begin{minipage}{.5\linewidth}
\centering
\subfloat[Euclidean metrics and C.$B$]{\label{2017eb}\includegraphics[width = 3.2in]{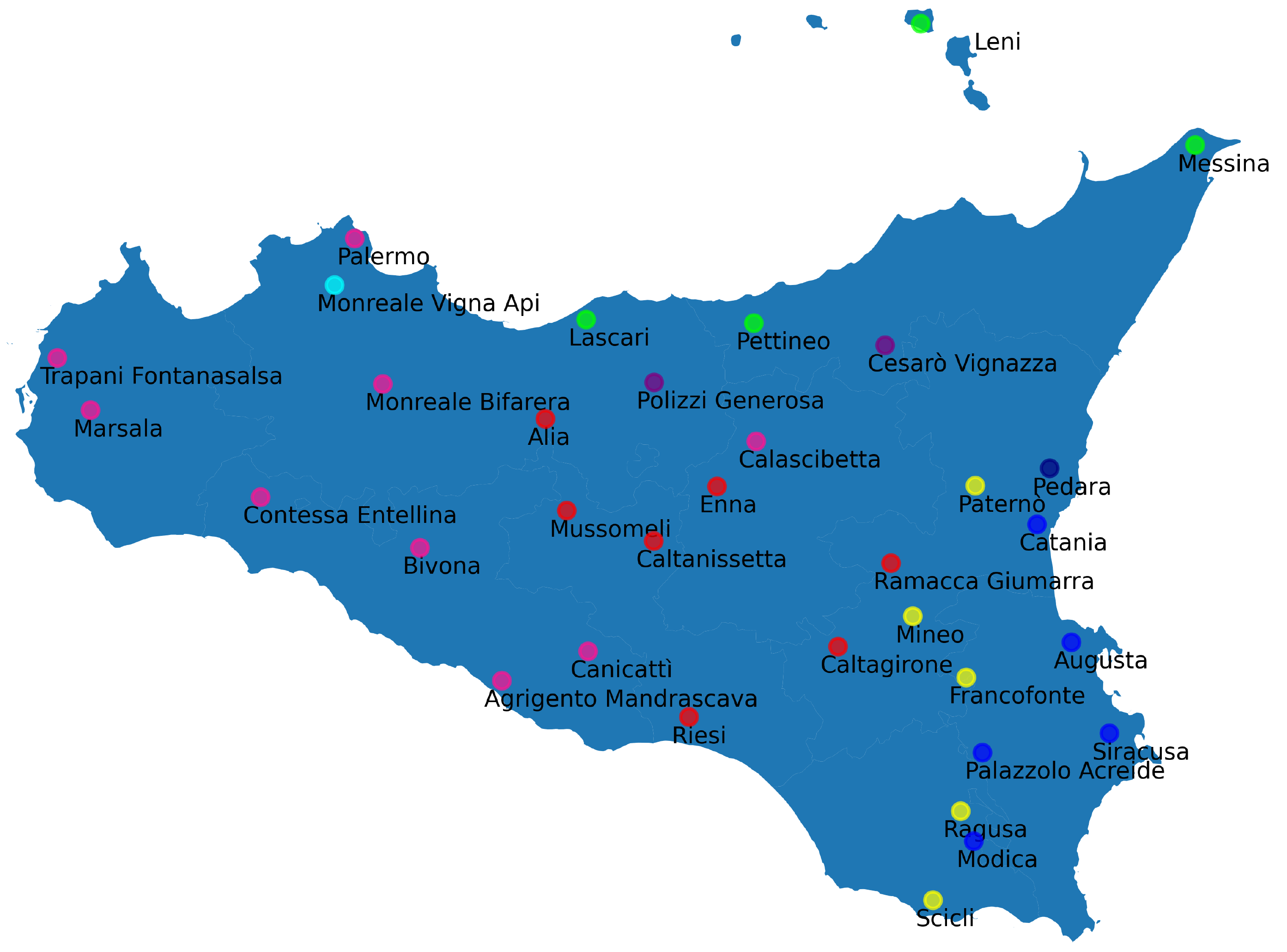}}
\end{minipage}%
\begin{minipage}{.5\linewidth}
\centering
\subfloat[Correlation metrics and C.$B$]{\label{2017cb}\includegraphics[width = 3.2in]{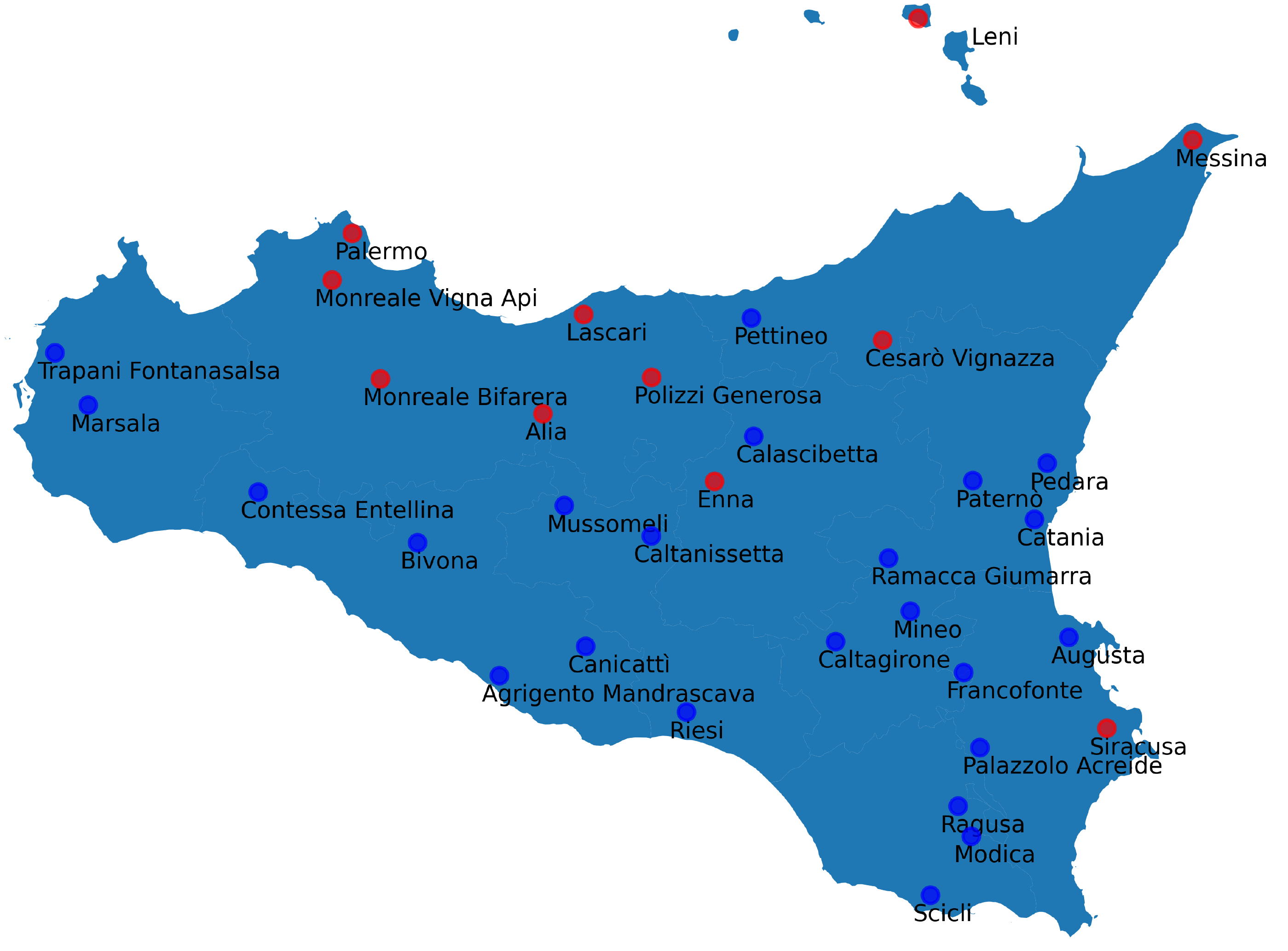}}
\end{minipage}\par\medskip
\caption{Annual case - 2017\\ \textbf{Panel a}: The two main clusters are reported in green and yellow. \textbf{Panel b}: The two clusters are reported in red and blue. \textbf{Panel c}: The six main clusters are reported in pink, red, yellow, blue, purple and green. \textbf{Panel d}: The two clusters are reported in red and blue.}
\label{annual2017}
\end{figure}

\begin{figure}[h!]
\begin{minipage}{.5\linewidth}
\centering
\subfloat[Euclidean metrics and C.$A$]{\label{2018ea}\includegraphics[width = 3.2in]{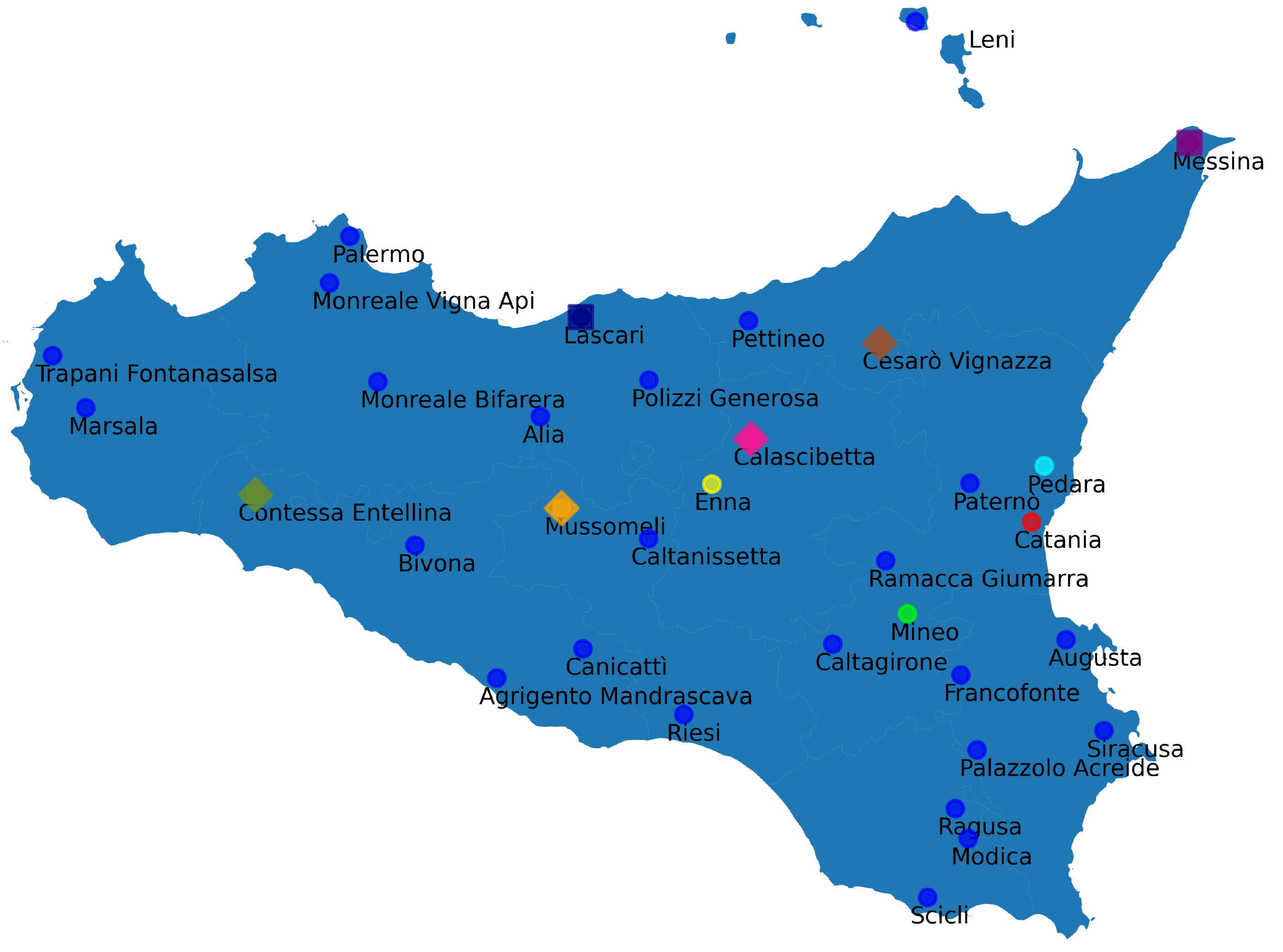}}
\end{minipage}%
\begin{minipage}{.5\linewidth}
\centering
\subfloat[Correlation metrics and C.$A$]{\label{2018ca}\includegraphics[width = 3.2in]{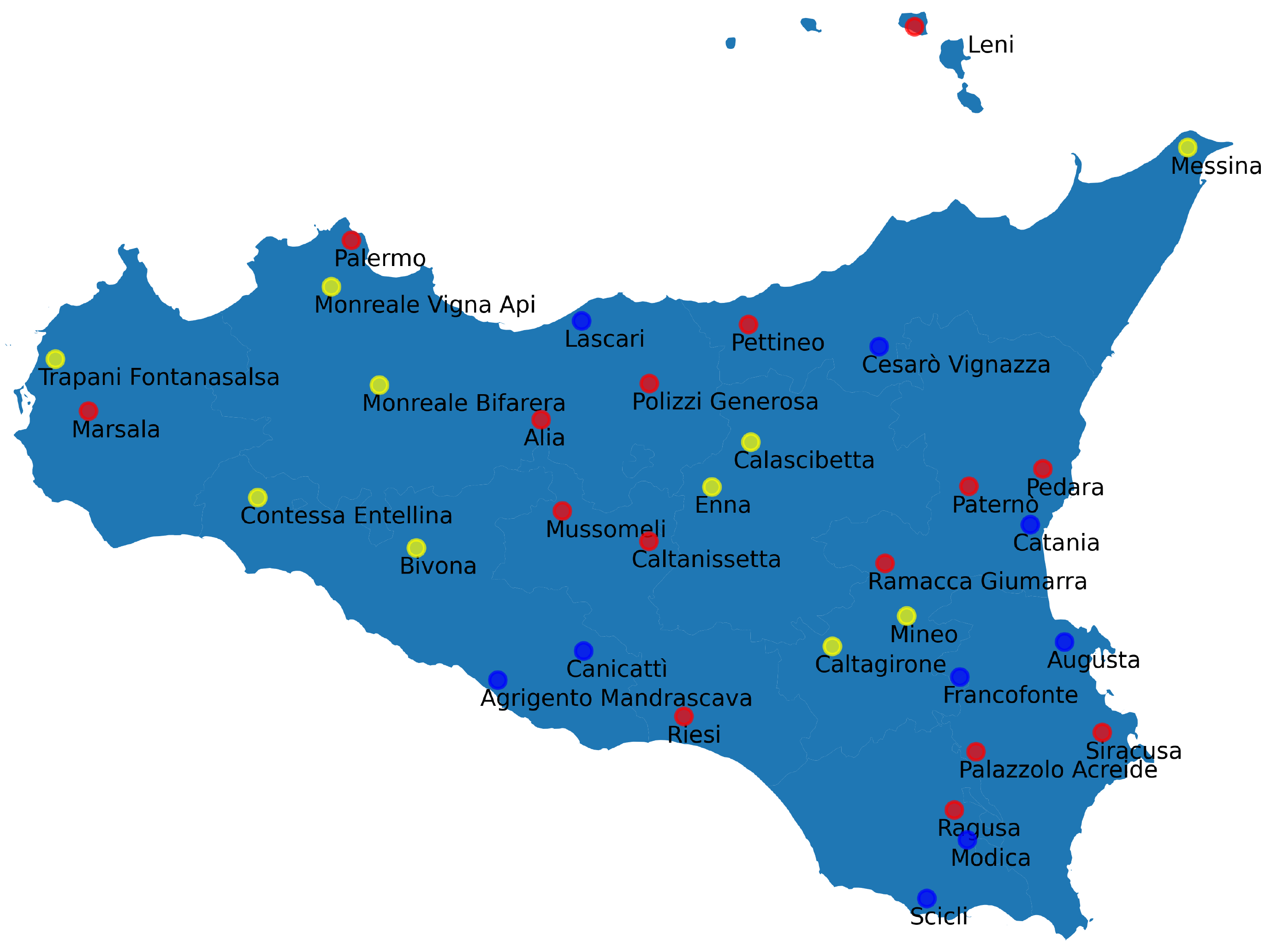}}
\end{minipage}\par\medskip
\begin{minipage}{.5\linewidth}
\centering
\subfloat[Euclidean metrics and C.$B$]{\label{2018eb}\includegraphics[width = 3.2in]{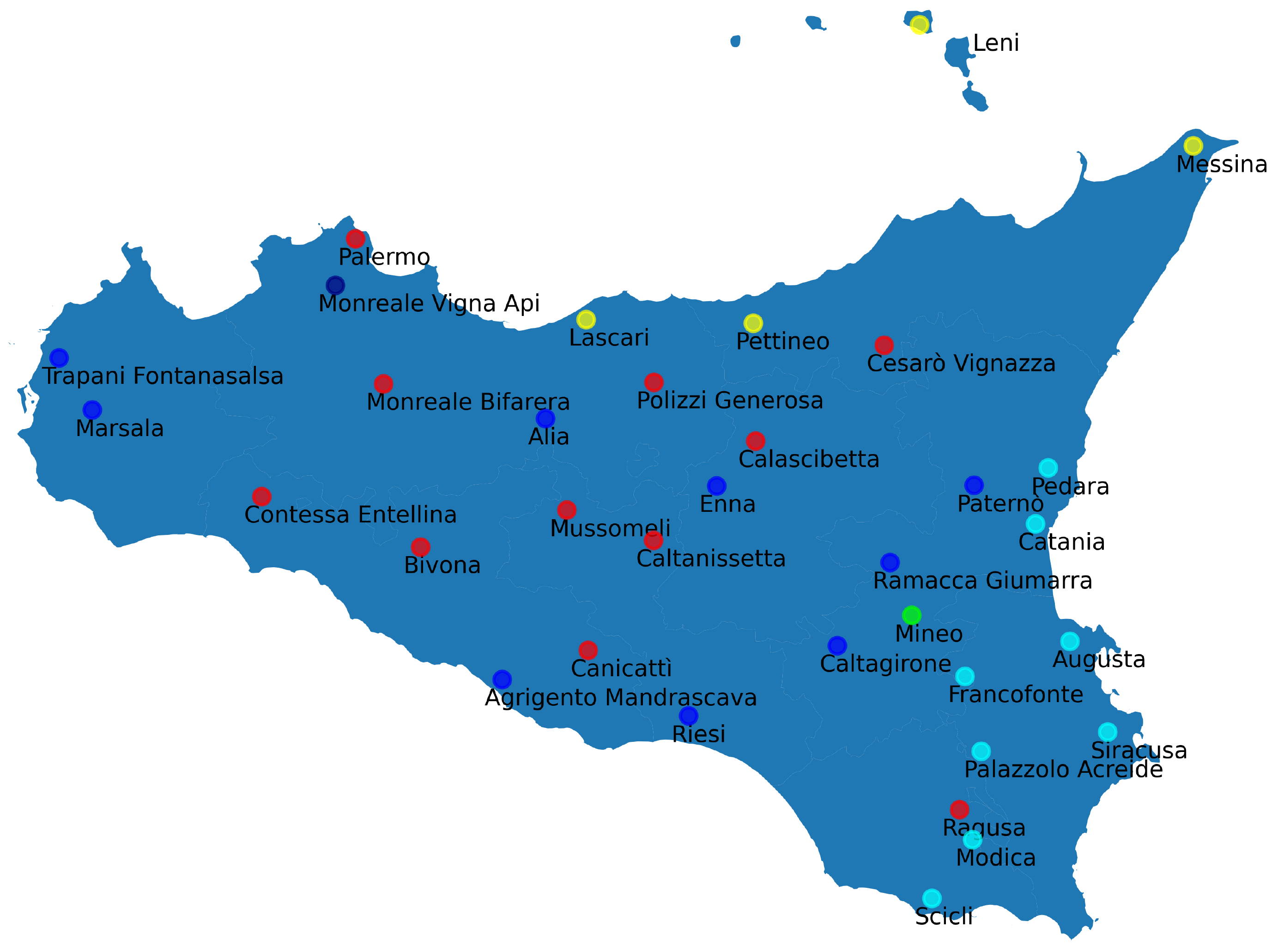}}
\end{minipage}%
\begin{minipage}{.5\linewidth}
\centering
\subfloat[Correlation metrics and C.$B$]{\label{2018cb}\includegraphics[width = 3.2in]{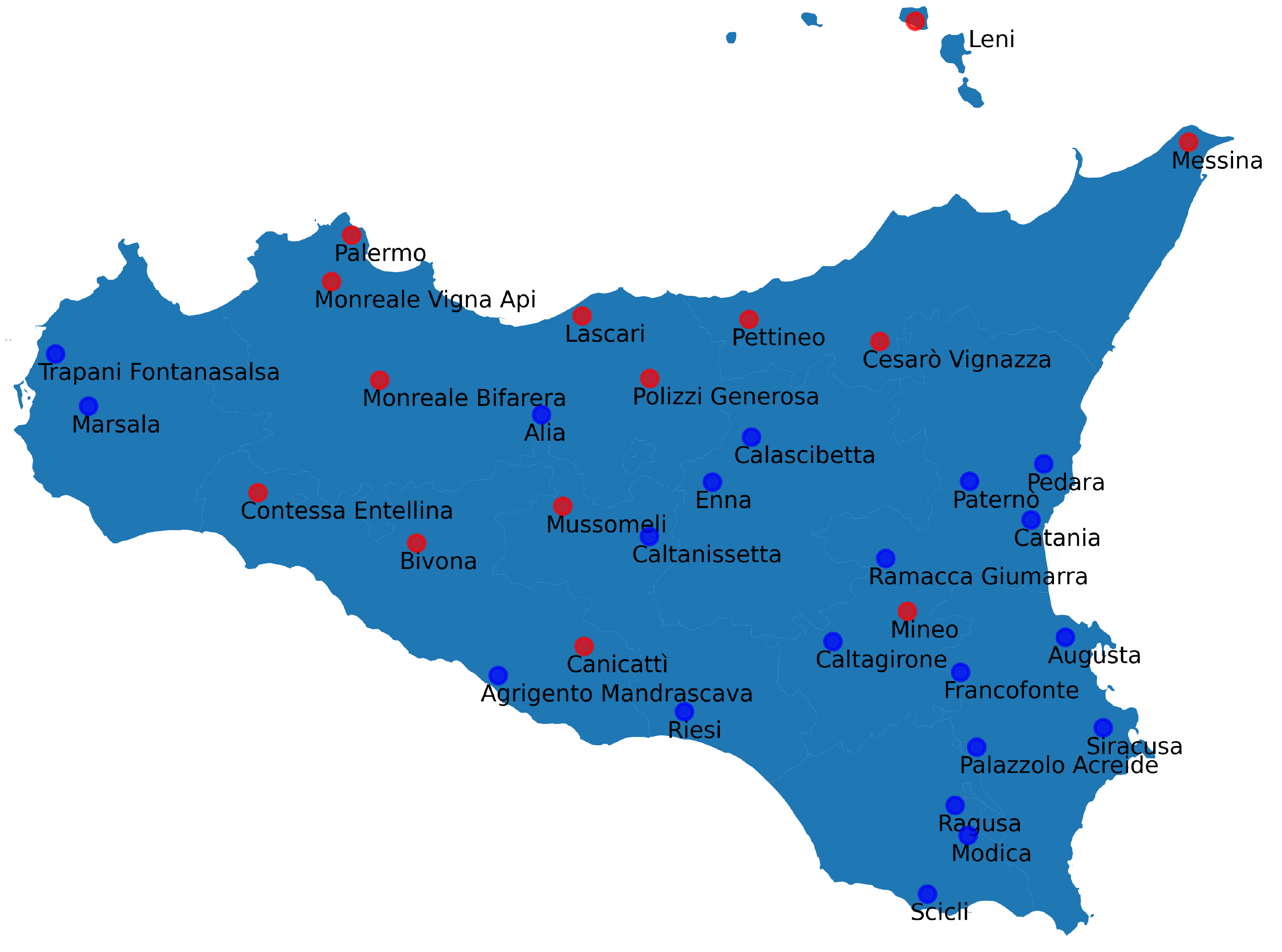}}
\end{minipage}\par\medskip
\caption{Annual case - 2018\\ \textbf{Panel a}: The principal cluster is reported in blue. \textbf{Panel b}: The three clusters are reported in red, blue and yellow. \textbf{Panel c}: The four main clusters are reported in red, blue, light blue and yellow.\textbf{Panel d}: The two clusters are reported in red and blue.}
\label{annual2018}
\end{figure}

\begin{figure}[h!]
\begin{minipage}{.5\linewidth}
\centering
\subfloat[Euclidean metrics and C.$A$]{\label{2019ea}\includegraphics[width = 3.2in]{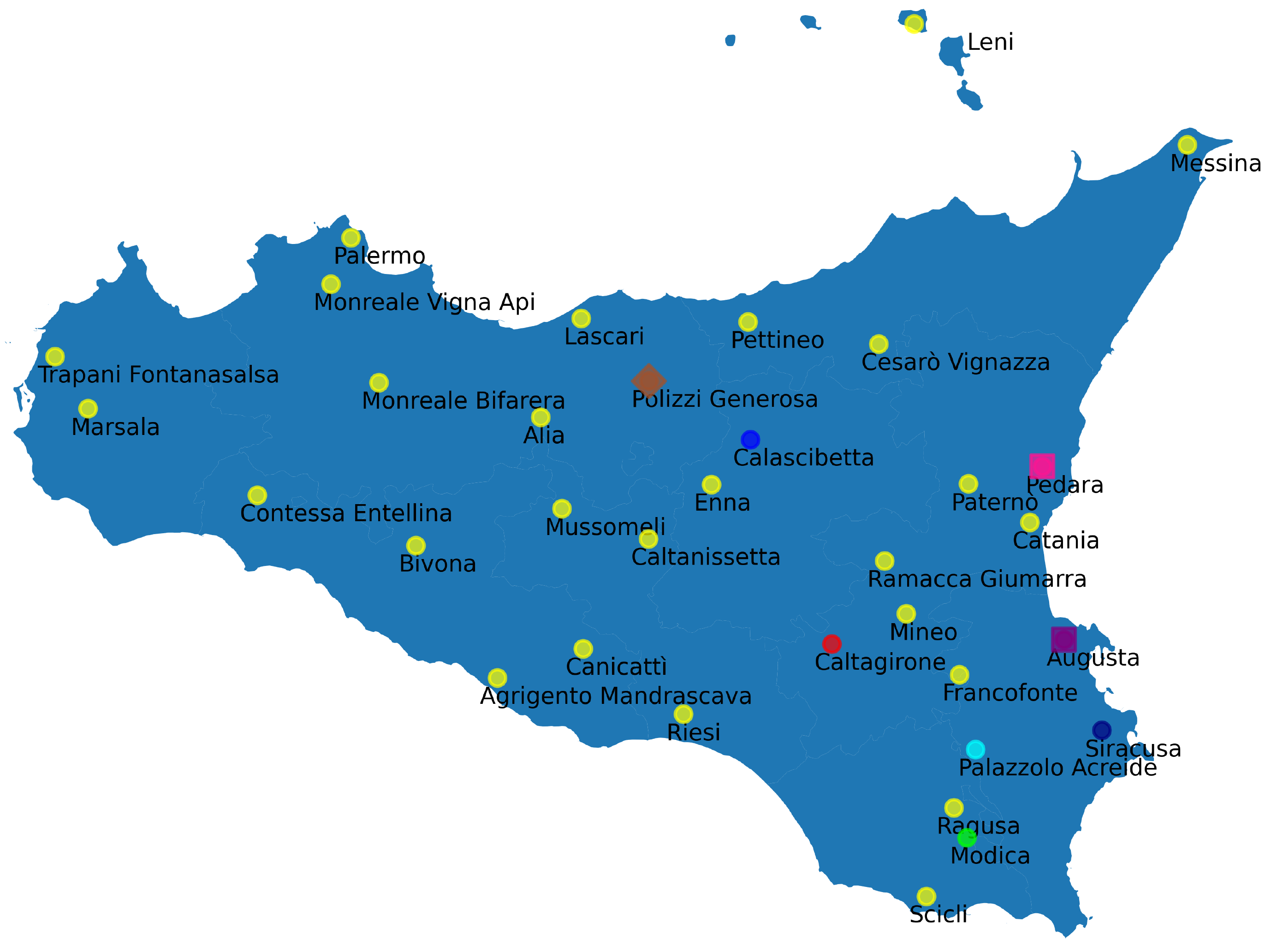}}
\end{minipage}%
\begin{minipage}{.5\linewidth}
\centering
\subfloat[Correlation metrics and C.$A$]{\label{2019ca}\includegraphics[width = 3.2in]{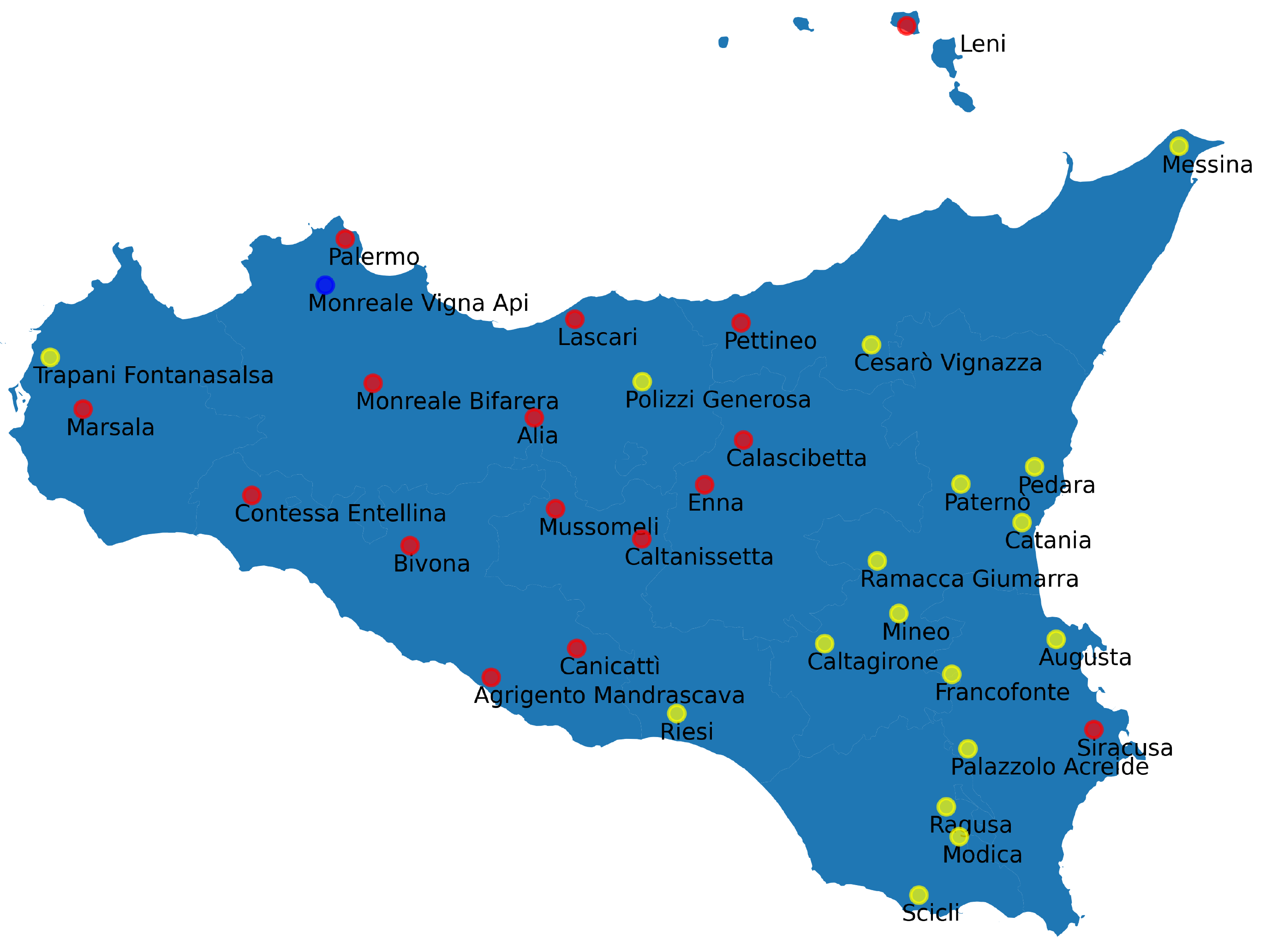}}
\end{minipage}\par\medskip
\begin{minipage}{.5\linewidth}
\centering
\subfloat[Euclidean metrics and C.$B$]{\label{2019eb}\includegraphics[width = 3.2in]{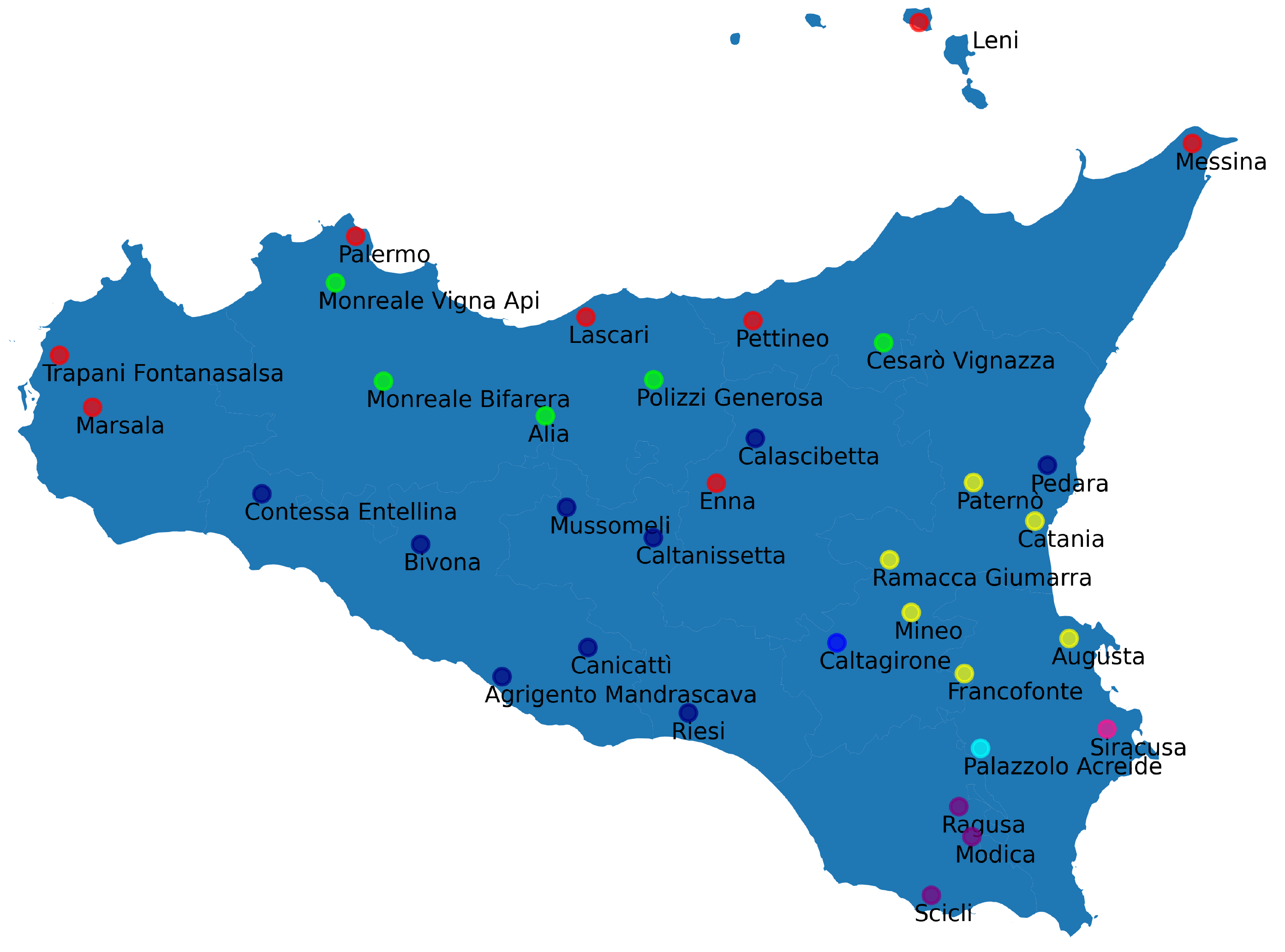}}
\end{minipage}%
\begin{minipage}{.5\linewidth}
\centering
\subfloat[Correlation metrics and C.$B$]{\label{2019cb}\includegraphics[width = 3.2in]{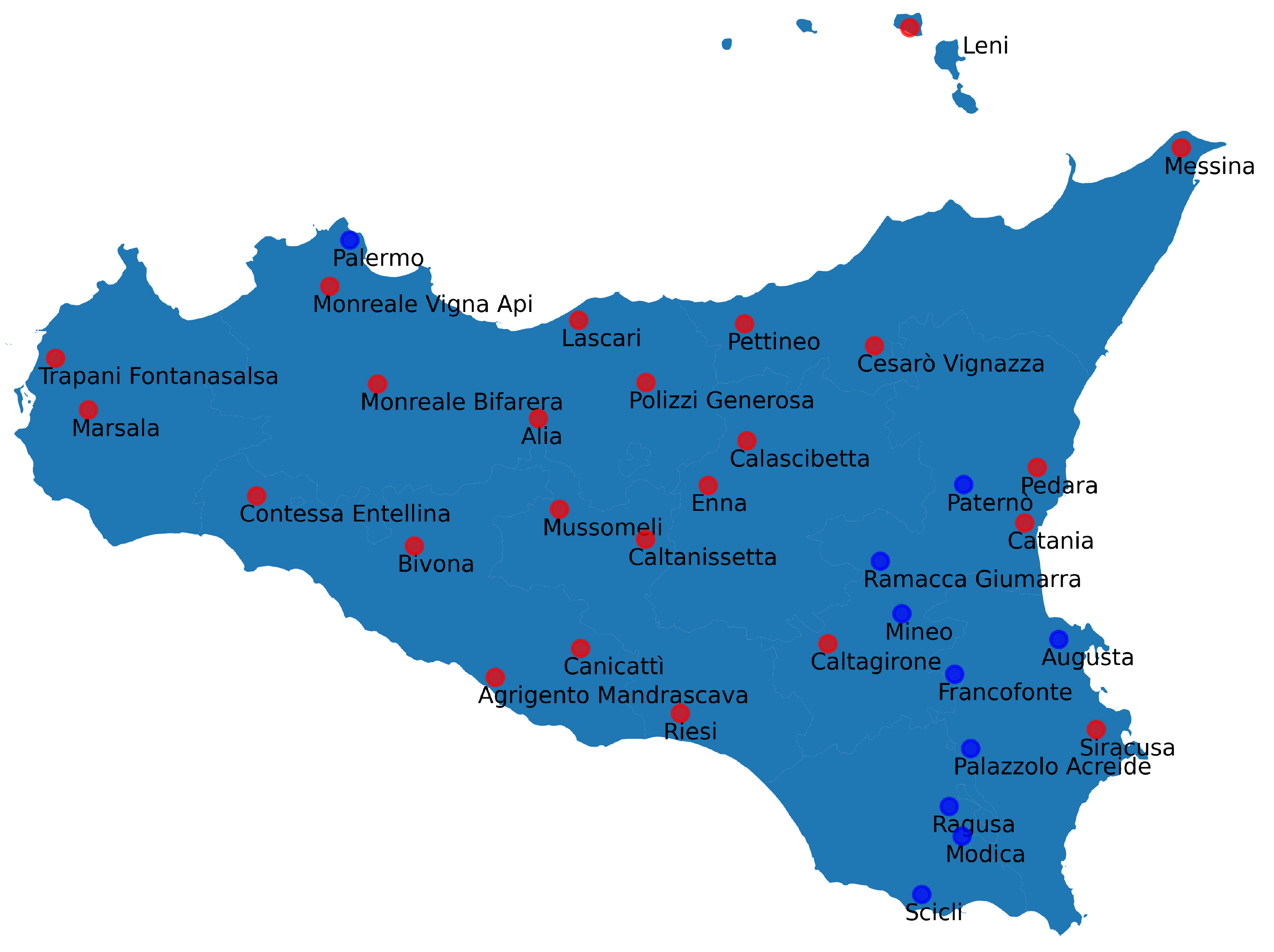}}
\end{minipage}\par\medskip
\caption{Annual case - 2019\\ \textbf{Panel a}: The principal cluster is reported in yellow. \textbf{Panel b}: The two main clusters are reported in red and yellow. \textbf{Panel c}: The five main clusters are reported in green, yellow, dark blue, purple and red. \textbf{Panel d}: The two clusters are reported in red and blue.}
\label{annual2019}
\end{figure}

\begin{figure}[h!]
\begin{minipage}{.5\linewidth}
\centering
\subfloat[Euclidean metrics and C.$A$]{\label{2020ea}\includegraphics[width = 3.2in]{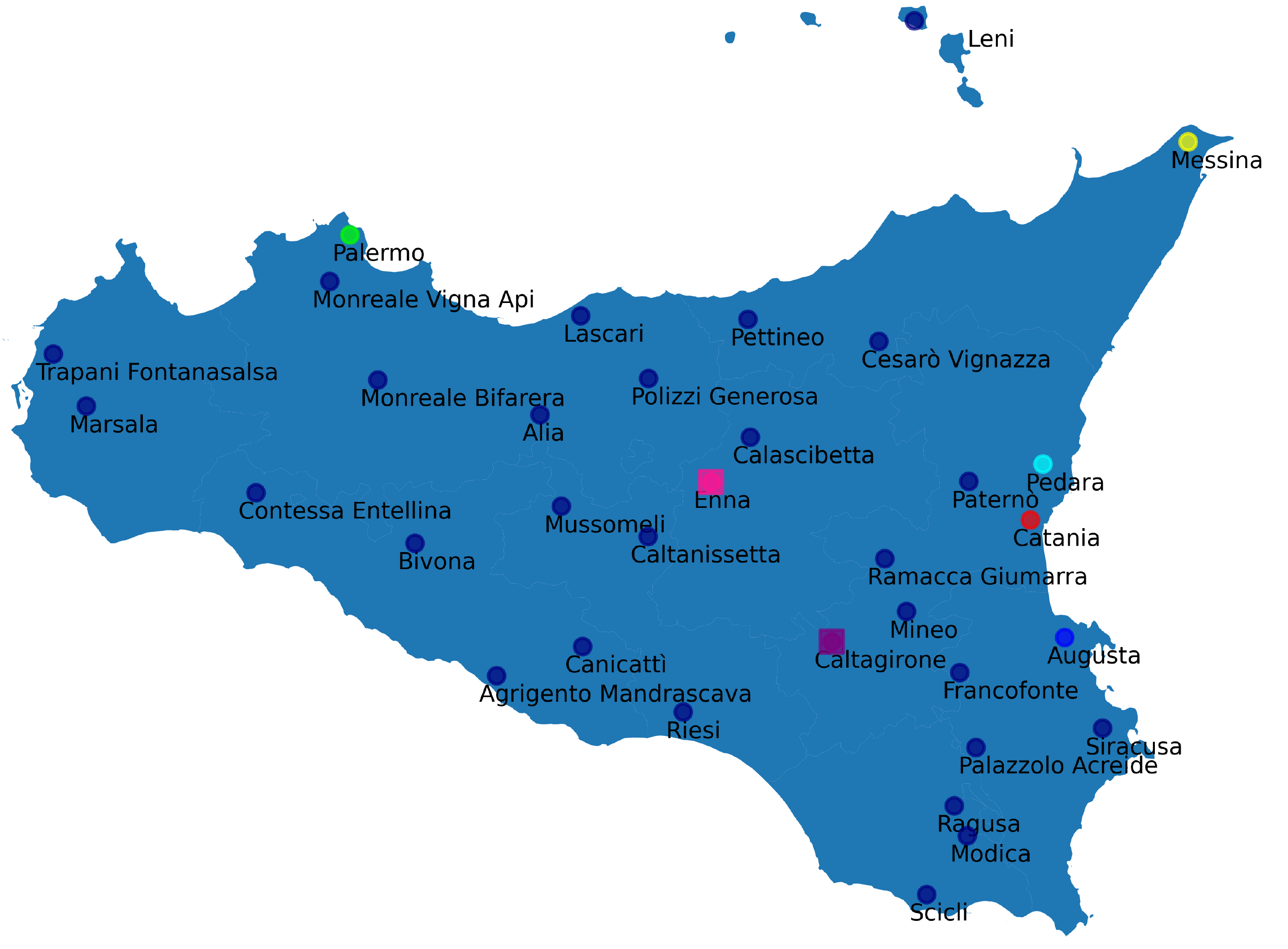}}
\end{minipage}%
\begin{minipage}{.5\linewidth}
\centering
\subfloat[Correlation metrics and C.$A$]{\label{2020ca}\includegraphics[width = 3.2in]{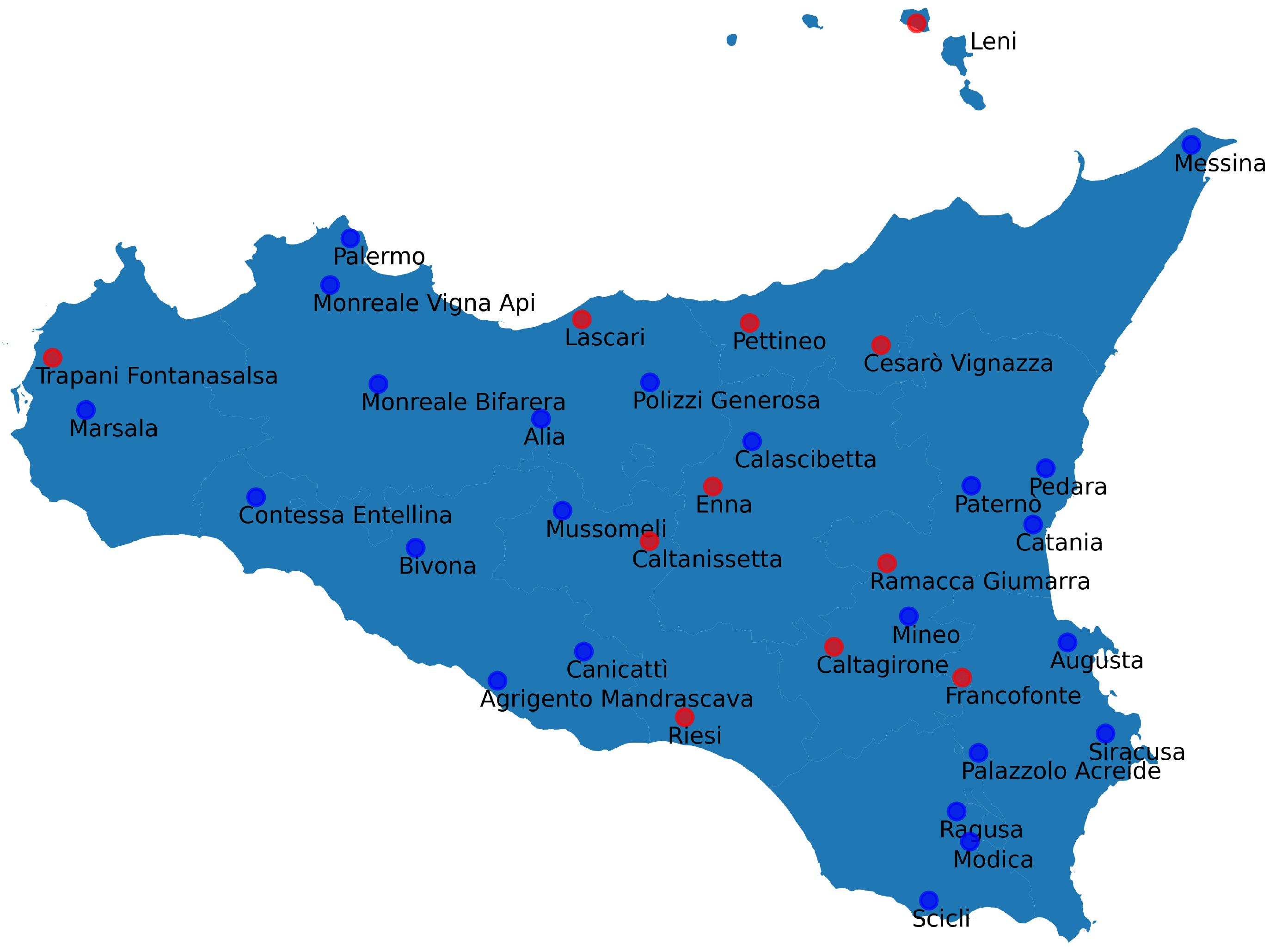}}
\end{minipage}\par\medskip
\begin{minipage}{.5\linewidth}
\centering
\subfloat[Euclidean metrics and C.$B$]{\label{2020eb}\includegraphics[width = 3.2in]{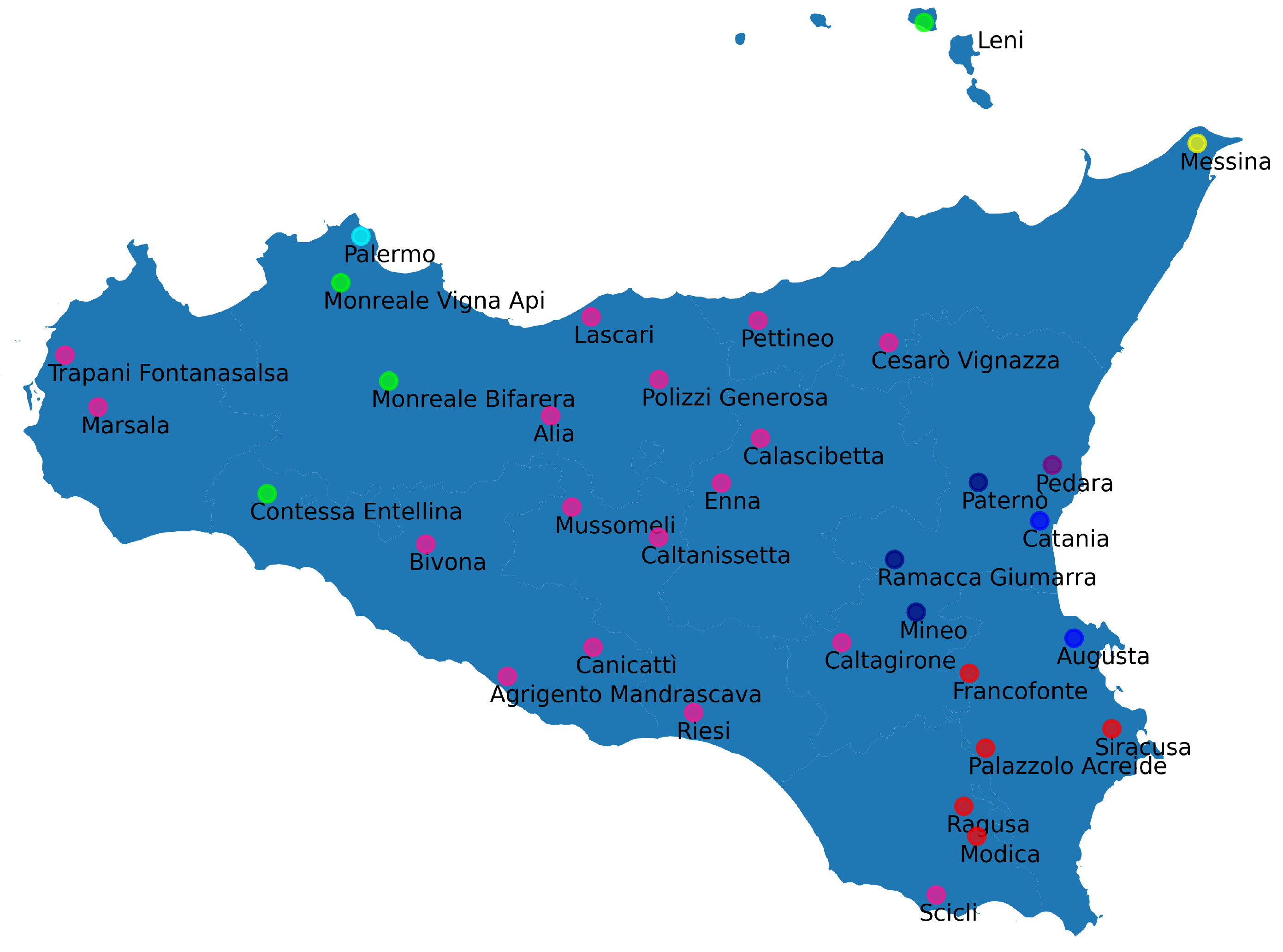}}
\end{minipage}%
\begin{minipage}{.5\linewidth}
\centering
\subfloat[Correlation metrics and C.$B$]{\label{2020cb}\includegraphics[width = 3.2in]{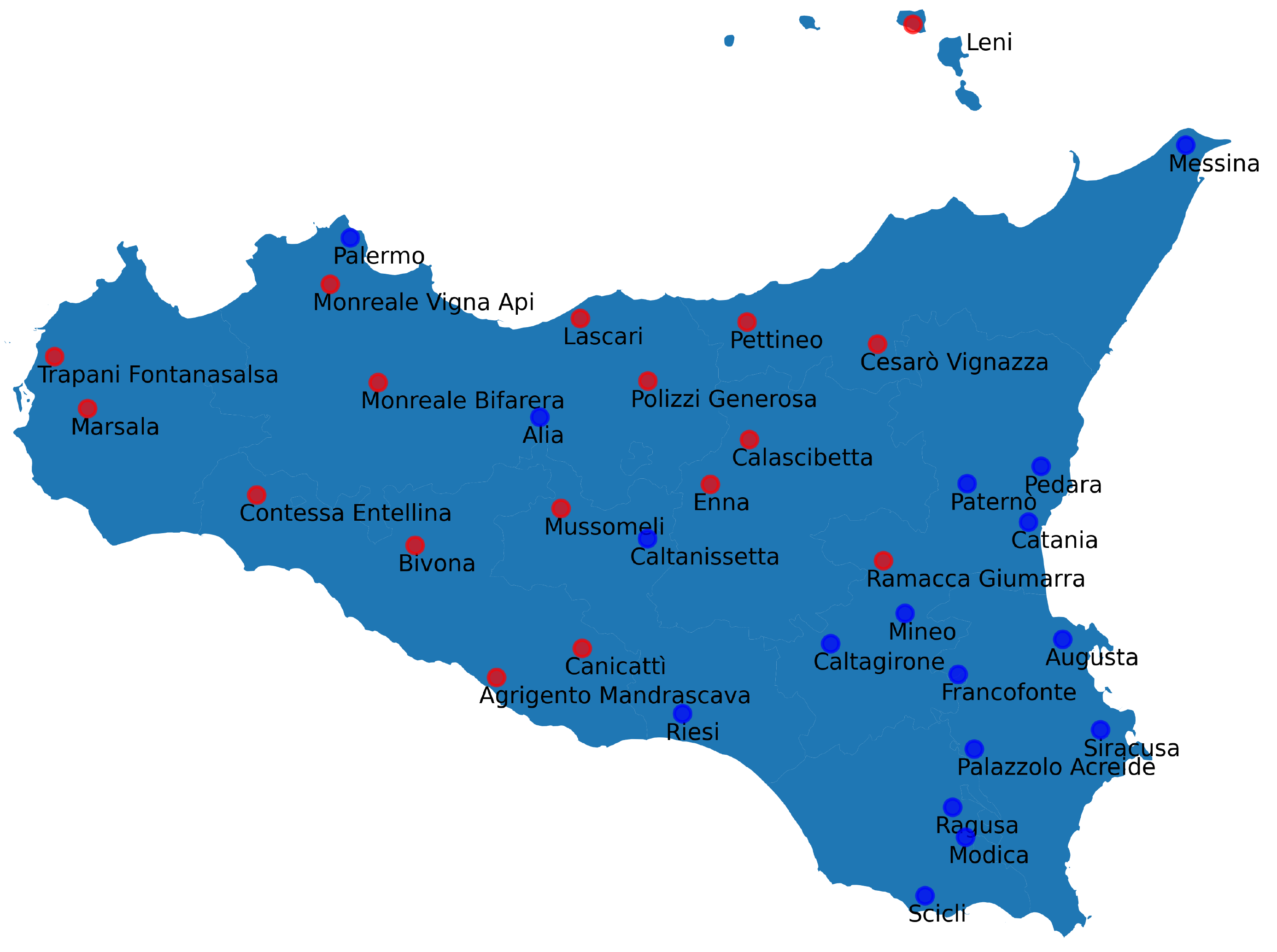}}
\end{minipage}\par\medskip
\caption{Annual case - 2020\\ \textbf{Panel a}: The main cluster is reported in dark blue. \textbf{Panel b}: The two clusters are reported in red and blue. \textbf{Panel c}: The five main clusters are reported in green, pink, red, blue and dark blue. \textbf{Panel d}: The two clusters are reported in red and blue.}
\label{annual2020}
\end{figure}

\begin{figure*}[h!]
\begin{minipage}[c]{.5\linewidth}
\centering
\subfloat[]{\label{2021:a}\includegraphics[width = 7.5cm]{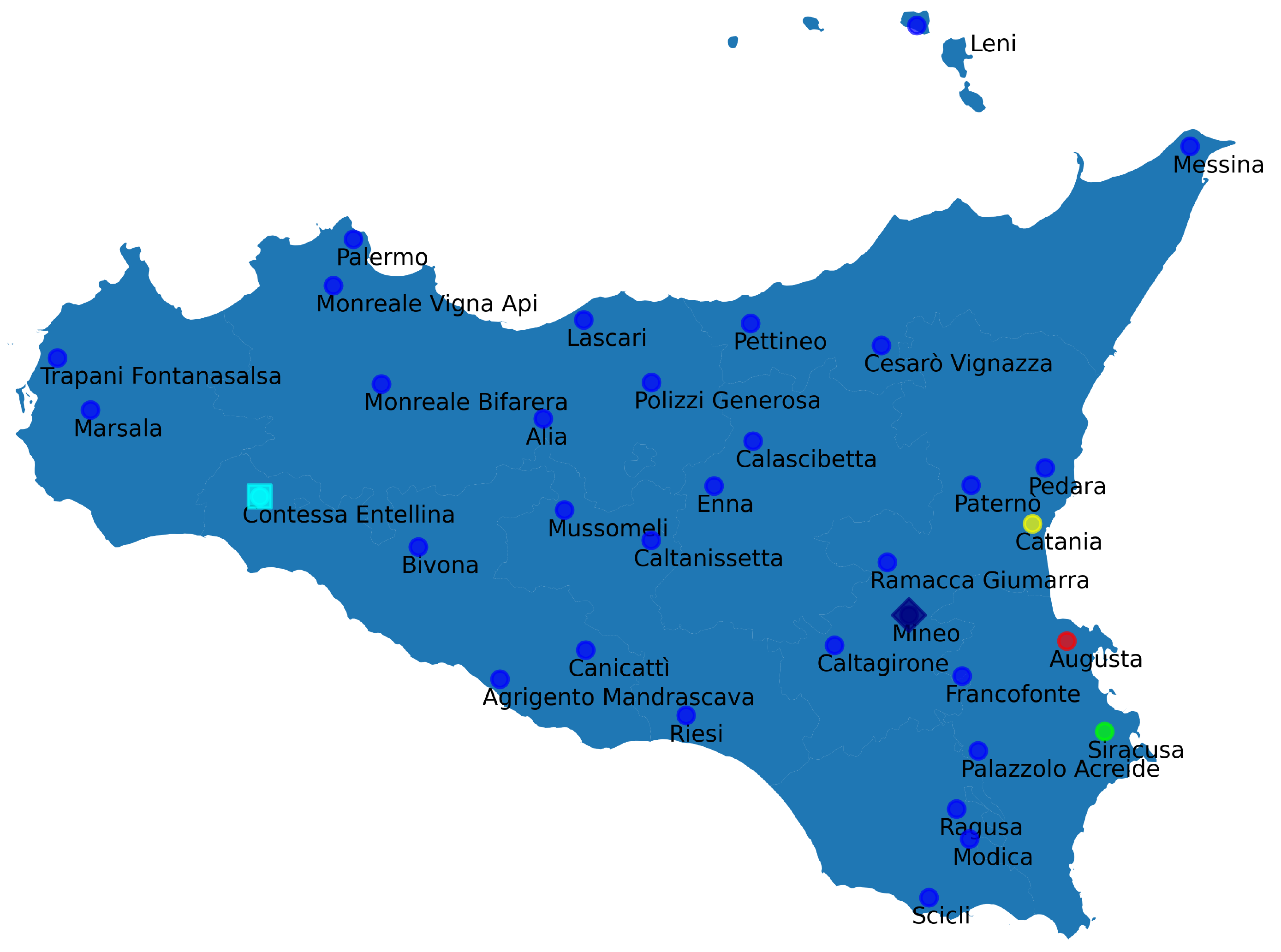}}
\end{minipage}
\begin{minipage}[c]{.5\linewidth}
\centering
\subfloat[]{\label{2021:c}\includegraphics[width = 7.5cm]{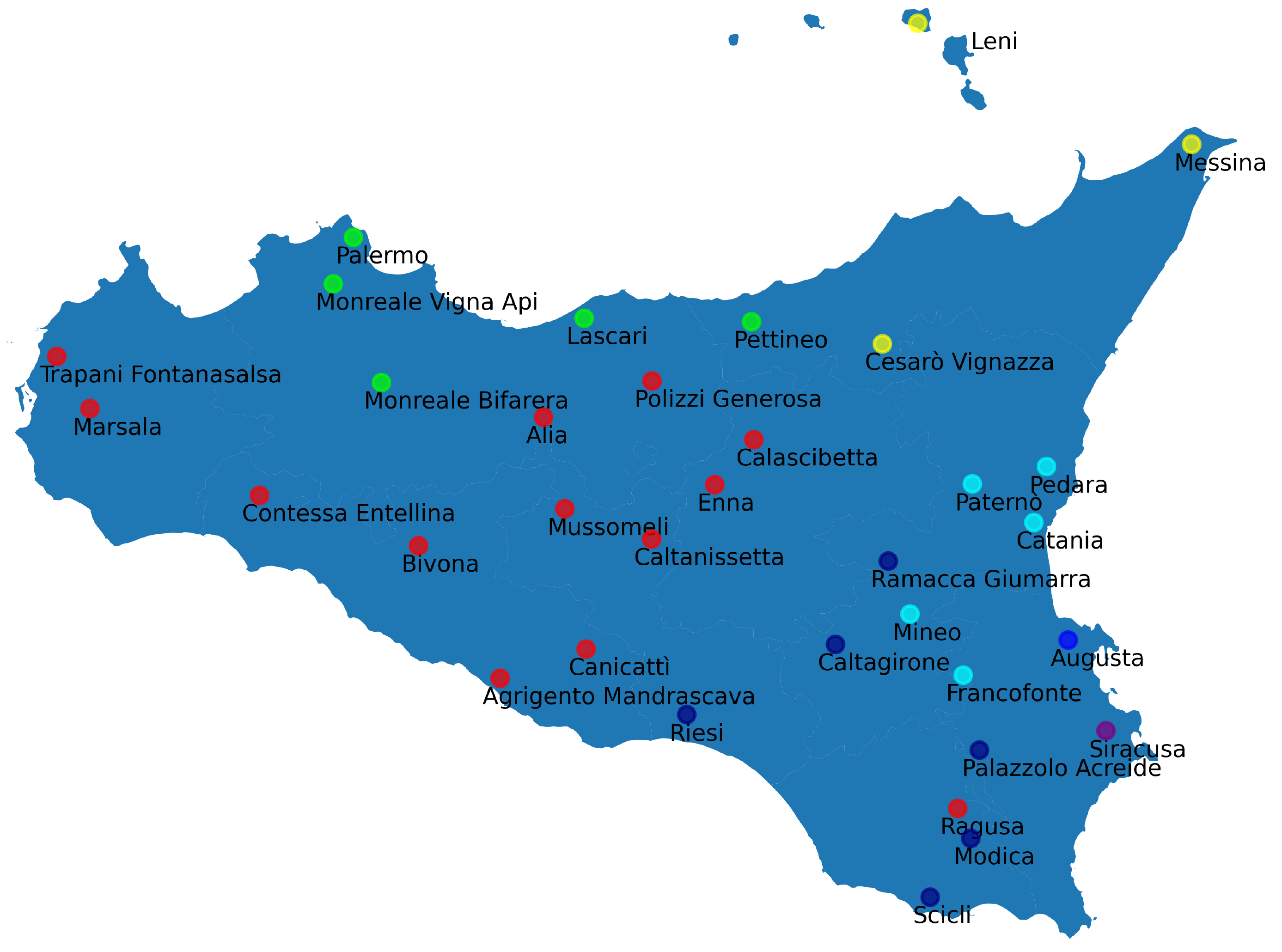}}
\end{minipage}
\begin{minipage}[c]{.5\linewidth}
\centering
\subfloat[]{\label{2021:b}\includegraphics[width = 9.2cm]{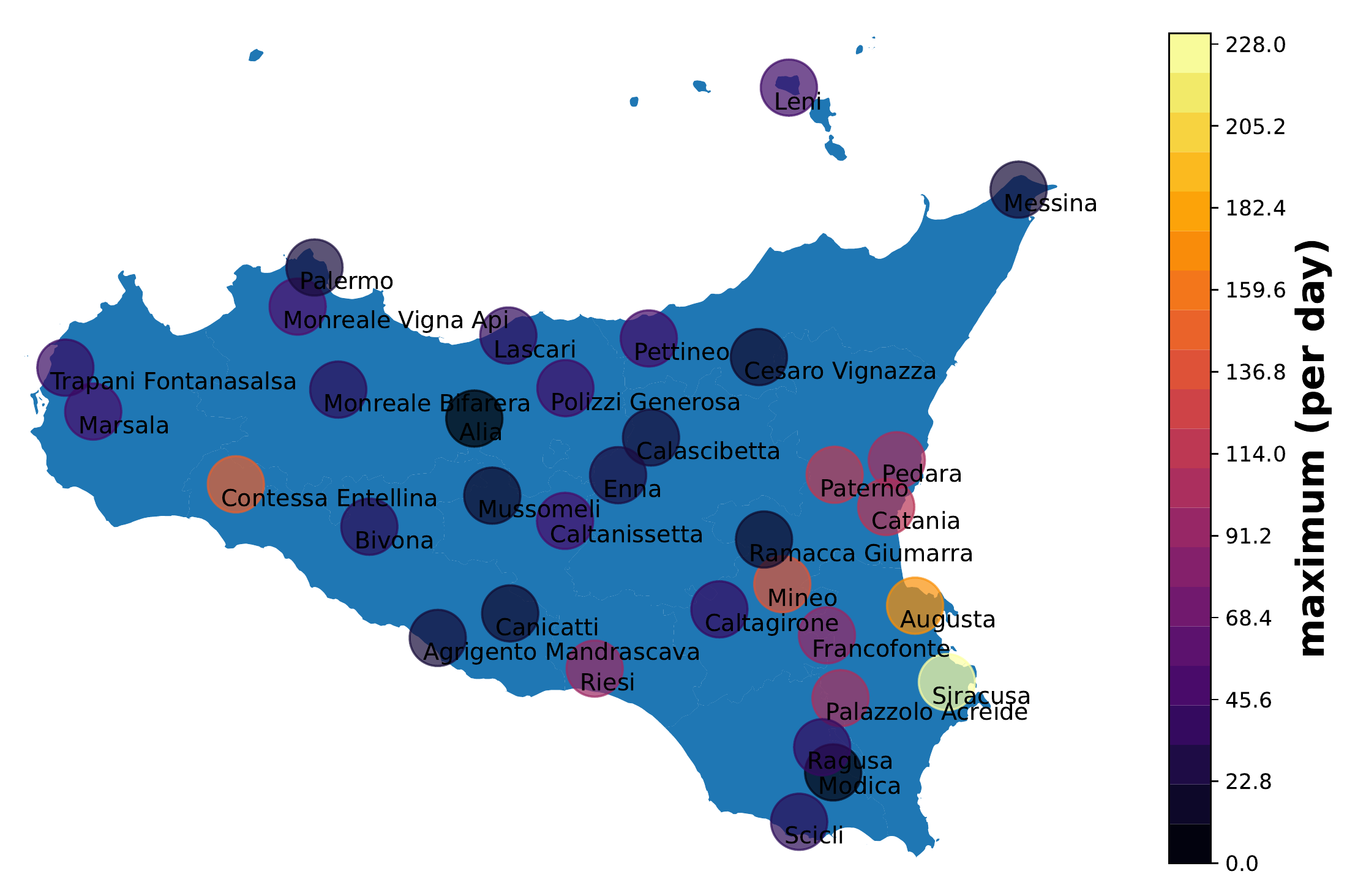}}\\
\end{minipage}
\begin{minipage}[c]{.5\linewidth}
\centering
\subfloat[]{\label{2021:d}\includegraphics[width = 7.5cm]{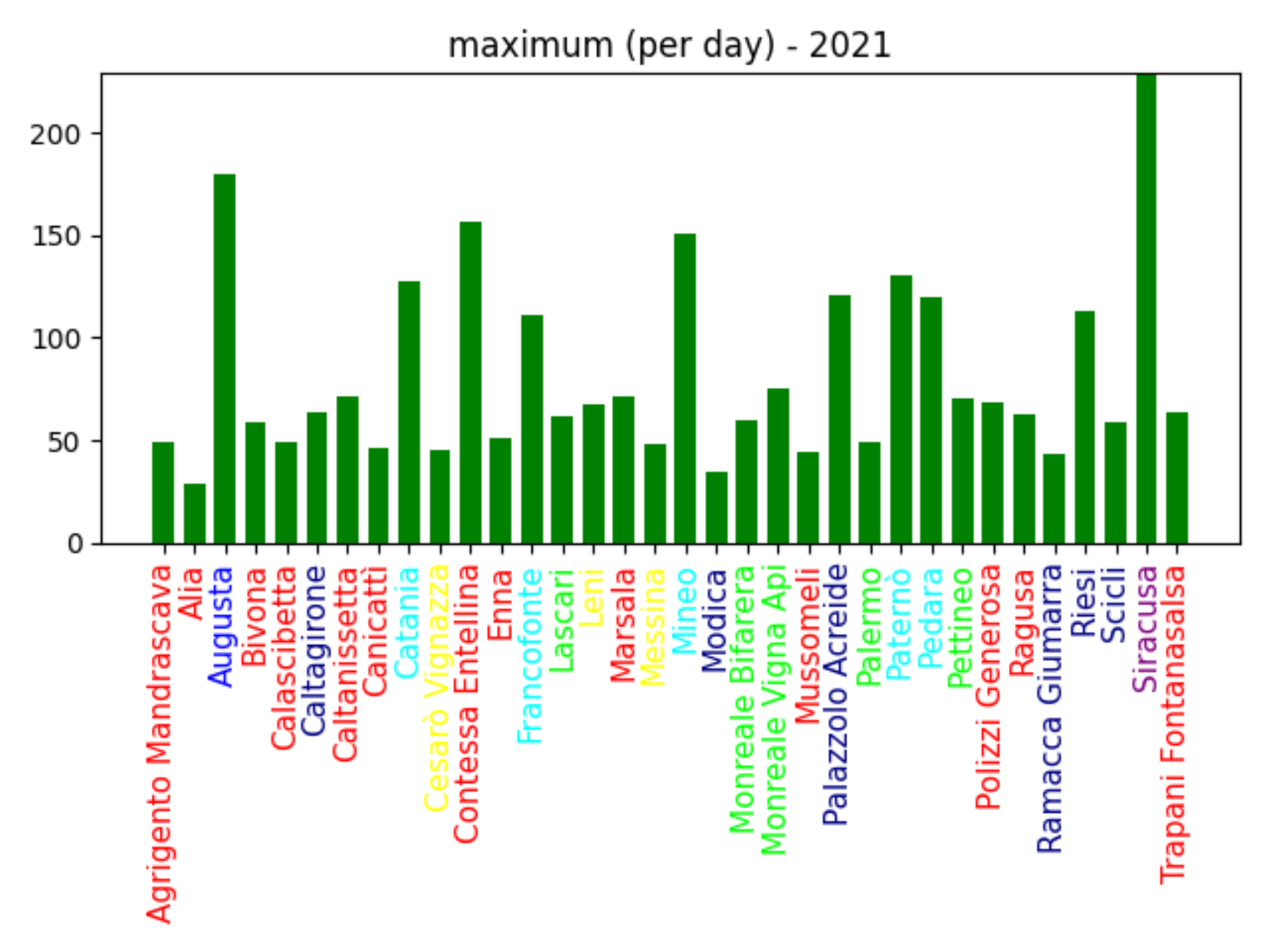}}
\end{minipage}
\caption[Annual case for 2021 - Euclidean metrics.]{Annual case for 2021 - Euclidean metrics.  Different colors represent different clusters, both in the maps and in the histogram. \textbf{(a)} C.$A$. The principal cluster is coloured blue. Square and diamond points indicate clusters obtained by the second and the third iteration of the algorithm, respectively. \textbf{(b)} C.$B$. The five principal clusters are coloured red, green, blue, light blue and yellow.
\textbf{(c)} Heat-map of the \textit{md} indicator for both C.$A$ and C.$B$.
\textbf{(d)} Histogram of the \textit{md} indicator with labels coloured as the C.$B$ clustering.}\label{fig:2021_1}
\end{figure*}
\begin{figure*}[h!]
\begin{minipage}{.5\linewidth}
\centering
\subfloat[]{\label{2021:c1}\includegraphics[width = 7.5cm]{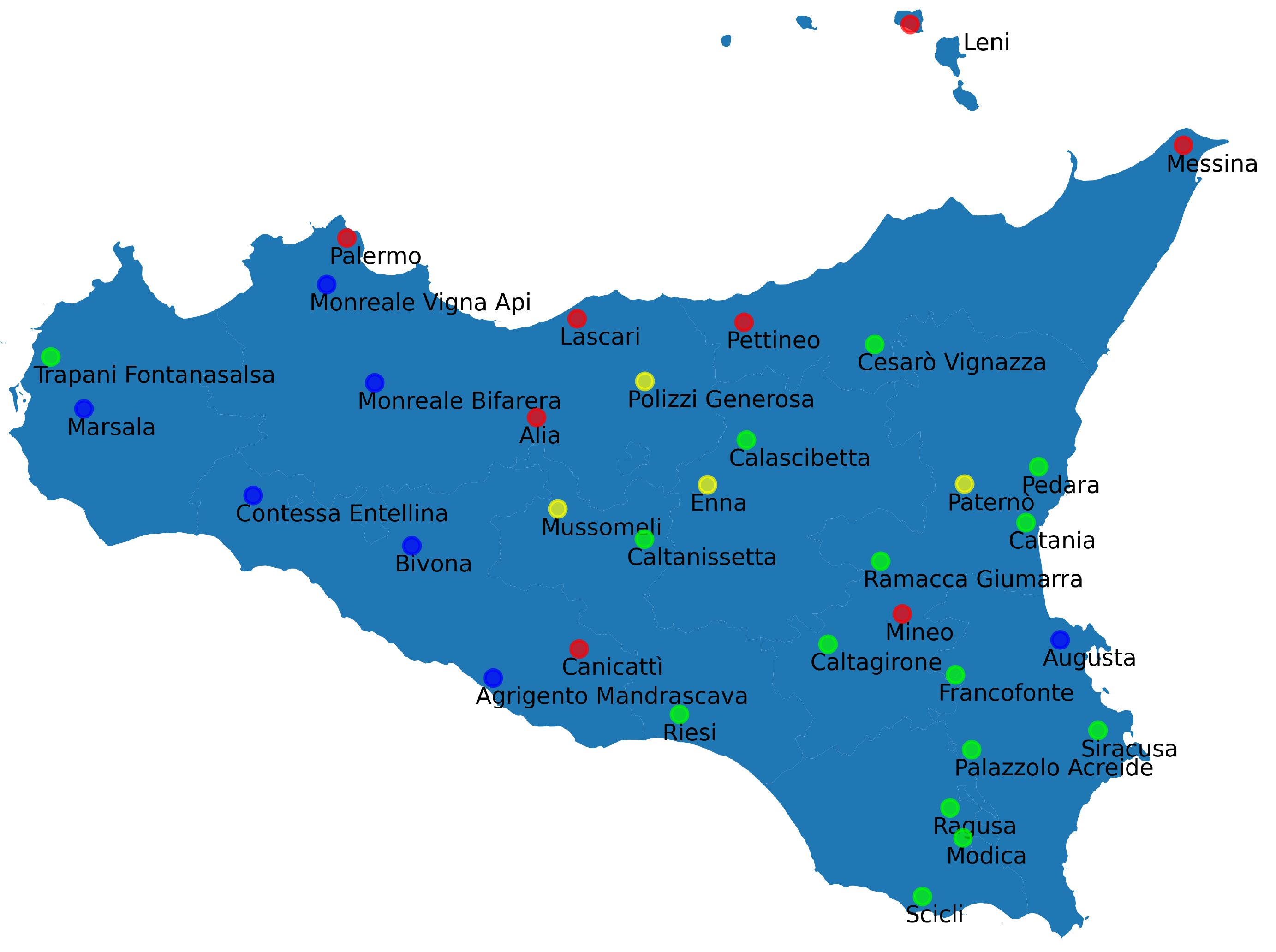}}
\end{minipage}%
\begin{minipage}{.5\linewidth}
\centering
\subfloat[]{\label{2021:c2}\includegraphics[width = 7.5cm]{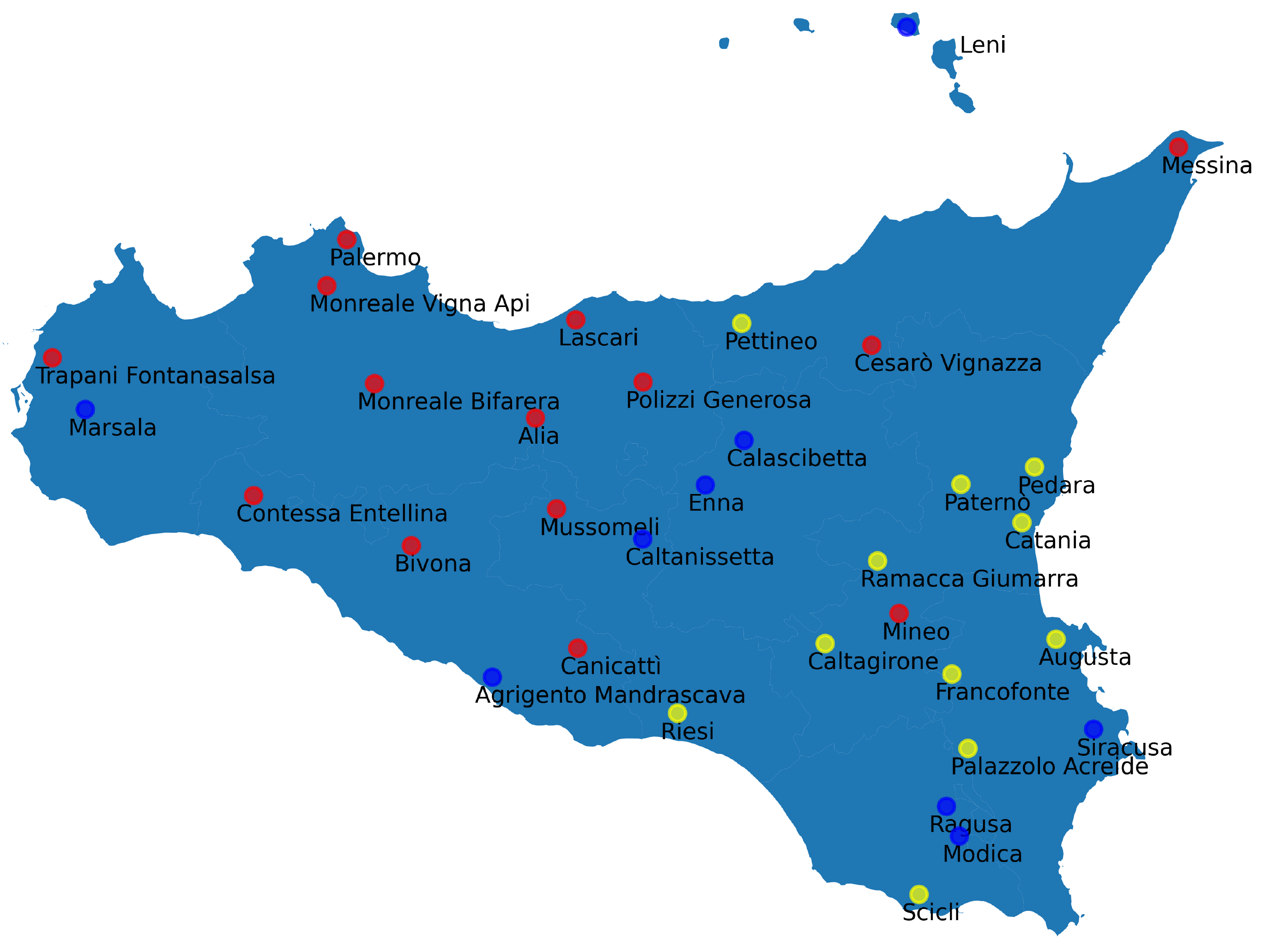}}
\end{minipage}
\caption[Annual case - 2021 - Correlation metrics.]{Annual case - 2021 - Correlation metrics.  \textbf{(a)} C.$A$. The colours of the four clusters are blue, red, yellow and green.  \textbf{(b)} C.$B$. The three clusters are coloured red, blue and yellow.}\label{fig:2021_2}
\end{figure*}
Figure \ref{fig:2021_1} shows the exceptional rainfall events occurred in East Sicily in 2021.
In fact, among the characterizing indicators obtained with the procedure explained in the main document, $md$ (maximum per day) is found to be particularly relevant. Panel \ref{2021:a} shows the presence of a principal cluster and some anomalies, for instance \textit{Catania}, \textit{Augusta} and \textit{Siracusa}. Differently, panel \ref{2021:c} presents several principal clusters distributed in the North, in the center, in the North-East, in the South-East and in the eastern center; in this case only \textit{Augusta} and \textit{Siracusa} are clustered as anomalous by the algorithm. Panels \ref{2021:b} and \ref{2021:d} represent the \textit{md} indicator geographically referenced and value-based, respectively.
Comparing independently panels \ref{2021:a} and \ref{2021:c}, with panel \ref{2021:b}, we observe a coincidence between the maximum values and the anomalies in the clusters. In particular, most of the anomalies in panel \ref{2021:a} represent the highest values of the \textit{md} indicator in panel \ref{2021:b} or in panel \ref{2021:d}. The same happens with panels \ref{2021:c} for the two anomalies of \textit{Augusta} and \textit{Siracusa} and for the locations in the \textit{light blue} cluster in panel \ref{2021:c}, which show the second highest values of the \textit{md} indicator, with  the only exception of \textit{Contessa Entellina}.
The same statement, not reported here, has been found also for the $mv$ (maximum daily variation) indicator. Therefore, in 2021 the anomalous clusters consist of the stations with the highest $md$ values. Moreover, East Sicily emerges as the most \textit{extreme} zone of the island.

On the other hand Figure \ref{fig:2021_2} shows that using the Correlation metrics, no coincidences between characterizing indicators and clusters are found for the year 2021. Actually, this happens in all of the other annual cases and in the full cases as well, in agreement with the fact that Correlation metric is less sensitive to outliers than the Euclidean one.